\documentclass[preprint,3p,12pt,sort&compress]{elsarticle}
\usepackage{mathrsfs}
\usepackage{amsmath}
\usepackage{stmaryrd}
\usepackage{bbding}
\usepackage{dcolumn}
\usepackage{graphicx}
\usepackage{amsfonts}
\usepackage{amssymb}
\usepackage{psfrag}
\usepackage{wrapfig}
\usepackage{subfigure}
\usepackage{makeidx}
\usepackage{bm}
\usepackage{epsf}
\usepackage{epsfig}
\usepackage{setspace}
\usepackage{graphicx}
\usepackage{epstopdf}
\usepackage{psfrag}
\usepackage{subfigure}
\usepackage{color}
\usepackage{enumerate}
\usepackage{verbatim} % comment-environment
\usepackage{booktabs}
\usepackage[utf8]{inputenc}

\usepackage{tikz}
\usetikzlibrary{positioning,calc}
\usetikzlibrary{arrows,shapes}
\tikzstyle{block} = [rectangle,rounded corners,thin,align=center,fill=green!20,draw=black!20]
\tikzstyle{line} = [-latex]

\newcommand{\abs}[1]{\left|#1\right|}
\newcommand{\menge}[1]{\left\lbrace#1\right\rbrace}

\newcommand{\rom}[1]{\uppercase\expandafter{\romannumeral #1\relax}}

\newcommand{\inner}[1]{\left< #1 \right>}

\usepackage[ruled,vlined]{algorithm2e}
\SetKwRepeat{Do}{do}{while}

\begin{document}

\title{Predicting continuum breakdown with deep neural networks}

\author{Tianbai Xiao\corref{cor}}
\ead{tianbaixiao@gmail.com}
\author{Steffen Schotthöfer}
%\ead{steffen.schotthoefer@kit.edu}
\author{Martin Frank}
%\ead{martin.frank@kit.edu}
\cortext[cor]{Corresponding author}
\address{Karlsruhe Institute of Technology, Karlsruhe, Germany}

\begin{abstract}
    The multi-scale nature of gaseous flows poses tremendous difficulties for theoretical and numerical analysis.
    The Boltzmann equation, while possessing a wider applicability than hydrodynamic equations, requires significantly more computational resources due to the increased degrees of freedom in the model.
    The success of a hybrid fluid-kinetic flow solver for the study of multi-scale flows relies on accurate prediction of flow regimes.
    In this paper, we draw on binary classification in machine learning and propose the first neural network classifier to detect near-equilibrium and non-equilibrium flow regimes based on local flow conditions.
    Compared with classical semi-empirical criteria of continuum breakdown, the current method provides a data-driven alternative where the parameterized implicit function is trained by solutions of the Boltzmann equation.
    The ground-truth labels are derived rigorously from the deviation of particle distribution functions and the approximations based on the Chapman-Enskog ansatz.
    Therefore, no tunable parameter is needed in the criterion.
    Following the entropy closure of the Boltzmann moment system, a data generation strategy is developed to produce training and test sets.
    Numerical analysis shows its superiority over simulation-based samplings.
    A hybrid Boltzmann-Navier-Stokes flow solver is built correspondingly with adaptive partition of local flow regimes.
    Numerical experiments including one-dimensional Riemann problem, shear flow layer and hypersonic flow around circular cylinder are presented to validate the current scheme for simulating cross-scale and non-equilibrium flow physics.
    The quantitative comparison with a semi-empirical criterion and benchmark results demonstrates the capability of the current neural classifier to accurately predict continuum breakdown. The code for the data generator, hybrid solver and the neural network implementation is available in the open source repositories~\cite{KITRT,xiao2021kinetic}.
\end{abstract}

\begin{keyword}
computational fluid dynamics, kinetic theory, Boltzmann equation, multi-scale method, deep learning
\end{keyword}

\maketitle

\section{Introduction}

Gases present a wonderfully diverse set of behaviors in different flow regimes.
Such regimes are often categorized according to the Knudsen number, which is defined as the ratio of molecular mean free path to a characteristic length scale.
With the variation of Knudsen number, the domain of flow physics can be qualitatively divided into continuum ($\rm{Kn}<0.001$), slip ($0.001<\rm{Kn}<0.1$), transition ($0.1<\rm{Kn}<10$), and free molecular regimes ($\rm{Kn}>10$) \cite{tsien1946superaerodynamics}.
The Knudsen number indicates a relative importance between individual particle transports and their collective dynamics.

Different governing equations are routinely established to describe the fluid motions at different scales.
As an example, in rarefied gas where Kn is of $O(1)$, the particle transport and collision processes are %loosely coupled and can thus be modeled by the operator splitting approach in the Boltzmann equation.
distinguishable and can thus be modeled by two independent operators in the Boltzmann equation.
In another limit with asymptotically small Kn, the Euler and Navier-Stokes equations are used to describe collective behaviors of fluid elements at a macroscopic level.
It is worth mentioning that there is no quantitative description for the scale of a fluid element.
Usually it refers to a macroscopically infinitesimal concept, where the flow variables inside the element can be considered as almost constant.%s with no fluctuations.
With a high amount of intermolecular collisions, the fluid inside an element is considered to be in local thermodynamic equilibrium.

Computational fluid dynamics focuses on numerical solution of the corresponding governing equations.
The direct Boltzmann solvers employ a discretized phase space to compute transport and collision terms respectively.
An alternative methodology is the direct simulation Monte Carlo (DSMC) method, which mimics the probability distribution function with a large amount of test particles and the collision term is calculated statistically.
On the other hand, the compressible Navier-Stokes solvers are mostly based on the Riemann solvers for inviscid flux and the central difference method for viscous terms. Only the macroscopic flow variables are tracked in the simulation. Compared with the kinetic methods, the computational cost of continuum fluid solvers is much lower.

Macroscopic and microscopic equations describe the physical evolution of the same substance and should correspond to each other.
The well-known Chapman-Enskog expansion bridges such a connection \cite{chapman1990mathematical}, where the Euler and Navier-Stokes equations can be derived from the asymptotic limits of expansion solutions of the Boltzmann equation.
Although the hydrodynamic equations are based on first-principle modeling, the Chapman-Enskog ansatz provides a rigorous criterion to define their validity.
In other words, the usage of hydrodynamic equations incorporates the assumption that the Chapman-Enskog solution plays as a proper approximation of particle distribution function.
However, this judgement cannot be verified in a macroscopic fluid simulation since the information of particle distribution functions has already been filtered in the coarse-grained modeling.
It is possible that the hydrodynamic equations are misused where they don't apply in scientific and engineering practice.

Different criteria have been proposed to predict the failure of continuum mechanics and construct the corresponding multi-scale numerical algorithms.
Some typical examples are listed below.
Bird \cite{bird1994molecular} proposed a parameter $\mathcal P=D(\ln \rho)/Dt/\nu$ for the DSMC simulation of expansion flows, where $\rho$ is gas density and $\nu$ is collision frequency, and the breakdown threshold of translational equilibrium is set as $\mathcal P=0.05$.
Boyd et al. \cite{boyd1995predicting} extended the above concept to a gradient-length-local Knudsen number $\mathrm{Kn}_{GLL}=\ell |\nabla I|/I$, where $\ell$ is the local molecular mean free path and $I$ is a scalar of interest, with the critical value being $\mathcal C=0.05$.
Garcia et al. \cite{garcia1998generation} proposed a breakdown parameter based on dimensionless stress and heat flux $\mathcal B=\max(|\tau_*|,|q_*|)$, with the switching criterion of $\mathcal B=0.1$.
Levermore et al. \cite{levermore1998moment} developed non-dimensional matrices from the moments of particle distribution function. The tuning parameter $\Delta \mathcal V$ is then defined as the deviation of the eigenvalues of this matrix from their equilibrium values of unity, with the critical value of 0.25.
The idea of all the above methods is to assemble components in the Chapman-Enskog expansion.
However, due to the fact that the ground-truth information of particle distribution is missing in a macroscopic fluid simulation, it is virtually impossible to employ the quantitative deviation between particle distributions from full Boltzmann solution and from Chapman-Enskog reconstruction directly.
It is difficult to prove that the above criteria can be universally applied to complex systems under different conditions of flow and geometry.

The rapid development of deep learning provides us a promising alternative for classification and regression tasks.
The relevant modeling and simulation strategies have been applied in fluid mechanics, e.g., building data-to-solution mapping \cite{raissi2020hidden,li2020variational,khoo2019switchnet}, constructing physics-informed neural networks \cite{raissi2019physics,mao2020physics,sun2020surrogate}, identifying sparse dynamical systems \cite{brunton2016discovering,rudy2017data,zhang2020data},
and solving kinetic equations \cite{xiao2021using,schotthofer2021structure,lou2021physics}.
In this paper, we turn to the application of binary classification.
The idea is to employ neural networks as surrogate models, which classify the most probable flow regime  based on local flow conditions.
The neural networks accept macroscopic quantities including velocity moments and their slopes serve as inputs, and return labels of flow regimes.
Following the principle of minimal entropy distributions, a data generation strategy is developed to sample particle distributions near and out of equilibrium in the training and test sets.
Based on kinetic solutions, the ground-truth labels are rigorously determined by the deviation between the particle distribution functions and the Chapman-Enskog solutions.
Therefore, a data-driven parameterized  function is defined implicitly by the neural network in the high-dimensional function space.
Based on the neural classifier, we develop a multi-scale hybrid method, which realizes a dynamic adaptation of flow regimes and fuses the continuum and kinetic solutions seamlessly.

The paper is organized as follows. In Sec. 2 we introduce some fundamental concepts in the kinetic theory of gases and the Chapman-Enskog expansion.
Sec. 3 presents the idea and design of the neural network architecture. 
Sec. 4 introduces the strategy for generating data in training and test set.
Sec. 5 details the numerical algorithm of the hybrid solver incorporated with the neural network classifier.
Sec. 6 contains several numerical experiments to validate the current method.
The last section is the conclusion.

%Collisional processes in the rarefied regime, while not a dominant feature, are important for understanding instabilities and damping processes in applications ranging from high-altitude turbulence and drag analysis for re-entry vehicles to fusion yield and stability in the tokamak.

\section{Kinetic Theory}\label{sec:theory}

The Boltzmann equation describes the time-space evolution of a one-particle distribution function $f(t,\mathbf x,\mathbf v)$ in dilute monatomic gas, i.e.,
\begin{equation}
    \partial_t f + \mathbf v \cdot \nabla_\mathbf x f = Q(f,f) = \int_{\mathbb{R}^{3}} \int_{\mathbb S^{2}}\left[f\left(\mathbf{v}^{\prime}\right) f\left(\mathbf{v}_{*}^{\prime}\right)-f(\mathbf{v}) f\left(\mathbf{v}_{*}\right)\right] \mathcal{B}(\cos \theta, g) d \mathbf \Omega d \mathbf{v}_{*},
    \label{eqn:boltzmann}
\end{equation}
where $\{\mathbf v, \mathbf v_*\}$ are the pre-collision velocities of two classes of colliding particles, and $\{\mathbf v', \mathbf v_*'\}$
are the corresponding post-collision velocities.
The collision kernel $\mathcal{B}(\cos \theta, g)$ measures the probability of collisions in different directions, where $\theta$ is the deflection angle and $g = |\mathbf g| = |\mathbf v - \mathbf v_*|$ is the magnitude of relative pre-collision velocity.
The solid angle $\mathbf \Omega$ is the unit vector along the relative post-collision velocity $\mathbf v' - \mathbf v_*'$, and the deflection angle satisfies the relation $\theta=\mathbf \Omega \cdot \mathbf g / g$.

The Boltzmann equation depicts a physical process with increasing physical entropy.
The H-theorem indicates that the entropy is a Lyapunov function for the Boltzmann equation and the logarithm of its maximizer must be a linear combination of the collision invariants $\psi=(1,\mathbf v,\mathbf v^2/2)^T$ \cite{bouchut2000kinetic}.
The equilibrium solution related to maximal entropy is the so-called Maxwellian distribution function,
\begin{equation}
    \mathcal M := \rho\left(\frac{m}{2\pi k T}\right)^{3/2} \exp(-\frac{m}{2kT} \left(\mathbf v - \mathbf V)^2 \right),
    \label{eqn:maxwellian}
\end{equation}
where $m$ is molecular mass, $\mathbf V$ is macroscopic fluid velocity, $T$ is temperature, and $k$ is the Boltzmann constant.

The macroscopic conservative flow variables can be obtained by taking moments from the particle distribution function over velocity space, i.e.,
\begin{equation}
    \mathbf{W}=\left(\begin{array}{c}
    \rho \\
    \rho \mathbf V \\
    \rho E
\end{array}\right)=\int f \psi d \mathbf v,
\label{eqn:moments}
\end{equation}
where $\rho E=\rho \mathbf V^2/2 + \rho e$, $e$ is the internal energy per unit mass, and $\psi$ is the vector of collision invariants.
For ideal gas, the internal energy is related with temperature as
\begin{equation}
    \rho e = \frac{3}{2}nkT,
\end{equation}
where $n=\rho/m$ is the number density.
Taking moments of the Boltzmann equation with respect to collision invariants yields the transport equations for conservative variables,
\begin{equation}
    \partial_t \mathbf W + \int_{\mathbb R^3}\psi \mathbf v \cdot \nabla_\mathbf x f d\mathbf v = 0,
    \label{eqn:moment}
\end{equation}
i.e.,
\begin{equation}
\begin{aligned}
    &\frac{\partial \rho}{\partial t}+\nabla \cdot(\rho \mathbf{V})=0, \\
    &\frac{\partial(\rho \mathbf{V})}{\partial t}+\nabla \cdot(\rho \mathbf{V} \otimes \mathbf{V})=\nabla \cdot \mathbf{P}, \\
    &\frac{\partial(\rho E)}{\partial t}+\nabla \cdot(\rho {E} \mathbf{V})=\nabla \cdot(\mathbf{P} \cdot \mathbf{V})-\nabla \cdot \mathbf{q},
\end{aligned}
\end{equation}
where $\otimes$ denotes dyadic product, and the stress tensor $\mathbf P$ and heat flux $\mathbf q$ are defined as,
\begin{equation}
\mathbf{P}=\int(\mathbf{v}-\mathbf{V})(\mathbf{v}-\mathbf{V}) f d \mathbf v, \quad
\mathbf{q}=\int \frac{1}{2}(\mathbf{v}-\mathbf{V})(\mathbf{v}-\mathbf{V})^{2} f d \mathbf v.
\end{equation}

It is clear that the flux terms in the above equations are one order higher than the leading terms, which leads to the well-known closure problem~\cite{Levermore1996MomentCH}.
Different closure strategies, i.e., different forms of the distribution function $f$, result in vastly different macroscopic transport equations.
In the following, we briefly show the methodology of Chapman-Enskog ansatz, where the Navier-Stokes equations can be derived from the asymptotic solution of the Boltzmann equation.
With the introduction of the following dimensionless variables
\begin{equation}
    \tilde{\mathbf x} = \frac{\mathbf x}{L_0},  \ \tilde t = \frac{t}{L_0/V_0}, \ \tilde{\mathbf v} = \frac{\mathbf v}{V_0}, \ \tilde f = \frac{f}{n_0 V_0^3},
\end{equation}
where $V_0=\sqrt{2kT_0/m}$ is the most probable molecular speed,
the Boltzmann equation can be reformulated as
\begin{equation}
    \partial_t \tilde f + \tilde{\mathbf v} \cdot \nabla_{\tilde {\mathbf x}} {\tilde f} = \frac{1}{\mathrm{Kn}}Q(\tilde f, \tilde f).
    \label{eqn:dimensionless boltzmann}
\end{equation}
The Knudsen number is defined as
\begin{equation}
    \mathrm{Kn}=\frac{V_0}{L_0 \nu_0}=\frac{\ell_0}{L_0},
\end{equation}
where $\ell_0$ and $\nu_0$ are the molecular mean free path and mean collision frequency in the reference state.
For brevity, we drop the tilde notation to denote dimensionless variables henceforth.

Based on a small Knudsen number $\mathrm{Kn}=\varepsilon$, the Chapman-Enskog expansion approximates the particle distribution function \cite{chapman1990mathematical} as,
\begin{equation}
    f \simeq f_\varepsilon = \sum_{n=0}^\infty \varepsilon^n f^{(n)}, \quad f^{(0)} := \mathcal M.
    \label{eqn:chapman enskog}
\end{equation}
Truncating the above expansion to the first non-trival order, substituting it into Eq.(\ref{eqn:dimensionless boltzmann}) and projecting the kinetic system onto hydrodynamic level, one can derive the Navier-Stokes equations.
Here we omit the tedious mathematical derivation and refer the reader to the literature \cite{koganrarefied}.
The detailed expansion solution for the Navier-Stokes regime writes
\begin{equation}
\begin{aligned}
    f_{\text{NS, Boltzmann}}=& \mathcal M\left[1- \frac{2\kappa}{5R p}\left(\frac{\mathbf c^{2}}{2 R T}-\frac{5}{2}\right) \mathbf c \cdot \nabla_{\mathbf x}(\ln T)\right.\\
    &\left.-\frac{\mu}{R T p}\left(\mathbf c \otimes \mathbf c-\frac{1}{3} \mathbf c^{2} \mathbf I \right) : \nabla_{\mathbf x} \mathbf V\right].
\end{aligned}
\label{eqn:ce boltzmann}
\end{equation}
The viscosity and heat conductivity are determined by specific molecule models.
For example, the viscosity coefficient for hard-sphere molecules takes
\begin{equation}
    \mu = \mu_\mathrm{0} \left(\frac{T}{T_\mathrm{0}}\right)^\omega,
\end{equation}
where the power index $\omega$ needs to be calibrated for different substances, and the heat conductivity is linked by the Prandtl number $\mathrm{Pr}=c_p \mu / \kappa$ where $c_p$ is specific heat of the gas at a constant pressure.

% BGK
%\begin{equation}
%    f := \mathcal M - \tau (\partial_t \mathcal M + \mathbf v \cdot \nabla \mathcal M)
%\end{equation}

\section{Neural Network based classification of the flow regime}\label{sec_nn}

The universal approximation theorem \cite{csaji2001approximation}, as a generalization of Stone-Weierstrass theorem \cite{de1959stone}, indicates that a neural network in its simplest form can approximate continuous functions on compact subsets of $\mathbb R^n$, provided that there are sufficient neurons under mild assumptions on the activation function.
Defined in latent space and driven by data, the neural network simplifies data representations for the purpose of finding patterns in supervised learning.
Such surrogate models can provide an alternative for semi-empirical criteria to classify the continuum breakdown regions of a flow field.

\begin{figure}
    \centering
	\includegraphics[width=0.99\textwidth]{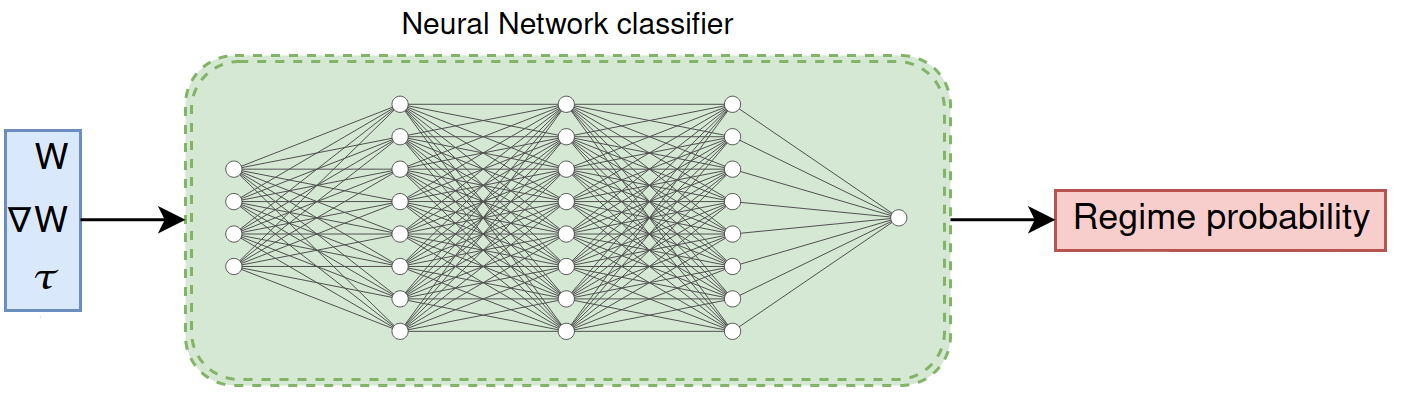}
    \caption{The neural network based regime classifier using macroscopic variables, their gradients and the collision time to predict the flow regime of the current grid cell.}
    \label{fig:nn}
\end{figure}

Following the spirit of Chapman-Enskog expansion, we build the neural network model as
\begin{equation}
    \hat{\mathcal R} = \mathrm{NN}_\theta(\mathbf U),
\end{equation}
where $\theta$ denotes the trainable parameters of the neural network.
As shown in Figure \ref{fig:nn}, the input of neural network $\mathbf U=(\mathbf W, \nabla_{\mathbf x} \mathbf W, \tau)$ is a combination of macroscopic variables, their slopes, and mean collision time.
The idea to constitute such function inputs is to draw on the Chapman-Enskog ansatz and provide the necessary information for the reconstruction of probable particle distribution functions.
The output $\hat{\mathcal R}$ is set to be a scalar, which denotes the likelihood for the current cell to be in non-equilibrium regime.
The neural network employs the sigmoid function as activation in the last layer, and thus the output satisfy $\hat{\mathcal R}\in[0,1]$ naturally.
With the floor function, the output takes binary values, where $1$ denotes rarefied (non-equilibrium) and $0$ denotes continuum (near-equilibrium) regime.

In the supervised learning task, the dataset consists of a set of inputs and ground-truth labels corresponding to the function $\mathbf U \mapsto \hat{\mathcal R}$.
For a given distribution function $f_\mathrm{ref}$, the flow regime label is defined as
\begin{equation}
\begin{aligned}
    & \mathcal R = \begin{cases}1, & d > \epsilon \\ 0, & d\leq \epsilon\end{cases}, \
    & d = \frac{||f_\mathrm{NS} - f_\mathrm{ref}||_2} {\rho},
\end{aligned}
\label{eqn:error}
\end{equation}
where $d$ denotes a normalized norm between the reference particle distribution function and the reconstructed Navier-Stokes distribution.
Following the Chapman-Enskog ansatz, the Navier-Stokes distribution function can be constructed using Eq.(\ref{eqn:ce boltzmann}).
Note the macroscopic quantities in the above two equations can be obtained by taking moments of reference distribution function as in Eq.(\ref{eqn:moments}), and the collision time  $\tau=1/\nu$ can be derived from kinetic theory.
Given the definition of labels in the dataset, the idea of the current neural network becomes clear. 
The data-driven approach builds an implicit function $\mathbf U \mapsto \hat{\mathcal R}$ in the high-dimensional functional space spanned by neural network parameters.
%With the macroscopic flow variables calculated from the reference kinetic solution $f_\mathrm{ref}$, we may understand the mechanism as the neural network reconstructs the most probable distribution function from moments level in the direction opposite to projection and compare it with the Chapman-Enskog solution.
The macroscopic flow variables, which are calculated from the reference kinetic solution $f_\mathrm{ref}$, are inputs to the neural network, and its prediction is the flow regime. Thus one may understand the neural networks internal mechanism as an implicit reconstruction of the most probable kinetic solution, which is then compared to the the Chapman-Enskog solution to determine the flow regime.
The surrogate model provided by neural network bridges macroscopic variables and flow regimes directly. 
Compared with classical criteria for continuum breakdown, no empirical and semi-empirical expansions are needed from asymptotic theory.

For this binary classification task, we employ the binary cross-entropy as loss function, i.e.,
\begin{equation}
    \mathcal {L}=-\frac{1}{N}
    \sum_{i=1}^{N} {\mathcal{R}}_{i} \cdot \log \hat{\mathcal R}_{i}+\left(1- {\mathcal{R}}_{i}\right) \cdot \log \left(1-\hat{\mathcal R}_{i}\right),
\end{equation}
where $\hat{\mathcal R}$ is the $i$-th scalar value in the model output, ${\mathcal{R}}$ is the corresponding target value, and the output size $N$ is the number of scalar values for the model output.
The cross entropy is equivalent to fitting the model using maximum likelihood estimation.  
The Kullback-Leibler divergence between the empirical distribution of training data and the distribution induced by the model is minimized.
The ADAM optimizer is used during all training processes.
The training and testing data is produced by sampling and processing prescribed kinetic solutions of particle distribution functions, and the validation set is generated with the help of kinetic simulation data from numerical cases.
%We leave the detailed strategy for generating dataset in the next section.
\section{Data Generation}
As presented in Eq.(\ref{eqn:error}), the information of exact particle distribution functions $f$ is needed to compute macroscopic quantities $\mathbf U$ and regime labels.
In the following we consider the space of $f$ as the sampling space under the constraint
\begin{equation}
    f\in F=\menge{ f(\mathbf v) \geq 0: \abs{ \int_{\mathbb R^3} f \psi_i d \mathbf{v}}<\infty, i=0,1,2},
    \label{eqn:constraint}
\end{equation}
i.e. the existence of the first $3$ moments $\{\rho,\rho \mathbf{V},\rho E\}$ and non-negativity of the particle distribution.
A strategy to sample data from $F$ usually creates a data-distribution $p_F$ implicitly, which influences the training and test performance of the neural network. As the goal of the classification network is to find the separation hyperplanes between the near-equilibrium and non-equilibrium regime, we need to systematically create a data-distribution $p_F$ that generates enough samples near the boundary between regimes. A naive strategy is to sample data by performing numerical simulations and storing the required data in a post-processing fashion.
The disadvantage of this is that $p_F$ can be heavily biased towards the dynamics of the chosen test cases and might not necessarily cover enough different regions of flow regimes. Furthermore, it comes with the computational expense of a full kinetic solver, that might compute the same solutions multiple times, e.g. in the farfield of a fluid simulation. 
In the following we demonstrate a sampling strategy to generate balanced data near and out of equilibrium.
\subsection{Sampling of particle distribution functions}
\begin{figure}
    \centering
    \includegraphics[width=0.47\textwidth]{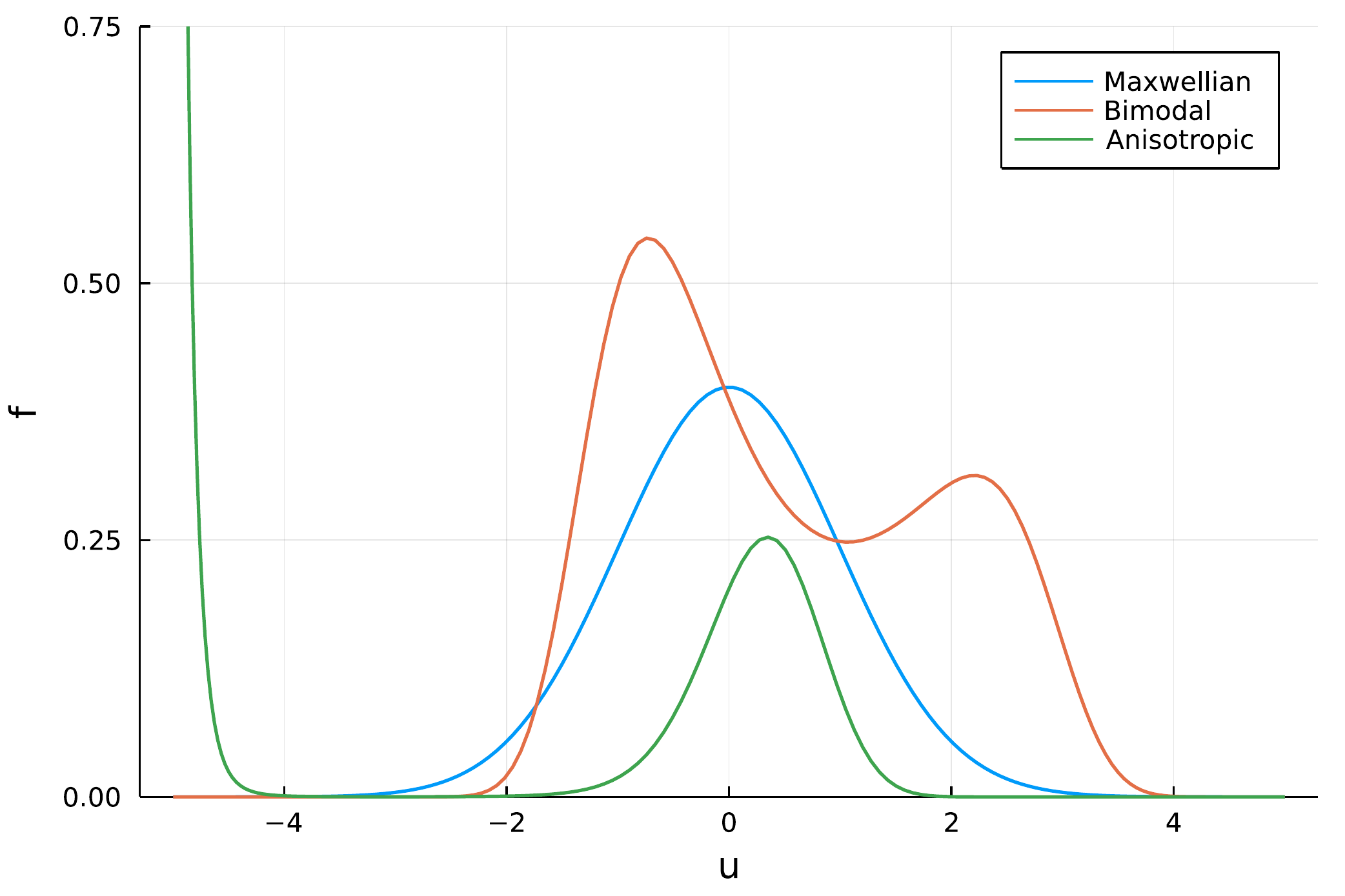}
    \caption{Sampling of particle distribution functions. The more a function deviates from the Maxwellian, the higher is the condition number of the corresponding entropy problem}
    \label{fig_entropy_sampling}
\end{figure}

The sampling of data leverages the entropy closure of the Boltzmann moment system. We briefly introduce the principle here and refer \cite{Levermore1996MomentCH} for details.
A general closure aims to reconstruct the particle distribution function $f$ from a vector of moments
\begin{align}
    \mathbf u = \int_{\mathbb R^3} f \mathbf m d\mathbf{v}\in\mathbb{R}^{N_m},
\end{align}
under the constraint
\begin{equation}
    f\in F_M=\menge{ f(\mathbf v) \geq 0: \abs{ \int_{\mathbb R^3} f m_i d \mathbf{v}}<\infty, i=0,\dots,N_m},
    \label{eqn:constraint_FM}
\end{equation}
where $\mathbf m(\mathbf v)\in\mathbb{R}^{N_m}$ is  a vector of velocity dependent basis functions. We choose the basis in a way that the first three moments coincide with the conservative variables of the Navier-Stokes equations in Eq.(\ref{eqn:moments}).
We thus rewrite $\mathbf m(\mathbf v)$ in the following form,
\begin{align}
    \mathbf m(\mathbf v) = (\psi_0,\psi_1,\psi_2, \tilde{\mathbf m}(\mathbf{v}))^T,
\end{align}
where $\tilde{\mathbf m}(\mathbf{v})$ can be arbitrary monomials and mixed polynomials up to degree $N_m$ and $\psi_i$ are the collision invariants of the Boltzmann equation. 
The minimal entropy closure employs an optimization problem to ensure uniqueness of the solution of the closure problem. The objective function of the optimization problem is denoted by the integrated mathematical entropy density $\eta$.
For the choice of entropy from Maxwell-Boltzmann statistics $\eta(f)=f\log(f)-f$ \cite{junk2000maximum}, the minimal entropy closure problem reads
\begin{align}\label{eq_entropyOCP} 
\min_{g\in F_m} \int_{\mathbb R^3} g\log(g)-g d\mathbf{v}\quad  \text{ s.t. }  \mathbf u=\int_{\mathbb R^3}\mathbf m gd\mathbf{v}.
\end{align}
If a solution of this optimization problem exists, it can be represented as
\begin{align}\label{eq_recons_density}
    f_{u}(\mathbf v) = \exp(\alpha_u \mathbf m(\mathbf v)),
\end{align}
where $\alpha_u\in\mathbb{R}^{N_m}$ is a the vector of Lagrange multipliers of the dual formulation of the optimization problem, which reads
\begin{align}\label{eq_entropyDualOCP}
    \alpha_u =  \underset{\alpha\in\mathbb{R}^{{N}_m}}{\text{argmax}} \menge{ \alpha\cdot \mathbf u - \inner{\exp(\alpha\cdot \mathbf m)}}.
\end{align}
The  set of all moments $u$ for which the minimal entropy problem in Eq.\eqref{eq_entropyOCP} has a solution is called the realizable set 
\begin{align}
    \mathscr{R} = \menge{\mathbf u: \mathbf u=\int_{\mathbb R^3} g \mathbf m d\mathbf v, g\in F_m}.
\end{align}
It should be noted that the minimal entropy problem has no solution at the boundary $\partial\mathscr{R}$ of the realizable set and its condition number $\sigma_H$ increases when approaching the boundary. The condition number of the minimal entropy closure at a moment $\mathbf u$ can be computed via the positive semi-definite Hessian of the dual problem
\begin{align}
    H_u = \int_{\mathbb R^3} \mathbf m\otimes \mathbf m \exp(\alpha_u\cdot \mathbf m) d\mathbf v. 
\end{align}
Reconstructed particle distributions with moments for which the minimal entropy closure has a low condition number are typically similar to the Maxwellian. Distribution functions corresponding to moments near $\partial\mathscr{R}$, where the minimal entropy problem is ill-conditioned, are highly anisotropic and have a high distance to a Maxwellian, which is illustrated in Fig.~\ref{fig_entropy_sampling}.
Further theories of realizability have been studied in detail \cite{Levermore, Curto_recursiveness,Junk,junk2000maximum,Pavan2011GeneralEA,schotthofer2022neural}.
The idea is to generate distribution functions which are solutions of the minimal entropy closure using Eq.\eqref{eq_recons_density}.
Specifically, we sample the corresponding Lagrange multipliers $\alpha_u$. Using the condition number of $H_u$ we can control the sampling of reference densities in near-equilibrium and non-equilibrium regime.
For example, the Maxwellian in Eq.\eqref{eqn:maxwellian} can be expressed  can be expressed with the following choice of $\alpha$,
\begin{equation}
\begin{aligned}
    &\mathcal{M} = \exp(\alpha\cdot \mathbf m), \ \alpha = (\alpha_0,\alpha_1,\alpha_2,\dots,\alpha_n)^T, \\
    &\alpha_0=\ln(\rho/(2\pi k T)^{3/2}) - \mathbf{V}^2/(2kT), \
    \alpha_1=\frac{\mathbf{V}}{kT}, \
    \alpha_2=-\frac{1}{2kT}, \
    \alpha_n=0 \ \text{for all}\  n>2 .
    \label{eqn_ansatz}
\end{aligned}
\end{equation}
The disturbance from the equilibrium state can be controlled for example by choosing $\alpha_n \not=0$ for $n>2$.  
For a fixed length $N_m$ of the Lagrange multiplier vector $\alpha_u$, we sample $\alpha_n$ for $n>0$ normally distributed with a prescribed standard deviation. The sampling mean is chosen according to the Lagrange multiplier, that recovers the Maxwellian above. 
Without loss of generality we assume that $u_0$ which corresponds to $\rho$ in terms of conservative variables equals one. 
For $\alpha_n\not=0$, $n>2$, in general, the computed particle density $\rho\not=1$.
To enforce the assumption, that $u_0=\rho=1$ we use for a given set of sampled coefficients $\alpha$ the ansatz
\begin{align}
   u_0= \rho = 1 = \int_{\mathbb R^3} \exp(\alpha\cdot \mathbf m) d\mathbf v.
\end{align}
Applying the natural logarithm to both sides of the equation, we get
\begin{align}\label{eq_alpha0}
    \alpha_0 = - \ln\left( \int_{\mathbb R^3} \exp((\alpha_1,\alpha_2,\dots)^T\cdot (\mathbf m_1,\mathbf m_2,\dots)^T) d\mathbf v\right).
\end{align}
The resulting sampling strategy is summarized in Algorithm~\ref{alg_samplingUniform}.
\begin{algorithm}\label{alg_samplingUniform}
\DontPrintSemicolon
\caption{Sampling of reference particle distribution functions }
\SetAlgoLined
\SetKwInOut{Input}{Input}
\Input{Spatial dimension $d$ and maximum moment order $N$\newline
Standard deviation $\sigma$ to sample $\alpha$\newline
Range of temperatures $[T_{min},T_{max}]$\newline
%Velocity space $V$\newline
Range of bulk velocities $[V_{min};V_{max}]$, where $V_{min}<V_{max}$\newline
Condition number threshold $c$
}
\KwResult{Balanced set of reference densities $\menge{f_{\text{ref},j}}_{i\in I}$}
\For {$i = 1,\dots,\vert I\vert$}{
    Sample $T\in[T_{min},T_{max}]$\;
    Sample $V\in[V_{min};V_{max}]$\;
    Compute mean for $\alpha_{u,1}$: $\frac{\mathbf{V}}{kT}$\;
    Compute mean for $\alpha_{u,2}$: $-\frac{1}{2kT}$\;
    \Do {$\sigma_{H_u(\alpha_u)}<c$}{
        Sample $\alpha_{u,n}$, $n=1,\dots,N_m $ \;
        Reconstruct $\alpha_{u,0}$\; %using Eq.~\eqref{eq_alpha0}\;
        Compute $H_u(\alpha_u)$\;
    }
    Compute  $f_{\text{ref},i} = \exp(\alpha_u \mathbf m)$ \;
 }
\end{algorithm}
\subsection{Assembly of the training data}
\begin{figure}
    \centering
    \subfigure[Equilibrium solutions in ghost cells, $\tau=0.0012$]{
		\includegraphics[width=0.47\textwidth]{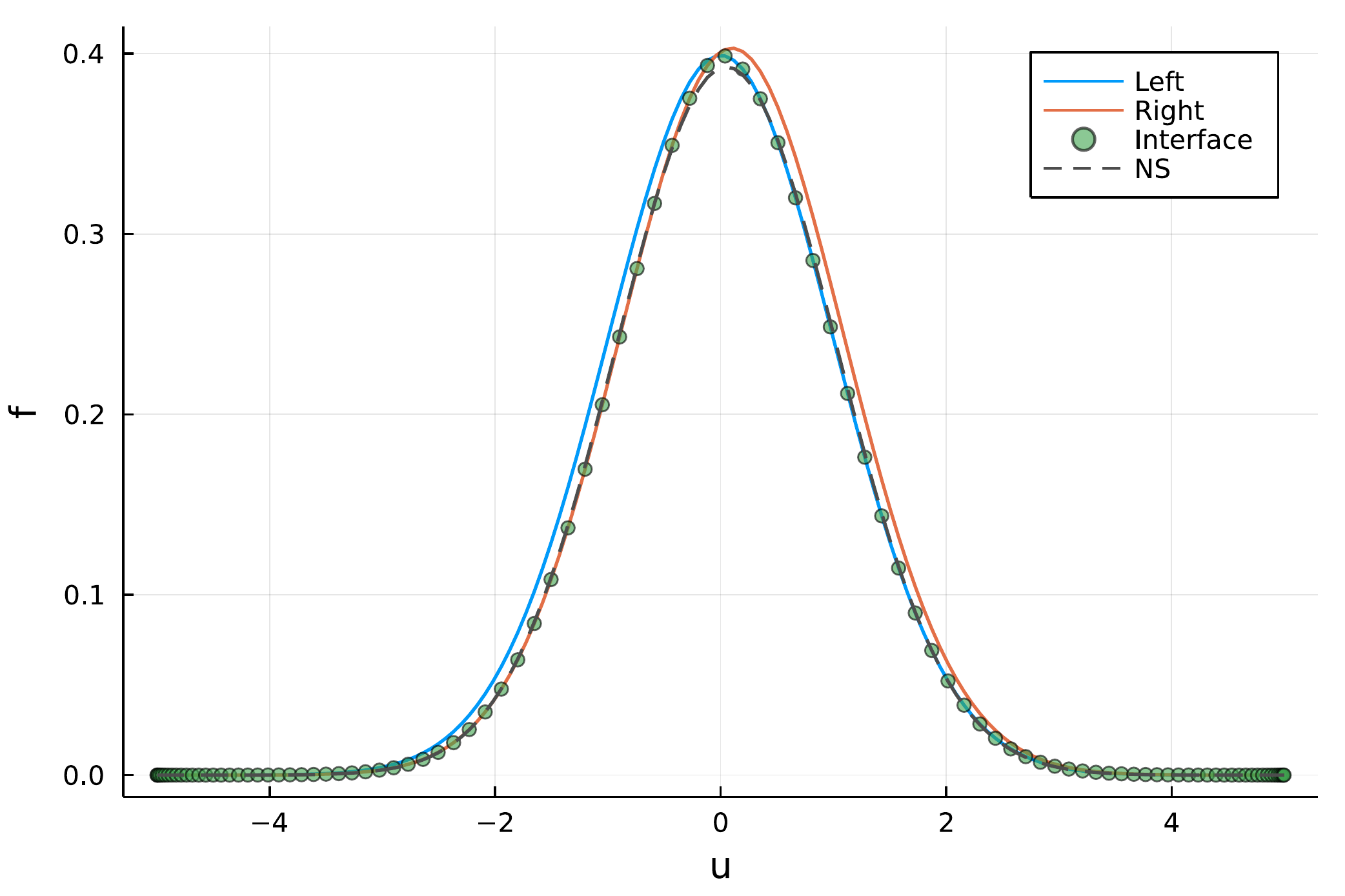}
	}
	\subfigure[Non-equilibrium solutions in ghost cells, $\tau=0.000511$]{
		\includegraphics[width=0.47\textwidth]{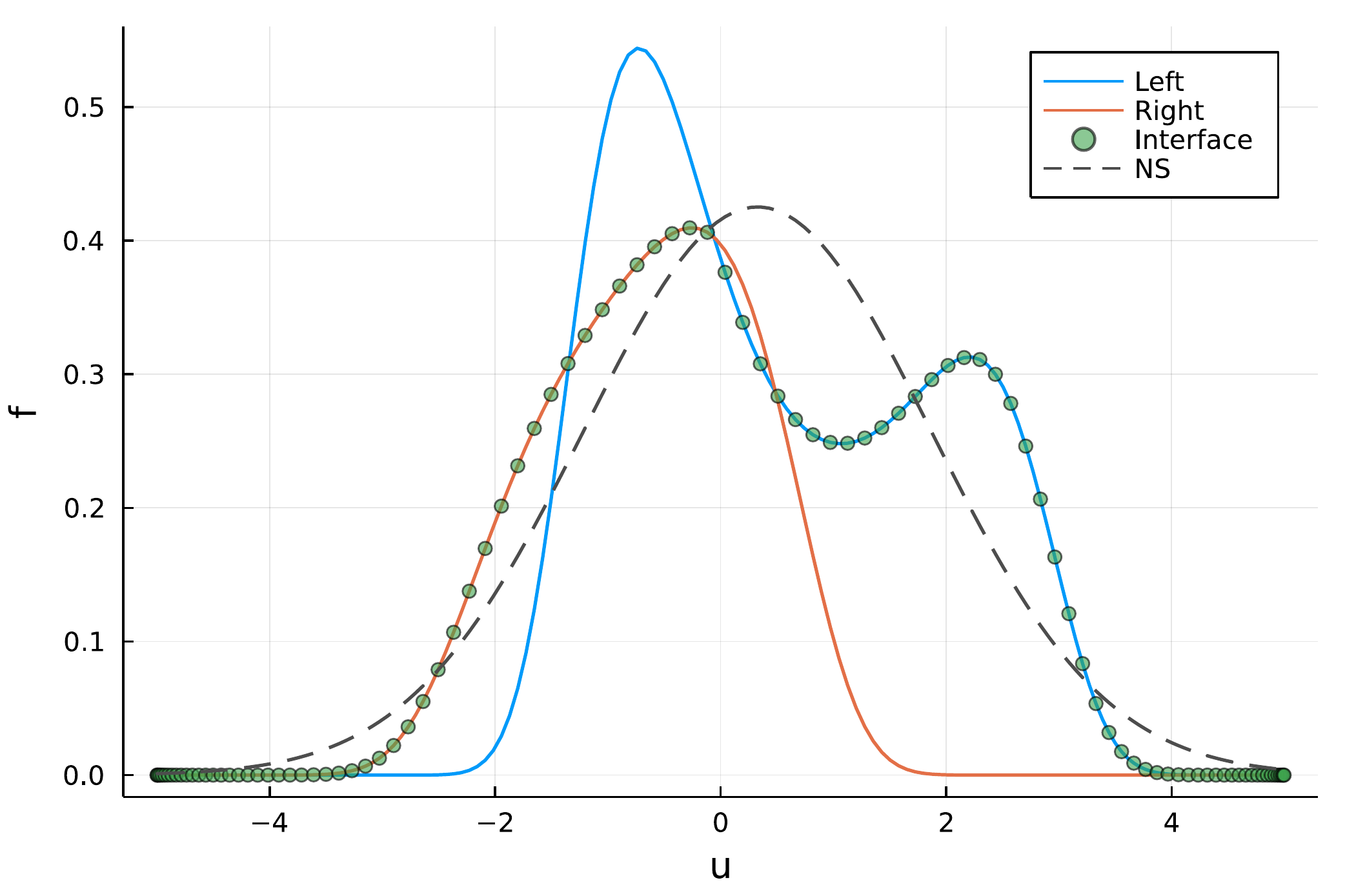}
	}
    \caption{Sampling of reference solution at the interface of two neighboring ghost cells and Chapman-Enskog reconstruction for $Kn=0.001$ and $dx=0.01$ in $1$ spatial dimension.}
    \label{fig_data_sampling}
\end{figure}

The input of neural network contains not only a set of conservative variables but also their gradients and local collision time.
The idea for data-generation is to combine two sampled distribution functions $\{f_L,f_R\}$ with two adjacent ghost cells, of which the positions $\{\mathbf x_L,\mathbf x_R\}$ as well as the unit normal vector $\mathbf{n}$ are randomly sampled.
Therefore, the reference particle distribution function at the interface can be approximated via an upwind reconstruction,
\begin{align}\label{eq_upwind}
   % f_{\text{ref}}(\mathbf v) =  \begin{cases}
    %  f_L(\mathbf v) & \text{if $\mathbf n\cdot \mathbf v>0$}\\
    %  f_R(\mathbf v) & \text{if $\mathbf n\cdot \mathbf v\leq0$}.
    %\end{cases} 
    f_{\text{ref}}(\mathbf v) = f_{L}(\mathbf v) H(\mathbf n\cdot \mathbf v) + f_{R}(\mathbf v) \left( 1-H(\mathbf n\cdot \mathbf v) \right),
\end{align}
where $H$ is the heaviside step function.
The conservative variables $\{\mathbf{W},\mathbf{W}_L,\mathbf{W}_R\}$ are obtained by taking moments of $f_{\text{ref}}$, and the gradients $\nabla_x \mathbf W$ are computed with a finite difference formula.
Figure~\ref{fig_data_sampling}a) displays the upwind approximation and Chapman-Enskog reconstruction in Eq.(\ref{eqn:ce boltzmann}) from the corresponding conservative variables at the interface of two ghost cells with near equilibrium distributions and Fig.~\ref{fig_data_sampling}b) the reconstruction of two non-equilibrium solutions. On sees, that in Fig.~\ref{fig_data_sampling}a) the Chapman-Enskog reconstruction is close to the upwind approximation, whereas in Fig.~\ref{fig_data_sampling}b) the respective distributions have a very different shape.

\begin{figure}%\label{fig_macro_data_gen}
    \centering
    \subfigure[Algorithmic sampler]{
		\includegraphics[width=0.47\textwidth]{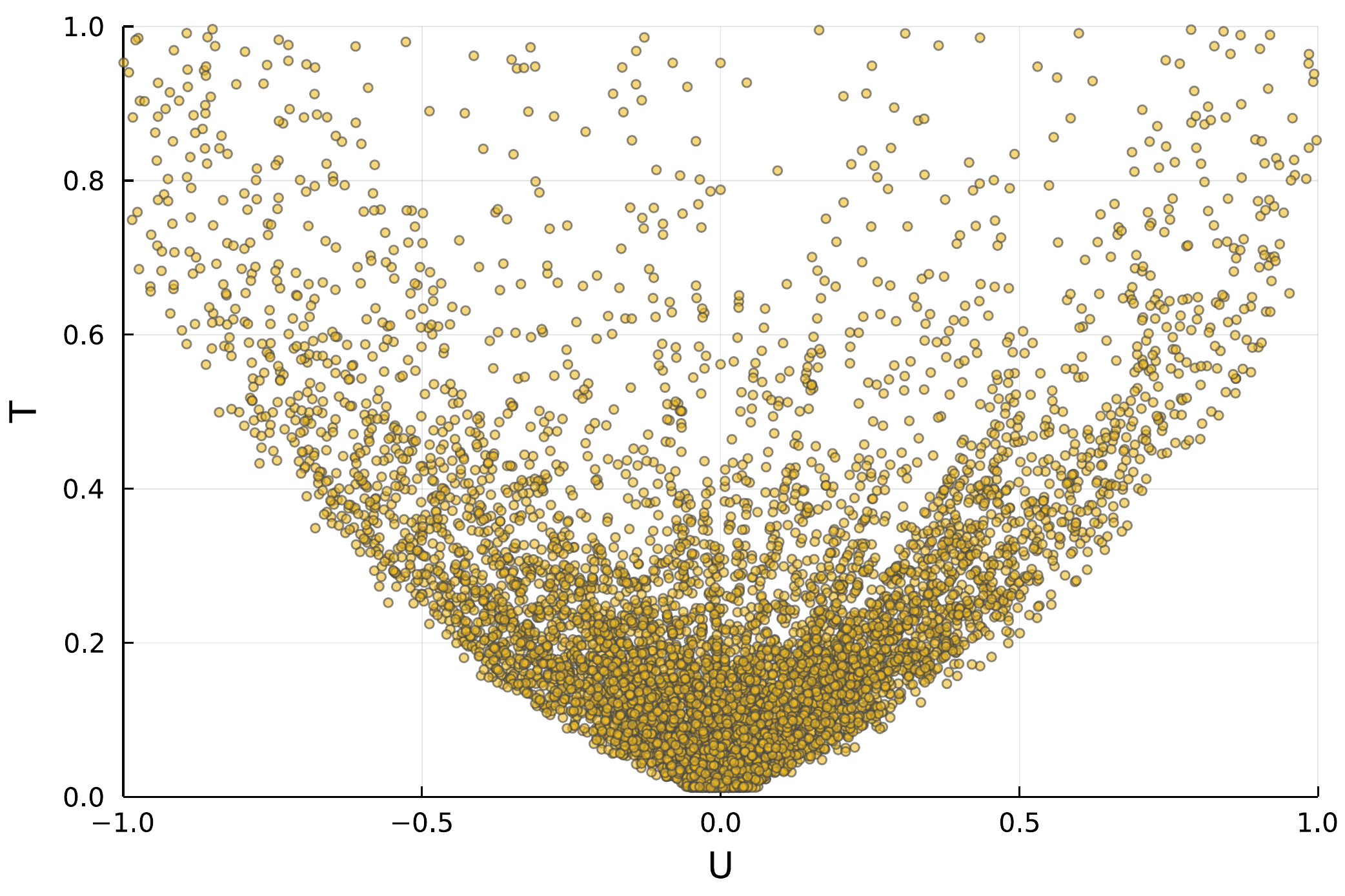}
	}
	\subfigure[Solver sampler]{
		\includegraphics[width=0.47\textwidth]{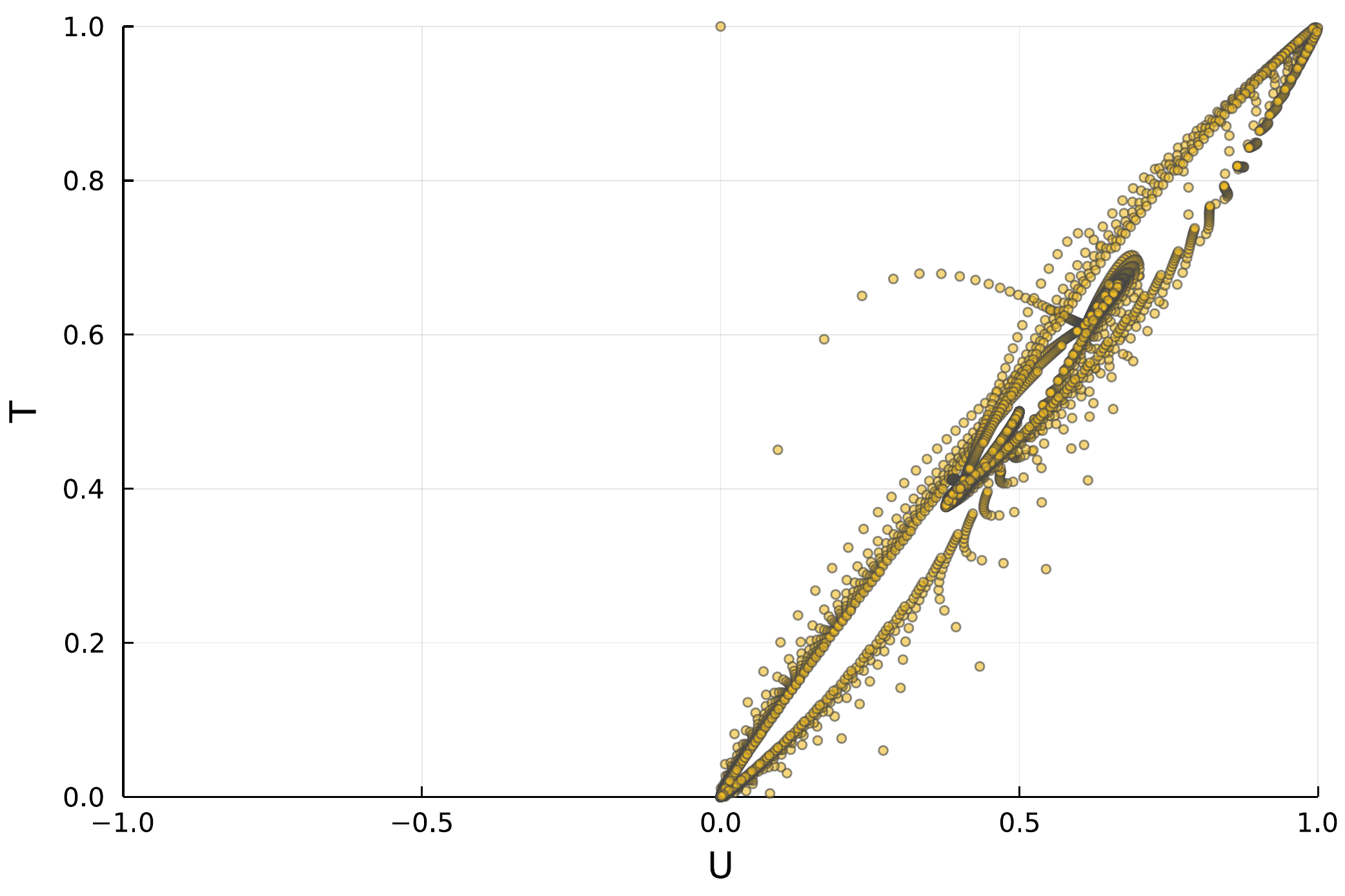}
	}
    \caption{Distributions of data points in $U$-$T$ phase diagram from the current algorithmic generator and sampled from Sod shock tube solution.}
    \label{fig:illustrator value}
\end{figure}

\begin{figure}%\label{fig_grad_macro_data_gen}
    \centering
    \subfigure[Algorithmic generator]{
		\includegraphics[width=0.47\textwidth]{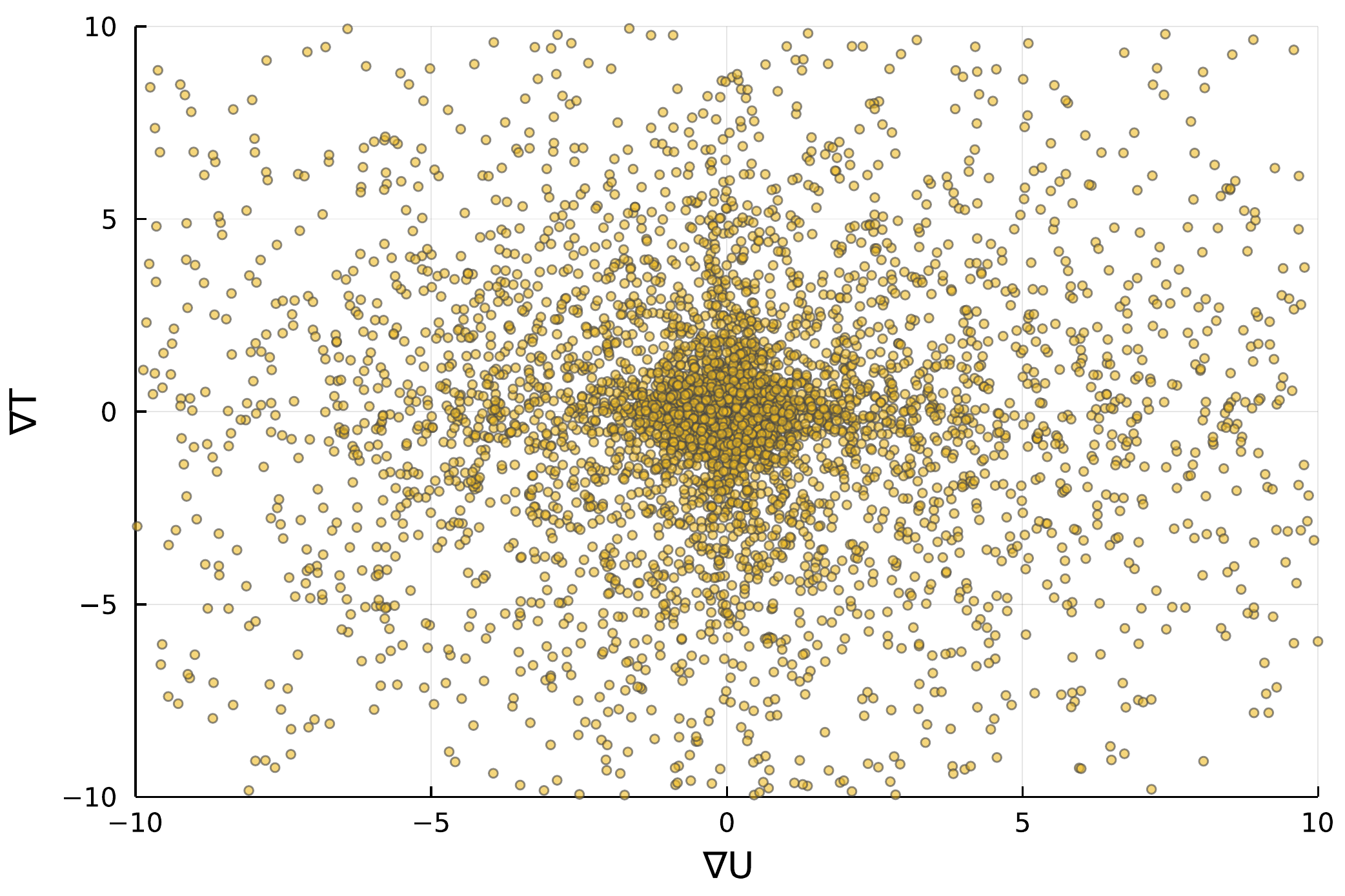}
	}
	\subfigure[Solver sampler]{
		\includegraphics[width=0.47\textwidth]{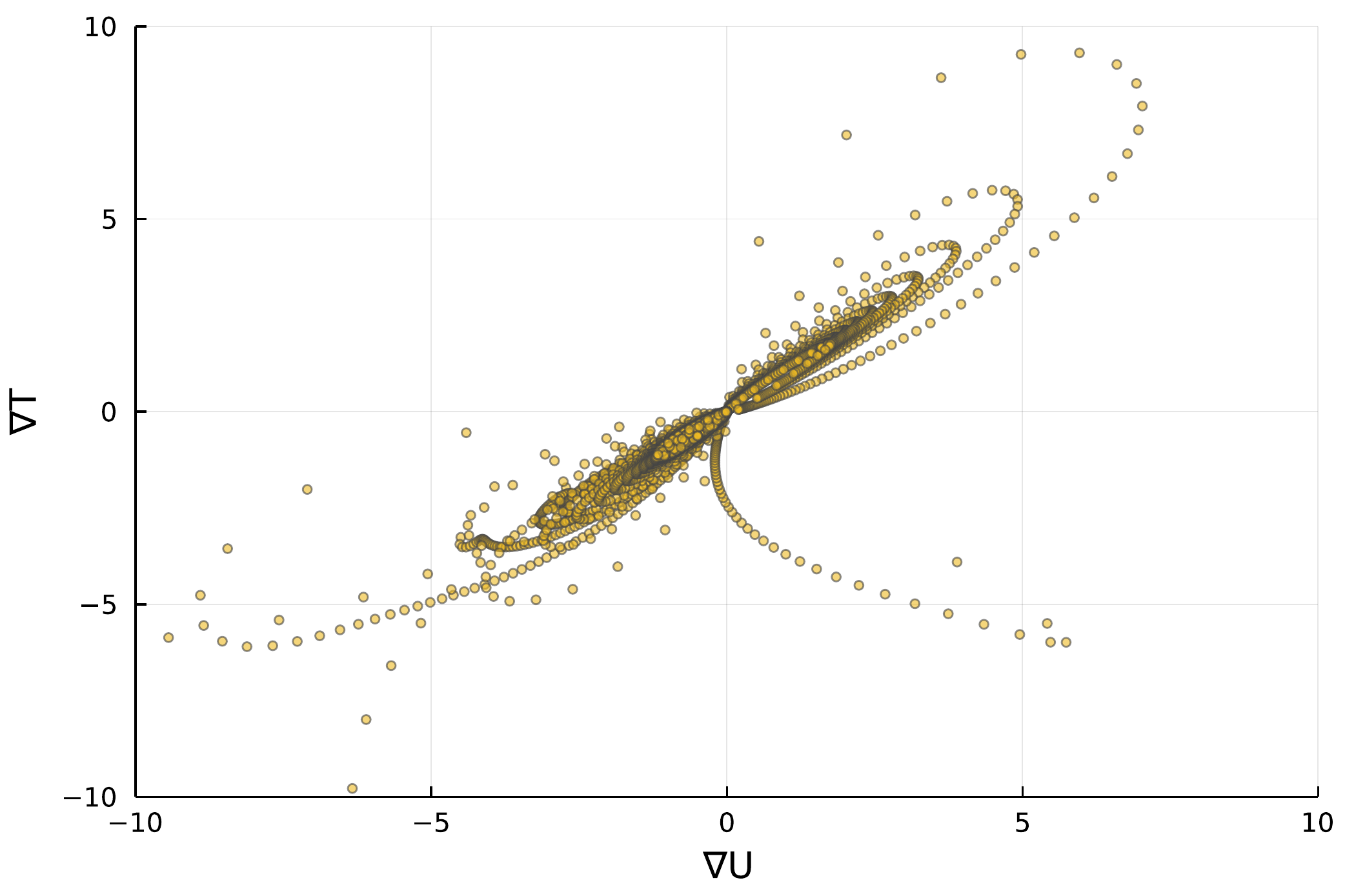}
	}
    \caption{Distributions of data points in $\nabla U$-$\nabla T$ phase diagram from the current algorithmic generator and sampled from Sod shock tube solution.}
	\label{fig:illustrator gradient}
\end{figure}

Using a randomly sampled Knudsen-number $Kn$ from a predefined range, we can compute the local collision time $\tau=1/\nu$ and obtain a completely assembled training data point $\mathbf U=(\mathbf W, \nabla_{\mathbf x} \mathbf W, \tau)$.
Finally we compute the label of the training data point by first computing $f_\mathrm{NS}$ using Eq.~\eqref{eqn:ce boltzmann} and then calculating the distance to the sampled reference solution $f_\mathrm{ref}$ using Eq.~\eqref{eqn:error}. The resulting sampling strategy is displayed in Alogrithm~\ref{alg_samplingMoments}.
\begin{algorithm}\label{alg_samplingMoments}
\DontPrintSemicolon
\caption{Sampling of labeled training data }
\SetAlgoLined
\SetKwInOut{Input}{Input}
\Input{Range of Knudsen numbers $[\text{Kn}_{min},\text{Kn}_{max}]$\newline
Range of particle densities $[\rho_{min},\rho_{max}]$\newline
Velocity space $V$\newline
Range of cell-center distances $[x_{min},x_{max}]$\newline
Range of particle densities $[\rho_{min},\rho_{max}]$
}
\KwResult{Training set with flow regime label$\menge{(U_j,R_j)}_{j\in J}$} 
Sample $\menge{f_{\text{ref},i}}_{i\in I}$ using Algorithm~\ref{alg_samplingUniform}\;
\For {$j = 1,\dots,\vert J\vert$}{
    Sample $f_L$ and $f_R$ from $\menge{f_{\text{ref},i}}_{i\in I}$\;
    Weight $f_L$ and $f_R$ with $\rho_L$ and $\rho_R$ sampled from $[\rho_{min},\rho_{max}]$\;
    Compute $\mathbf{W}_L$ and $\mathbf{W}_R$\;
    Sample the distance $[\delta_x,\delta_y]^T$ between ghost cell centers $x_L,x_R\in[x_{min},x_{max}]$\;
    Compute the unit normal vector $\mathbf{n}$ of the cell interface\;
    Compute $f_{\text{ref}}$ using Eq.~\eqref{eq_upwind}\;
    Compute $\mathbf{W}_{\text{ref}}$ from $f_{\text{ref}}$\;
    Compute $\nabla_x \mathbf{W}_{\text{ref}}$ with finite differences from  $\mathbf{W}_L$ and $\mathbf{W}_R$ and $[\delta_x,\delta_y]^T$\;
    Compute $\tau=\mu/p$\;
    Compute $f_\mathrm{NS}$ using Eq.~\eqref{eqn:ce boltzmann}\;
    Compute the regime label $\hat{\mathcal{R}}_j$ using Eq.~\eqref{eqn:error}\;
    Store $U_j=[\mathbf{W_{\text{ref}}},\nabla_x \mathbf{W_{\text{ref}}},\tau]$ and $\hat{\mathcal{R}}_j$\;
 }
\end{algorithm}
To illustrate the superiority of the current data generation strategy, we compare the data distributions resulting from Algorithm~\ref{alg_samplingMoments} to the data gathered from the simulation results of standard Sod shock tube problem with a full Boltzmann simulation.
Details of the setup can be found in Sec.~\ref{sec_sod_shock}.
Fig.~\ref{fig:illustrator value}(a) shows the macroscopic variables generated by the data generator using Algorithm~\ref{alg_samplingMoments} and Fig.~\ref{fig:illustrator value}(b) displays the generation from the simulation results. The results have been normalized via $\tilde{\mathbf{W}}=\mathbf{W}/\rho$ is displayed. It is evident, that the samples from the kinetic solver have a strong bias towards positive bulk velocity.
Temperature and velocity are strongly correlated. In contrast, the algorithmic sampler generates a wide range of macroscopic variables with different ranges of $\mathbf{U}$ and $T$.
Besides, the generated gradients of the macroscopic variables $\nabla_x\textbf{W}$ are shown in Fig.~\ref{fig:illustrator gradient}. The data sampled by the genreator is shown in Fig.~\ref{fig:illustrator gradient}(a) and exhibits a distribution that is concentrated around the origin, without strong bias towards a specific direction, whereas the data generated by the solver in Fig.~\ref{fig:illustrator gradient}(b) displays again a strong bias and fails to cover most parts of the domain. Furthermore, the presented sampling strategy does not require the computational expense of full simulations, possibly with multiple initial conditions. Computational resources for the data-sampler can be found in~\cite{KITRT}.
\section{Solution Algorithm}
In this section, we present the numerical implementation of the adaptive scheme based on the neural classifier.
The solution algorithm is built on top of a finite volume method.
\subsection{Kinetic solver}\label{sec:boltzmann}
Given the notation of cell-averaged particle distribution function in the physical element $\mathbf \Omega_i$ and velocity element $\mathbf \Omega_j$,
\begin{equation}
    f_{i,j}^n=\frac{1}{\mathbf \Omega_{i}(\mathbf x) \mathbf \Omega_{j}(\mathbf v)} \int_{\mathbf \Omega_i} \int_{\mathbf \Omega_j} f(t^n,\mathbf x,\mathbf v) d\mathbf x d \mathbf v,
\end{equation}
the update algorithm of finite volume scheme writes
\begin{equation}
    f_{i,j}^{n+1}=f_{i,j}^{n}-\frac{1}{\mathbf \Omega_{i}} \sum_{r \in \partial \mathbf \Omega_i} \int_{t^n}^{t^{n+1}} \mathbf{F}^f_{r,j} \cdot \boldsymbol{n}_{r} S_{r} d t+\int_{t^n}^{t^{n+1}} Q_j(f_i, f_i) d t,
    \label{eqn:pdf update}
\end{equation}
where $\mathbf n_r$ is the unit normal vector of surface $r$ that points outside of the element $\mathbf \Omega_i$, and $S_r$ is the surface area.
The interface flux of distribution function $\mathbf{F}^f$ can be computed via an upwind reconstruction,
\begin{equation}
    \mathbf F^f_{i+1/2,j} = \mathbf v_j \left( f_{L} H(\mathbf v_j \cdot \mathbf n_{i+1/2}) + f_{R} \left( 1-H(\mathbf v_j \cdot \mathbf n_{i+1/2}) \right) \right),
    \label{eqn:kinetic flux}
\end{equation}
where $H$ is the heaviside step function, and the status on the left and right sides of the interface are reconstructed via
\begin{equation}
\begin{aligned}
    & f_L = f_{i,j} + \nabla f_{i,j} \cdot (\mathbf x_{i+1/2} - \mathbf x_i), \\
    & f_R = f_{i+1,j} + \nabla f_{i+1,j} \cdot (\mathbf x_{i+1/2} - \mathbf x_{i+1}).
\end{aligned}
\end{equation}
Inside each element, the collision term $Q(f,f)$ is computed by the fast spectral method \cite{xiao2021kinetic}.
The discrete Fourier transform is employed to solve the convolution in the spectral domain efficiently.
We refer to \cite{mouhot2006fast} for detailed formulation of this method.
\subsection{Navier-Stokes solver}\label{sec:ns}
We define the average conservative flow variables in an element as
\begin{equation}
    \mathbf W_{i}^n=\frac{1}{\mathbf \Omega_{i}(\mathbf x)} \int_{\mathbf \Omega_i} \mathbf W(t^n,\mathbf x) d\mathbf x,
\end{equation}
and the finite volume algorithm writes
\begin{equation}
    \mathbf W_{i}^{n+1}=\mathbf W_{i}^{n}-\frac{1}{\mathbf \Omega_{i}} \sum_{r \in \partial \mathbf \Omega_i} \int_{t^n}^{t^{n+1}} \mathbf{F}^W_{r} \cdot \boldsymbol{n}_{r} S_{r} d t.
    \label{eqn:macro update}
\end{equation}
A key step for solving conservation laws is to compute the fluxes $\mathbf F^W$ of conservative variables.
Here, we employ the Chapman-Enskog solution from the BGK-type relaxation model \cite{bhatnagar1954model} to construct numerical fluxes.
The relaxation model writes
\begin{equation}
    \partial_t f + \mathbf v \cdot \nabla_{\mathbf x} f = \nu (\mathcal E - f).
\end{equation}
The equilibrium distribution $\mathcal E$ can be chosen as the Maxwellian in Eq.(\ref{eqn:maxwellian}) or its variants \cite{shakhov1968generalization,holway1966new},
and $\nu$ is the collision frequency.
The above equation can be written into the following successive form
\begin{equation}
    f = \mathcal E - \tau D_t \mathcal E + \tau D_t (\tau D_t \mathcal E) + \cdots,
\end{equation}
where $D_t$ denotes total derivative operator and $\tau=1/\nu$.
The above equation has the same structure as Eq.(\ref{eqn:chapman enskog}), and thus the first-order truncation of Chapman-Enskog expansion writes \cite{ohwada2004kinetic},
\begin{equation}
    f \simeq \mathcal E - \tau (\partial_t \mathcal E + \mathbf v \cdot \nabla_{\mathbf x} \mathcal E).
\end{equation}
In the solution algorithm, we follow the Chapman-Enskog expansion and construct the particle distribution function at interface $\mathbf x_{i+1/2}$ with an upwind approach,
\begin{equation}
\begin{aligned}
    &f_L = \mathcal E_{L} \left(1-\tau\left(\mathbf a_L \cdot \mathbf v+b_L\right)\right), \\
    &f_R = \mathcal E_{R} \left(1-\tau\left(\mathbf a_R \cdot \mathbf v+b_R\right)\right),
\end{aligned}
\label{eqn:interface f}
\end{equation}
where $\{\mathcal E_L,\mathcal E_R\}$ are the equilibrium distributions computed from reconstructed macroscopic variables, i.e.,
\begin{equation}
\begin{aligned}
    &\mathbf W_L = \mathbf W_{i,j} + \nabla \mathbf W_{i,j} \cdot (\mathbf x_{i+1/2} - \mathbf x_i), \\
    &\mathbf W_R = \mathbf W_{i+1,j} + \nabla \mathbf W_{i+1,j} \cdot (\mathbf x_{i+1/2} - \mathbf x_{i+1}).
\end{aligned}
\end{equation}
In a well-resolved region, the relation $\mathbf W_L=\mathbf W_R$ holds, and Eq.(\ref{eqn:interface f}) deduces to standard Chapman-Enskog expansion naturally.
The spatial derivatives of the particle distribution function $\mathbf a_{L,R}$ is related to macroscopic slopes via
\begin{equation}
    \int \mathbf a_{L,R} \mathcal E_{L,R}\psi d\mathbf v=\nabla_{\mathbf x} \mathbf W_{L,R},
\end{equation}
where $\psi=(1,\mathbf v,\mathbf v^2/2)^T$ are the collision invariants.
Then $\mathbf a_{L,R}$ can be obtained by solving a linear system \cite{xubook}.
Then the time derivative $b_{L,R}$ can be obtained through the compatibility condition of the BGK model, i.e.,
\begin{equation}
    \int \nu (\mathcal E - f) \psi d\mathbf v = 0,
\end{equation}
which yields
\begin{equation}
    \int b_{L,R} \mathcal E_{L,R}\psi d\mathbf v
    =-\int (a_{L,R} \cdot \mathbf v) \mathcal E_{L,R}\psi d\mathbf v.
\end{equation}
After the coefficients for spatial and time variations are determined, the interface fluxes for macroscopic variables can be obtained by taking moments over particle velocity space, i.e.,
\begin{equation}
    \mathbf F^W_{i+1/2,j} = \int \mathbf v \left( f_{L} H(\mathbf v \cdot \mathbf n_{i+1/2}) + f_{R} \left( 1-H(\mathbf v \cdot \mathbf n_{i+1/2}) \right) \right) \psi d\mathbf v,
\end{equation}
where $H$ is the heaviside step function.
Since the equilibrium state is based on Gaussian distribution, the above integral can be evaluated analytically.
It is remarkable that the above numerical method can be understood as a simplification of gas-kinetic scheme \cite{xubook,xiao2020velocity}.

\subsection{Adaptation strategy}
\begin{figure}[htb!]
	\centering
	{
		\begin{tikzpicture}[thick]
		\node[rectangle] (bx) {};
		\node[rotate = 0] at (-3.5, -0.4) {$\underbrace{\hspace{6.9cm}}$};
		\node[rotate = 0] at (+3.5, -0.4) {$\underbrace{\hspace{6.9cm}}$};
		\node at ($(bx)+(+0,+1.0)$) {adaptation face};
		\node at ($(bx)+(-4,+1.0)$) {continuum face};
		\node at ($(bx)+(+4,+1.0)$) {kinetic face};
		\draw[line] ($(bx)+(0,+0.8)$) -- ($(bx)+(+0,+0.3)$);
		\draw[line] ($(bx)+(-4,+0.8)$) -- ($(bx)+(-4,+0.3)$);
		\draw[line] ($(bx)+(+4,+0.8)$) -- ($(bx)+(+4,+0.3)$);
		\node at ($(bx)+(-3.5,-0.9)$) {near-equilibrium};
		\node at ($(bx)+(+3.5,-0.9)$) {non-equilibrium};
		\draw ($(bx)+(0,-0.2)$) -- ($(bx)+(0,+0.2)$);
		\draw ($(bx)+(-2,-0.2)$) -- ($(bx)+(-2,+0.2)$);
		\draw ($(bx)+(-4,-0.2)$) -- ($(bx)+(-4,+0.2)$);
		\draw ($(bx)+(-6,-0.2)$) -- ($(bx)+(-6,+0.2)$);
		\draw ($(bx)+(+2,-0.2)$) -- ($(bx)+(+2,+0.2)$);
		\draw ($(bx)+(+4,-0.2)$) -- ($(bx)+(+4,+0.2)$);
		\draw ($(bx)+(+6,-0.2)$) -- ($(bx)+(+6,+0.2)$);
		%lower part
		\draw[line] ($(bx)+(-7,+0)$) -- ($(bx)+(+7,+0)$);
		\draw[line] ($(bx)+(-6,-2)$) -- ($(bx)+(+6,-2)$);
		\node at ($(bx)+(-3.5,-1.7)$) {Navier-Stokes};
		\node at ($(bx)+(+3.5,-1.7)$) {Boltzmann};
		\draw[line] ($(bx)+(-1.5,-1.7)$) -- ($(bx)+(+1.5,-1.7)$);
		\end{tikzpicture}
	}
	\caption{Schematic of the adaptive scheme for multi-scale flow.}
	\label{fig:hybrid}
\end{figure}
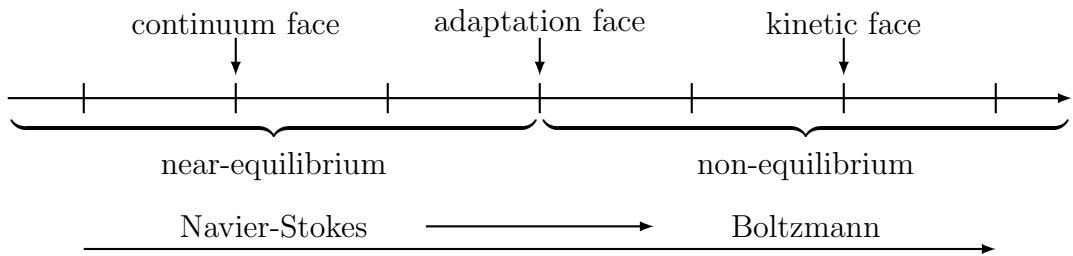

The Boltzmann and Navier-Stokes solvers can be combined to solve multi-scale flow problems efficiently with an adaptive continuous-discrete velocity transformation.
The work paradigm is shown in Fig. \ref{fig:hybrid}.
For a near-equilibrium flow region, the particle distribution function is formulated analytically from the Chapman-Enskog expansion.
Therefore, only the macroscopic flow variables are needed to store and iterate by the Navier-Stokes solver in Eq.(\ref{eqn:macro update}).
For non-equilibrium flows, the solution algorithm allocates the localized velocity quadrature to track the evolution of particle distribution function in Eq.(\ref{eqn:pdf update}).

A core task of the hybrid solver lies in the dynamic adaptation of time-varying flow regimes at different locations.
At every time step $t^n$, the spatial derivatives of the updated macroscopic variables are evaluated via $\nabla_{\mathbf x} \mathbf W=(\nabla_{\mathbf x} \mathbf W_L+\nabla_{\mathbf x} \mathbf W_R)/2$, where $\nabla_{\mathbf x} \mathbf W_{L,R}$ are the difference values between to neighboring cells.
The collision time is evaluated by $\tau=\mu/p$.
Therefore, the complete information needed for the neural network to predict the flow regime has been obtained.
As shown in Fig. \ref{fig:hybrid}, we have two types of cells, i.e. the non-equilibrium one holding discrete solution of distribution function and the near-equilibrium one with Navier-Stokes variables, and three types of cell interfaces based on the flow regimes, i.e.,
\begin{itemize}
    \item kinetic face: two neighboring cells are in non-equilibrium flow regime;
    \item continuum face: two neighboring cells of the face are in near-equilibrium flow regime;
    \item adaptation face: two neighboring cells of the face lie in different flow regimes.
\end{itemize}

The solution algorithm in type 1/2 cells is straightforward following the section \ref{sec:boltzmann} and \ref{sec:ns}.
At the adaptation face, both macroscopic and microscopic fluxes are evaluated to update the solutions in the left and right cells.
This is uniformly done by computing the kinetic flux in Eq.(\ref{eqn:kinetic flux}), where its velocity moments results macroscopic fluxes, i.e.,
\begin{equation}
    \mathbf F^W = \int \mathbf F^f \psi d\mathbf v \simeq \sum_j^{N_q} w_j \mathbf F_j \psi ,
\end{equation}
where $N_q$ is the number of quadrature points and $w_j$ the quadrature weights.
To utilize the above equation, a local velocity mesh is generated within
\begin{equation}
\mathbf v\in \left[-|\mathbf V_0|-4\sqrt{RT_0},\ |\mathbf V_0|+4\sqrt{RT_0} \right],
\end{equation}
where $\{\mathbf V_0, T_0\}$ are reference velocity and temperature, and $R$ is the gas constant.
The velocity grid is chosen such that more than $99\%$ of values of the Maxwellian distribution fall into its range.
In a continuum cell at $t^n$ which has discrete solution of distribution function at $t^{n-1}$, the memory can be freed by deallocations in static languages, e.g. C and Fortran, and by setting to be "None" type in dynamic languages, e.g. Python and Julia.
In a kinetic cell with no former record of discretized distribution function, the solution is reconstructed from the Chapman-Enskog expansion in Eq.(\ref{eqn:ce boltzmann}) in the continuum cell, and then used for flux evaluation.
This way, a hybrid continuum-kinetic solver has been set up, where no buffer zone is required to transit solutions.

\section{Numerical Experiments}

In this section, we conduct numerical experiments of several multi-scale flow problems to validate the neural classifier and the corresponding adaptive solver.
All the variables are nondimensionalized following the paradigm introduced in Sec. \ref{sec:theory}.
%We introduce the following dimensionless variables
%\begin{equation}
%    \tilde{\mathbf x} = \frac{\mathbf x}{L_0},  \ \tilde t = \frac{t}{L_0/V_0}, \ \tilde{\mathbf v} = \frac{\mathbf v}{V_0}, \ \tilde f = \frac{f}{n_0 V_0^3}
%\end{equation}
%where $V_0=\sqrt{2kT_0/m}$ is the most probable molecular speed.
%For brevity, we drop the tilde notation to denote dimensionless variables henceforth.
The hard-sphere gas model is employed in all cases.
We choose the gradient-length-local Knudsen number $\mathrm{Kn}_{GLL}$ \cite{boyd1995predicting} as reference and provide some quantitative comparisons to predict continuum breakdown.
It is worth mentioning that we are not here to censure this methodology, but rather to choose a widely accepted criterion as benchmark to point out potential possibilities of our new method. The computational resources of the hybrid solver can be found in~\cite{xiao2021kinetic}.

\subsection{Sod shock tube}\label{sec_sod_shock}

The first numerical experiment is the Sod shock tube, where the longitudinal processes dominate the flow motion in the one-dimensional Riemann problem.
The particle distribution function is initialized  as a Maxwellian, which corresponds to the following macroscopic variables
\begin{equation*}
\left(\begin{array}{c}
\rho \\
U \\
T \\
\end{array}\right)_{{t=0,L}}=\left(\begin{array}{c}
1 \\
0 \\
2 \\
\end{array}\right), \quad
\left(\begin{array}{c}
\rho \\
U \\
T \\
\end{array}\right)_{{t=0,R}}=\left(\begin{array}{c}
0.125 \\
0 \\
1.6 \\
\end{array}\right).
\end{equation*}
To test the capability of the current scheme to solve multi-scale flow problems, simulations are performed with different reference Knudsen numbers ranging in $\mathrm{Kn}=[0.0001,0.01]$.
The detailed computation setup is listed in Table \ref{tab:sod}.
\begin{table}[htbp]
	\caption{Computational setup of Sod shock tube problem.} 
	\centering
	\begin{tabular}{lllllll} 
		\hline
		$t$ & $x$ & $N_x$ & $\mathbf v$ & $N_u$ & $N_v$ & $N_w$ \\ 
		$[0,0.15]$ & $[0,1]$ & $200$ & $[-8,8]^3$ & 64 & 32 & 32 \\ 
		\hline 
		Quadrature & Kn & CFL & Integrator & Boundary \\ 
		Rectangular & $[0.0001,0.01]$ & $0.5$ & Euler & Dirichlet  \\
		\hline
	\end{tabular} 
	\label{tab:sod}
\end{table}

\begin{figure}
    \centering
	\subfigure[$\rm Kn=0.0001$]{
		\includegraphics[width=0.31\textwidth]{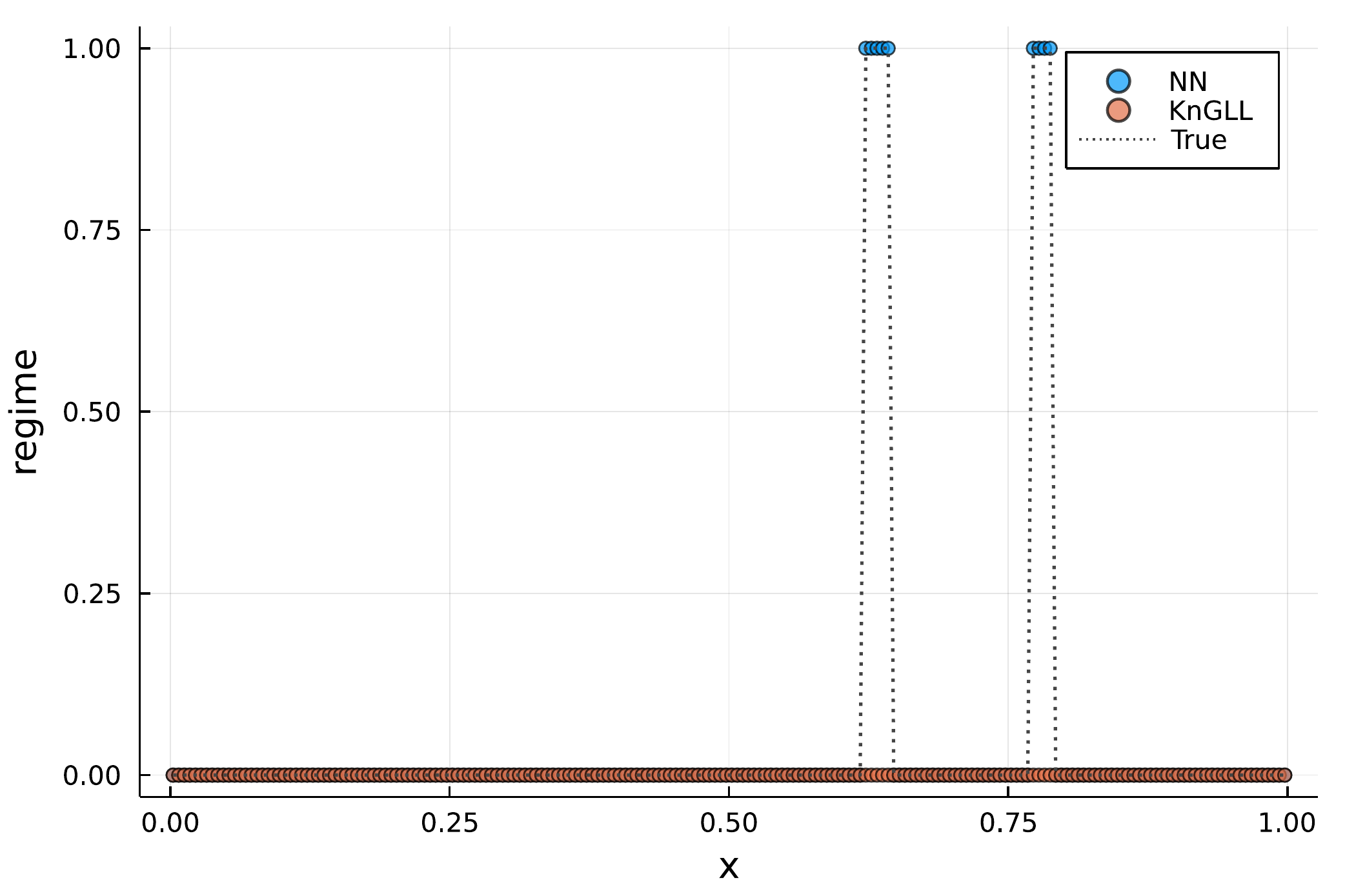}
	}
	\subfigure[$\rm Kn=0.001$]{
		\includegraphics[width=0.31\textwidth]{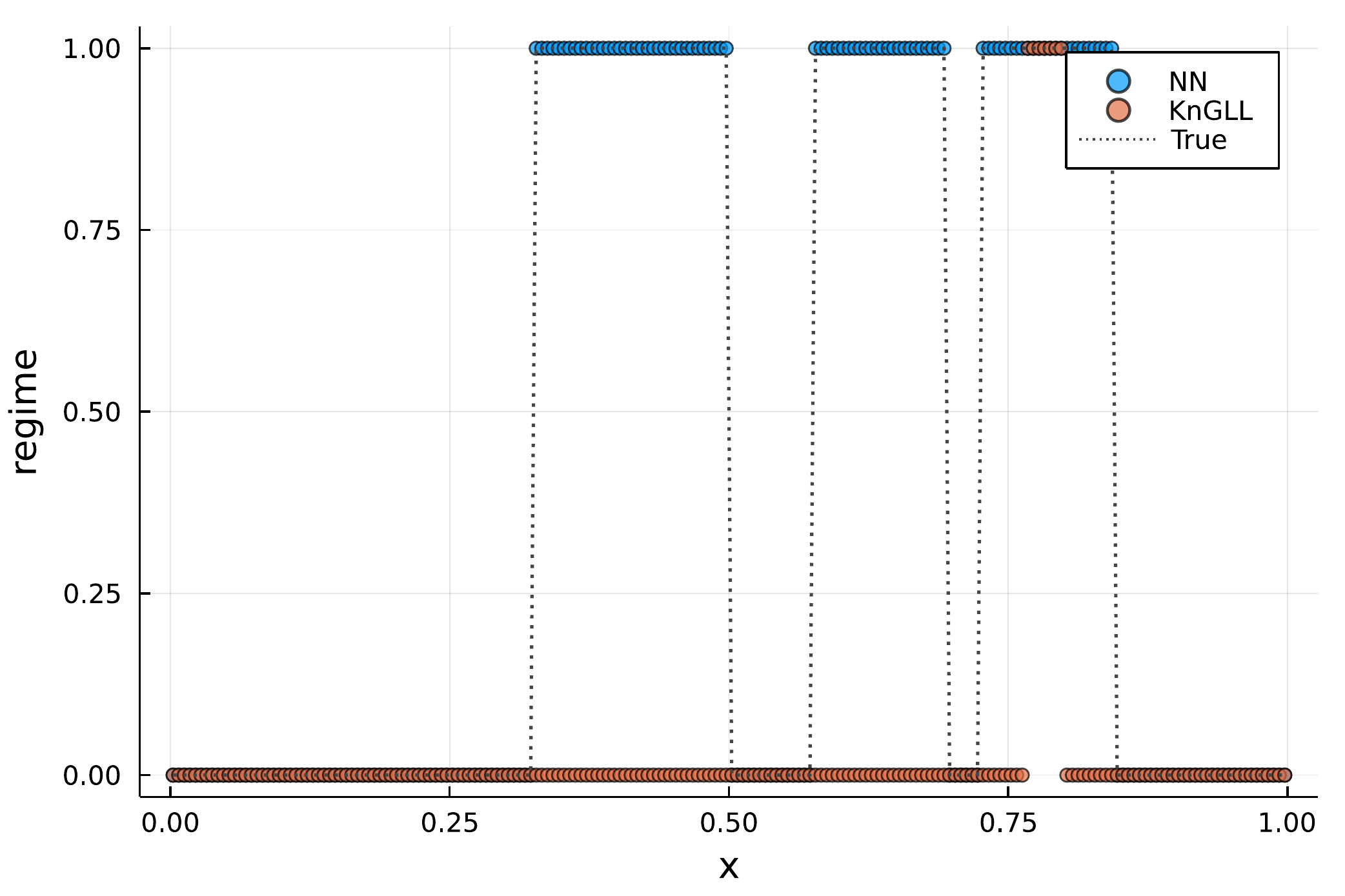}
	}
	\subfigure[$\rm Kn=0.01$]{
		\includegraphics[width=0.31\textwidth]{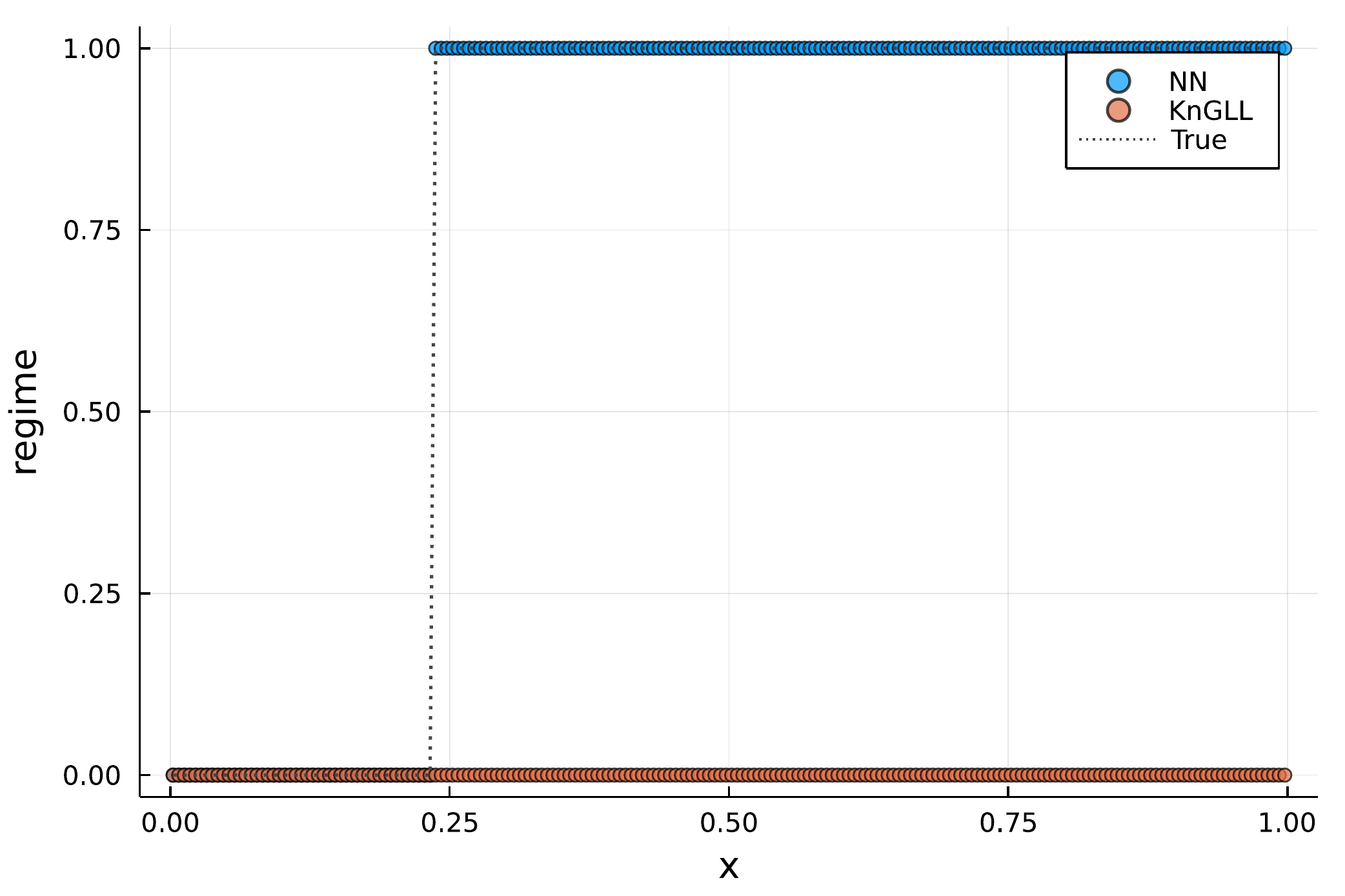}
	}
    \caption{Prediction of flow regimes from fully kinetic solutions at $t=0.15$ in the Sod shock tube with different criteria (0 denotes near-continuum, 1 denotes non-equilibrium).}
    \label{fig:sod regime}
\end{figure}

We first conduct a full kinetic simulation with the Boltzmann equation.
Based on the kinetic solution, the partition of flow regimes based on different criteria is shown in Fig. \ref{fig:sod regime}.
The ground-truth regime is obtained from the $L^2$ error between the particle distribution and its Chapman-Enskog reconstructed value in Eq.(\ref{eqn:error}).
It is clear that localized flow structures, including rarefaction wave, contact discontinuity and shock wave, contribute as sources of non-equilibrium effect.
In the reminding near-equilibrium regions the Chapman-Enskog expansion is able to approximate real particle distributions.
With the increasing Knudsen number, the kinetic regime enlarges due to the increasing rarefied gas effect.

\begin{figure}
    \centering
    \subfigure[Density]{
		\includegraphics[width=0.47\textwidth]{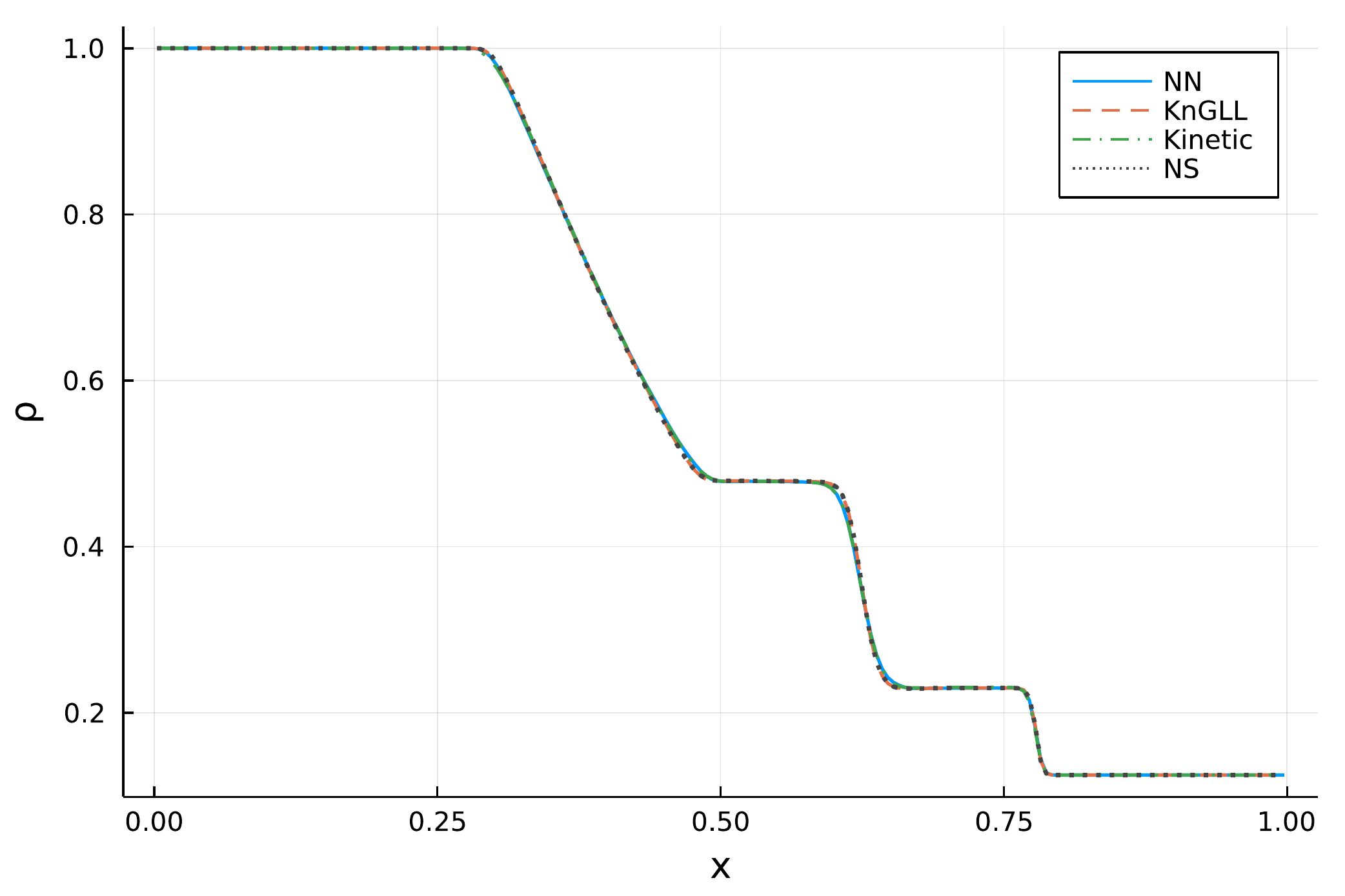}
	}
	\subfigure[Temperature]{
		\includegraphics[width=0.47\textwidth]{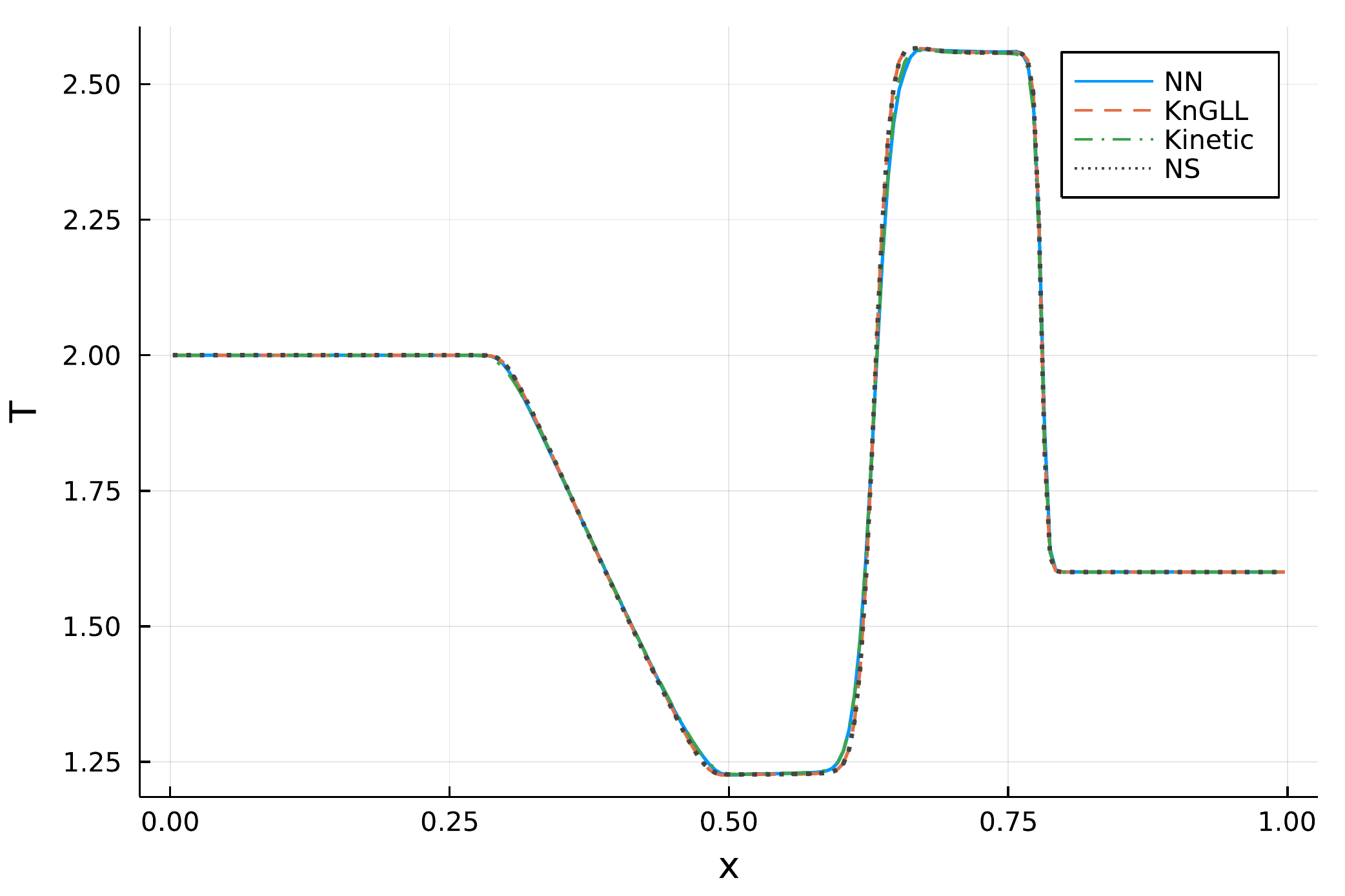}
	}
	\subfigure[Density (zoom-in)]{
		\includegraphics[width=0.47\textwidth]{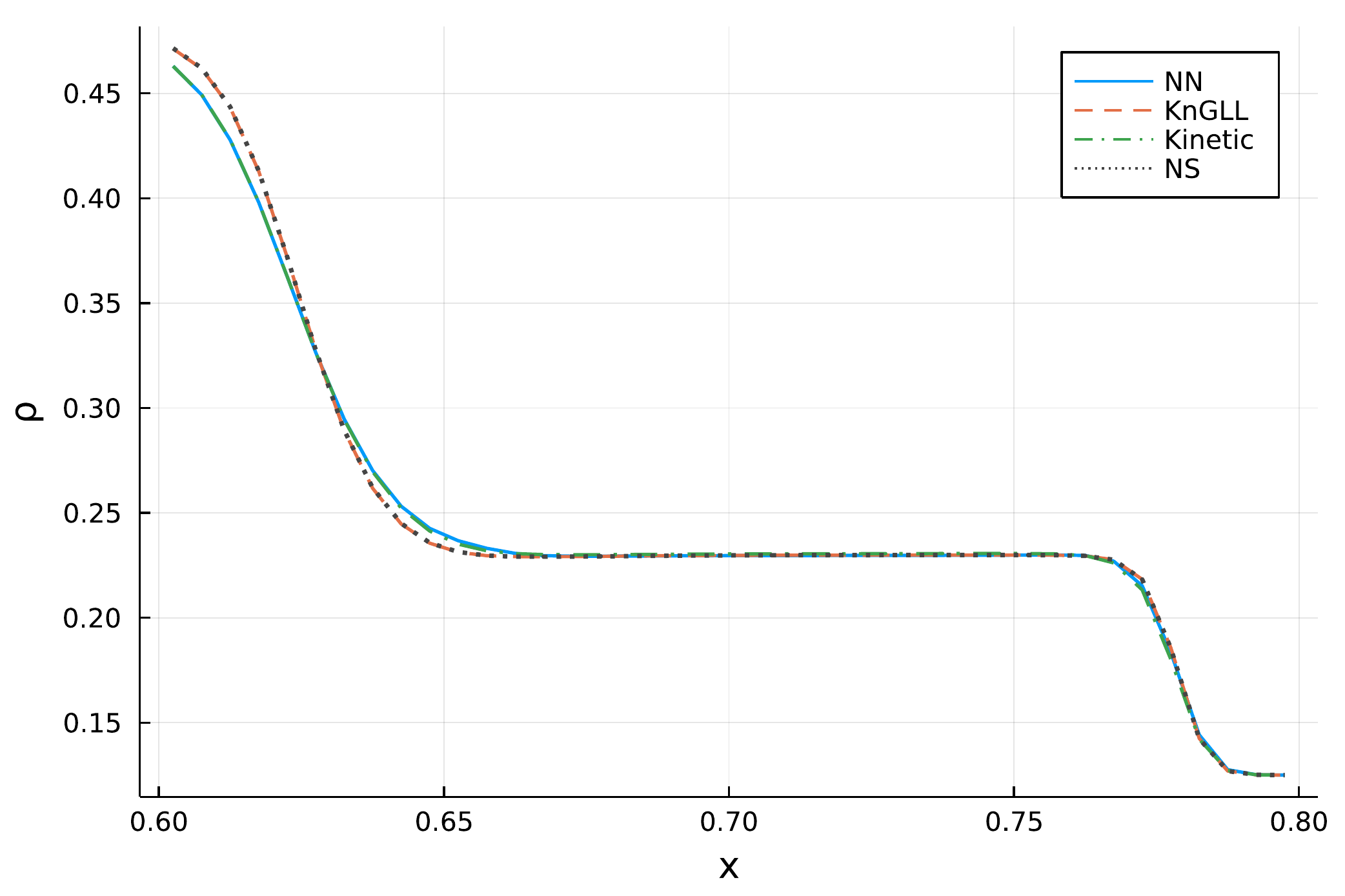}
	}
	\subfigure[Temperature (zoom-in)]{
		\includegraphics[width=0.47\textwidth]{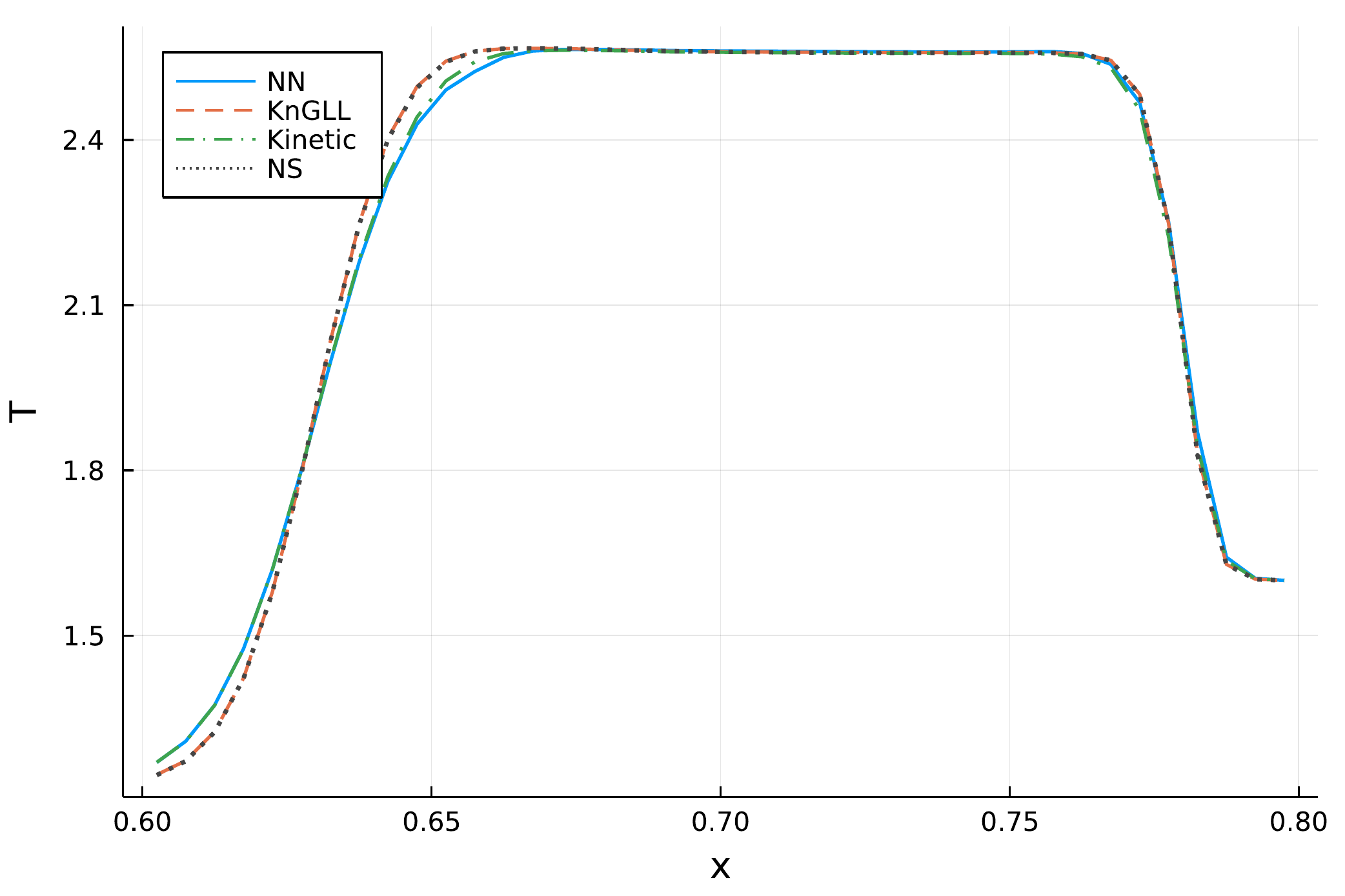}
	}
    \caption{Profiles of density and temperature in the shock tube at $t=0.15$ under $\rm{Kn}=0.0001$.}
    \label{fig:sod kn4}
\end{figure}

\begin{figure}
    \centering
    \subfigure[Density]{
		\includegraphics[width=0.47\textwidth]{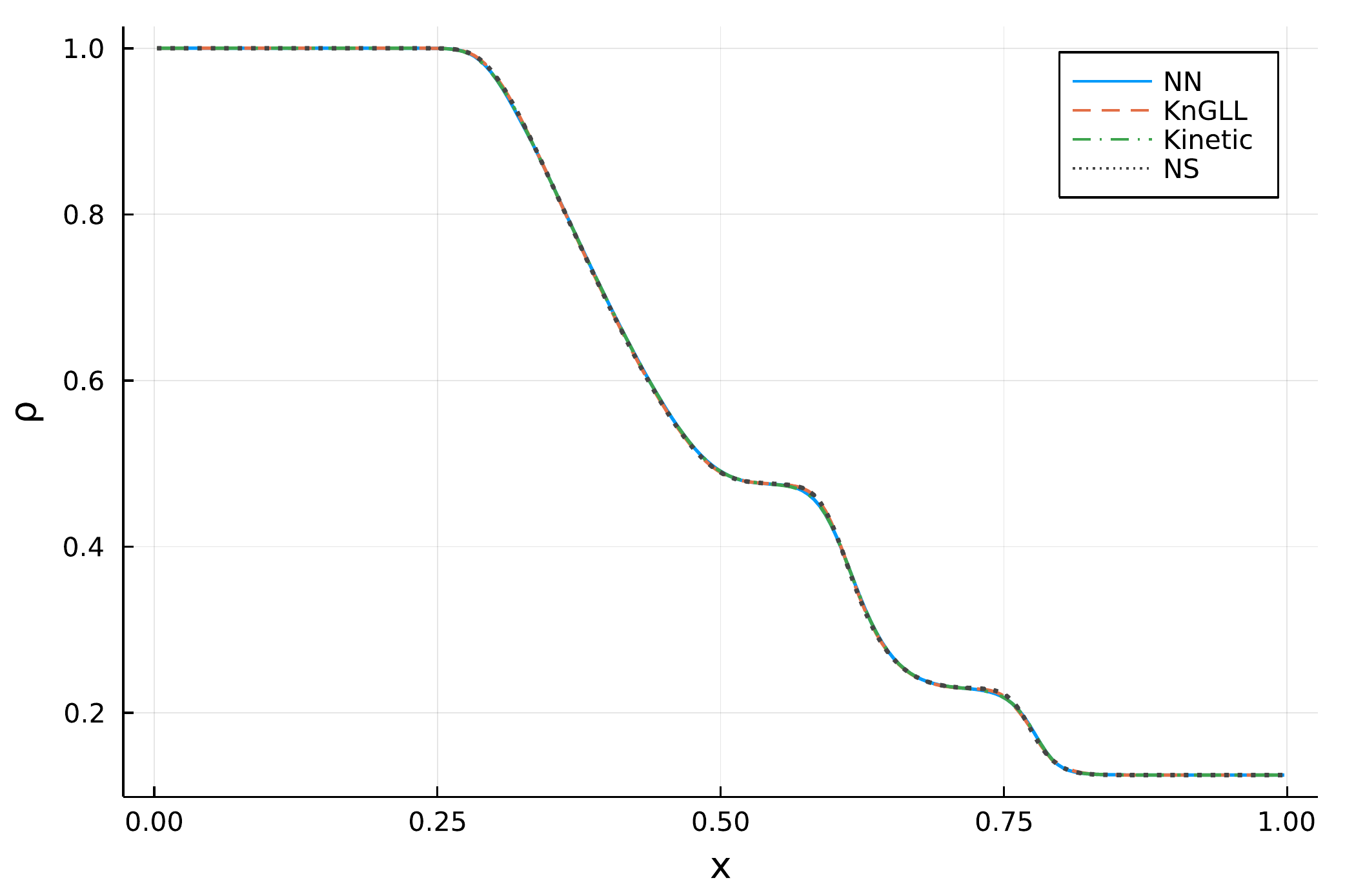}
	}
	\subfigure[Temperature]{
		\includegraphics[width=0.47\textwidth]{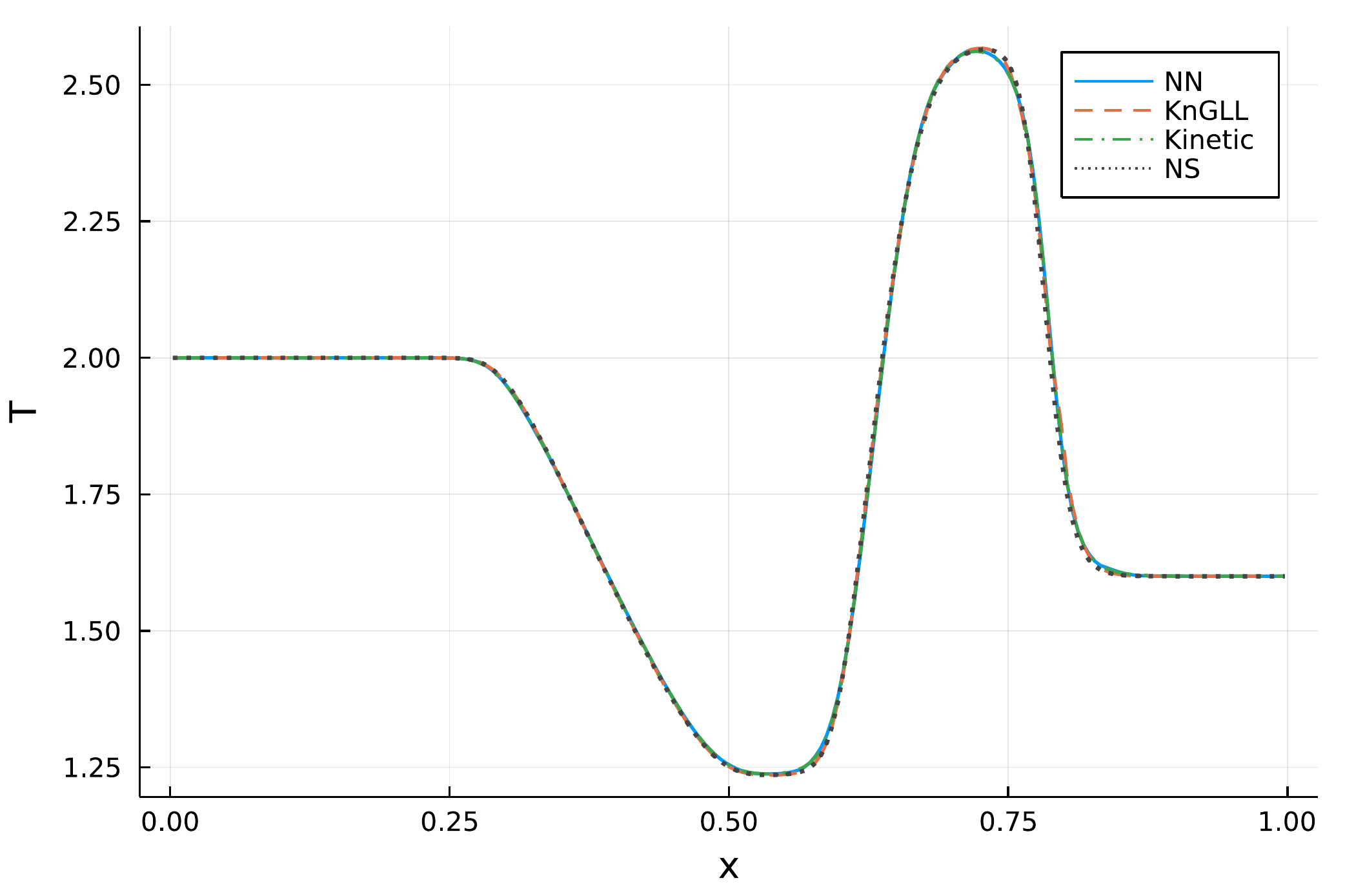}
	}
	\subfigure[Density]{
		\includegraphics[width=0.47\textwidth]{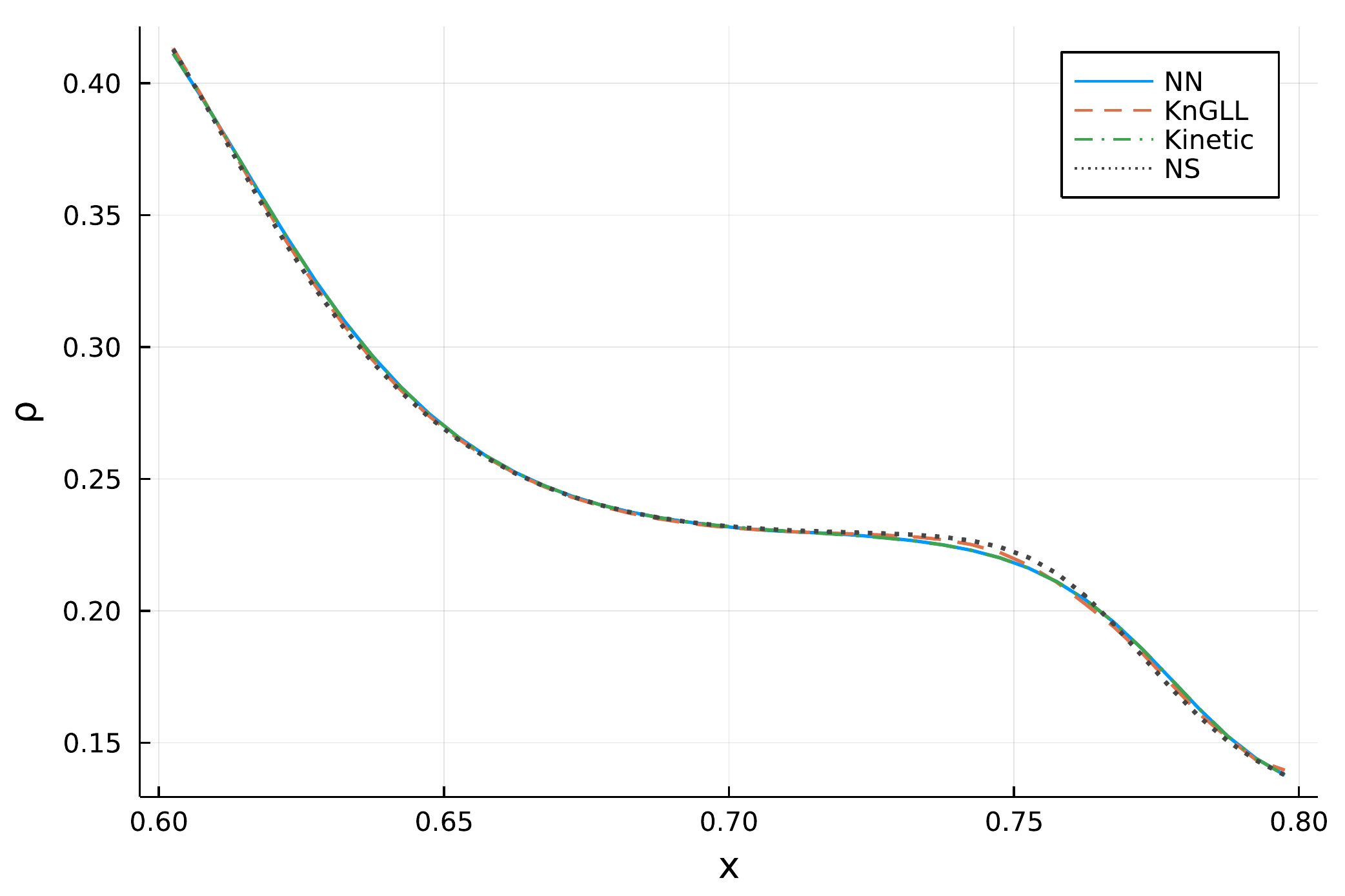}
	}
	\subfigure[Temperature]{
		\includegraphics[width=0.47\textwidth]{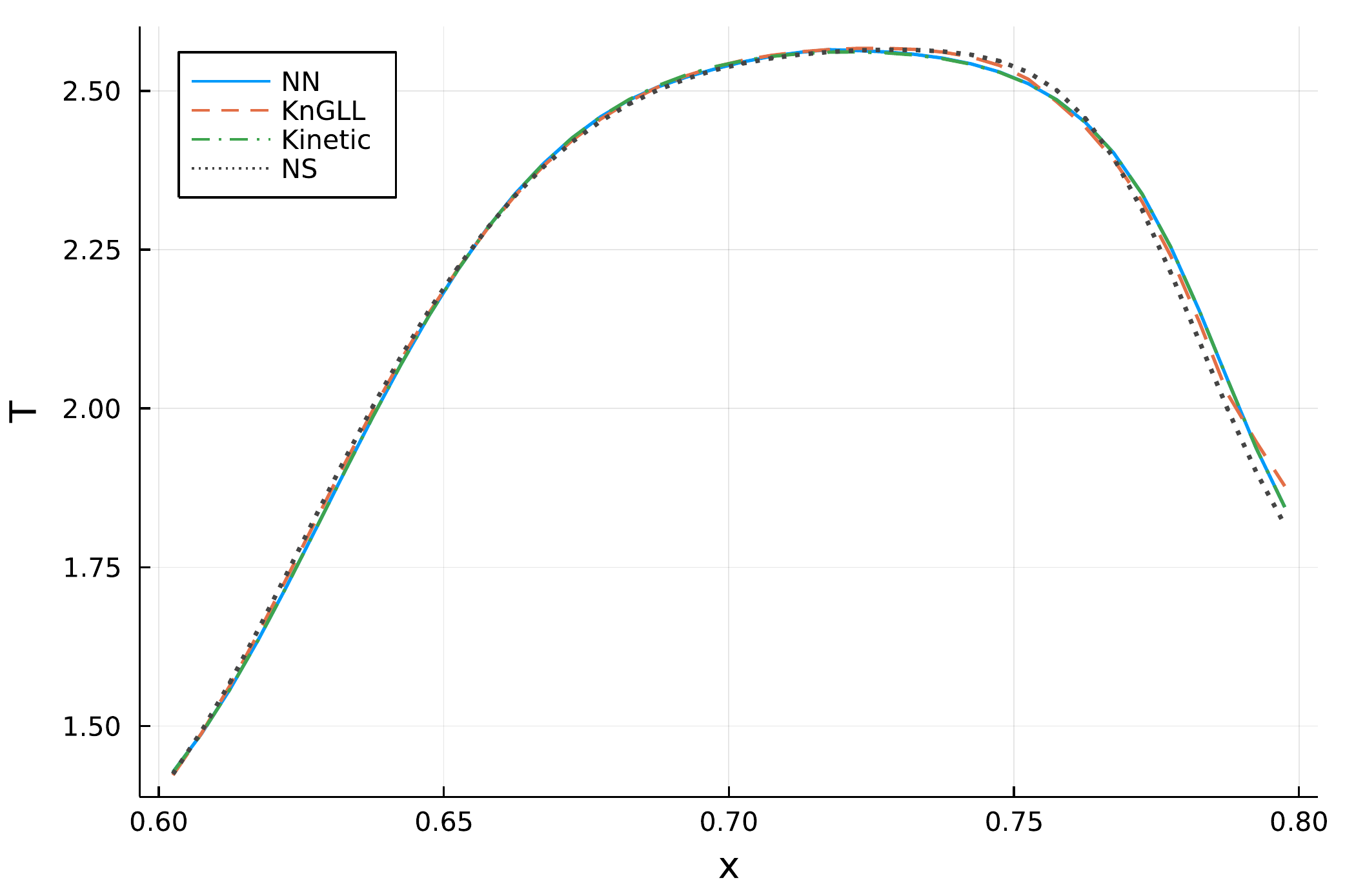}
	}
    \caption{Profiles of density and temperature in the shock tube at $t=0.15$ under $\rm{Kn}=0.001$.}
    \label{fig:sod kn3}
\end{figure}

\begin{figure}
    \centering
    \subfigure[Density]{
		\includegraphics[width=0.47\textwidth]{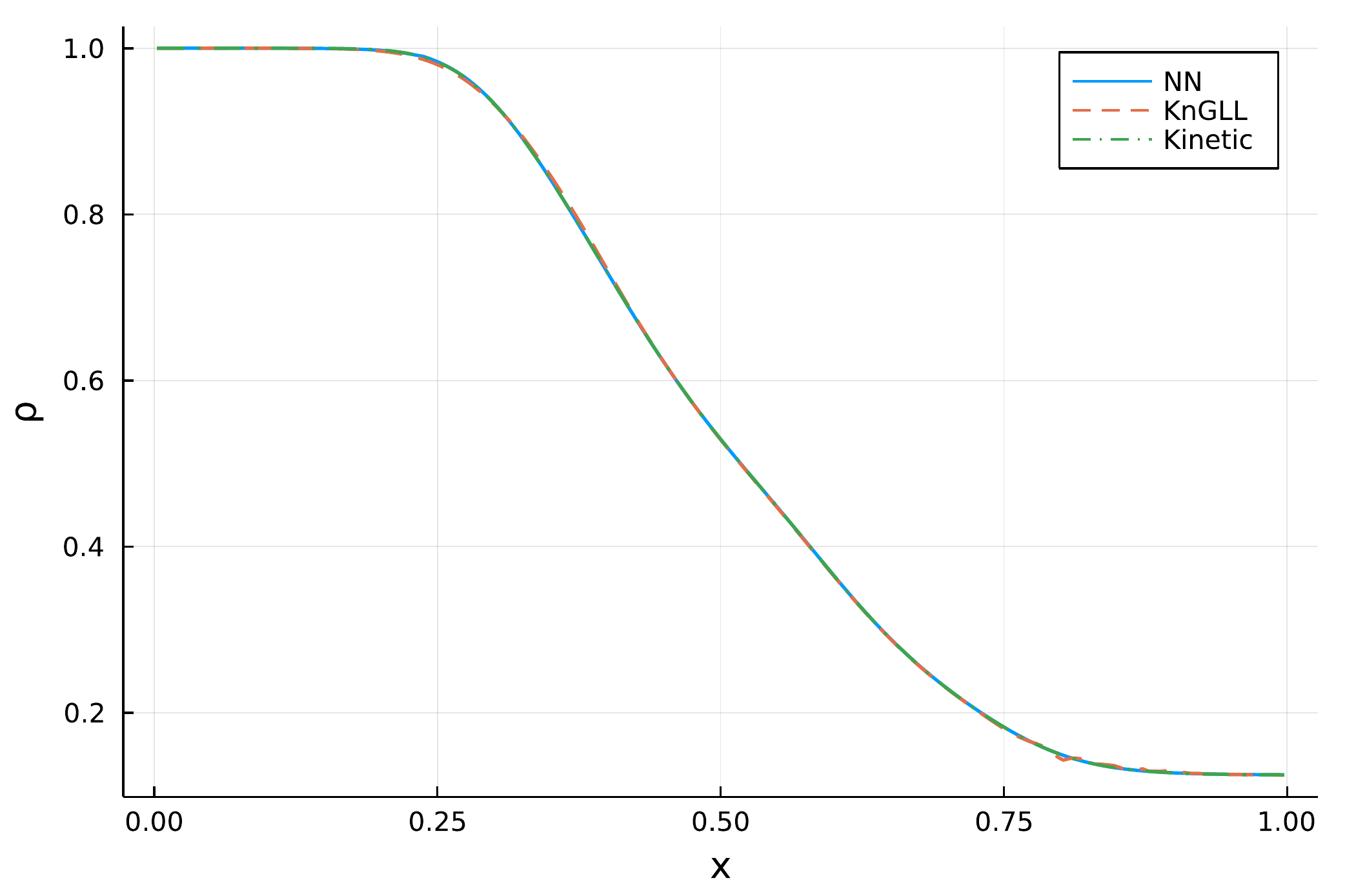}
	}
	\subfigure[Temperature]{
		\includegraphics[width=0.47\textwidth]{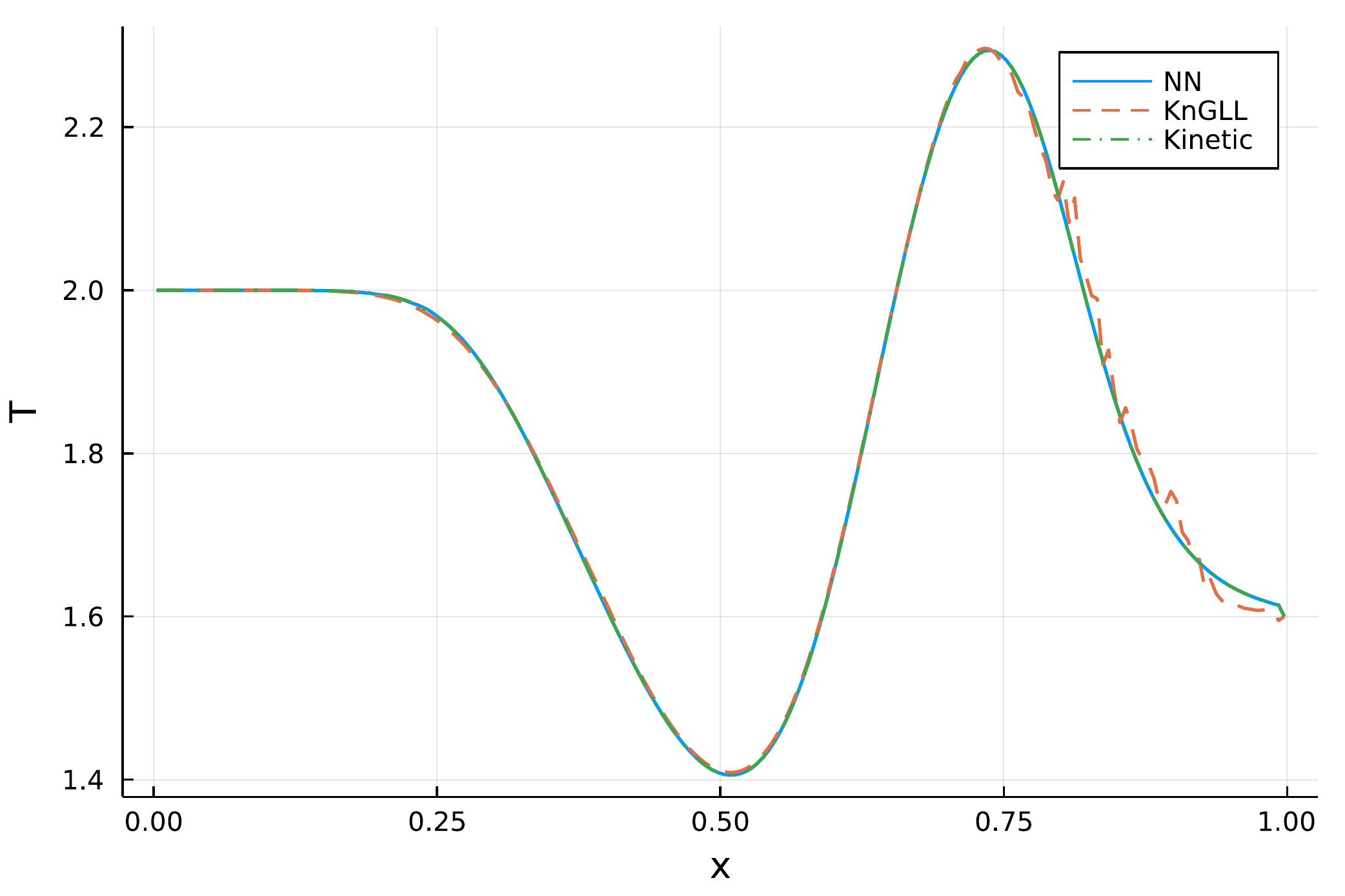}
	}
	\subfigure[Density]{
		\includegraphics[width=0.47\textwidth]{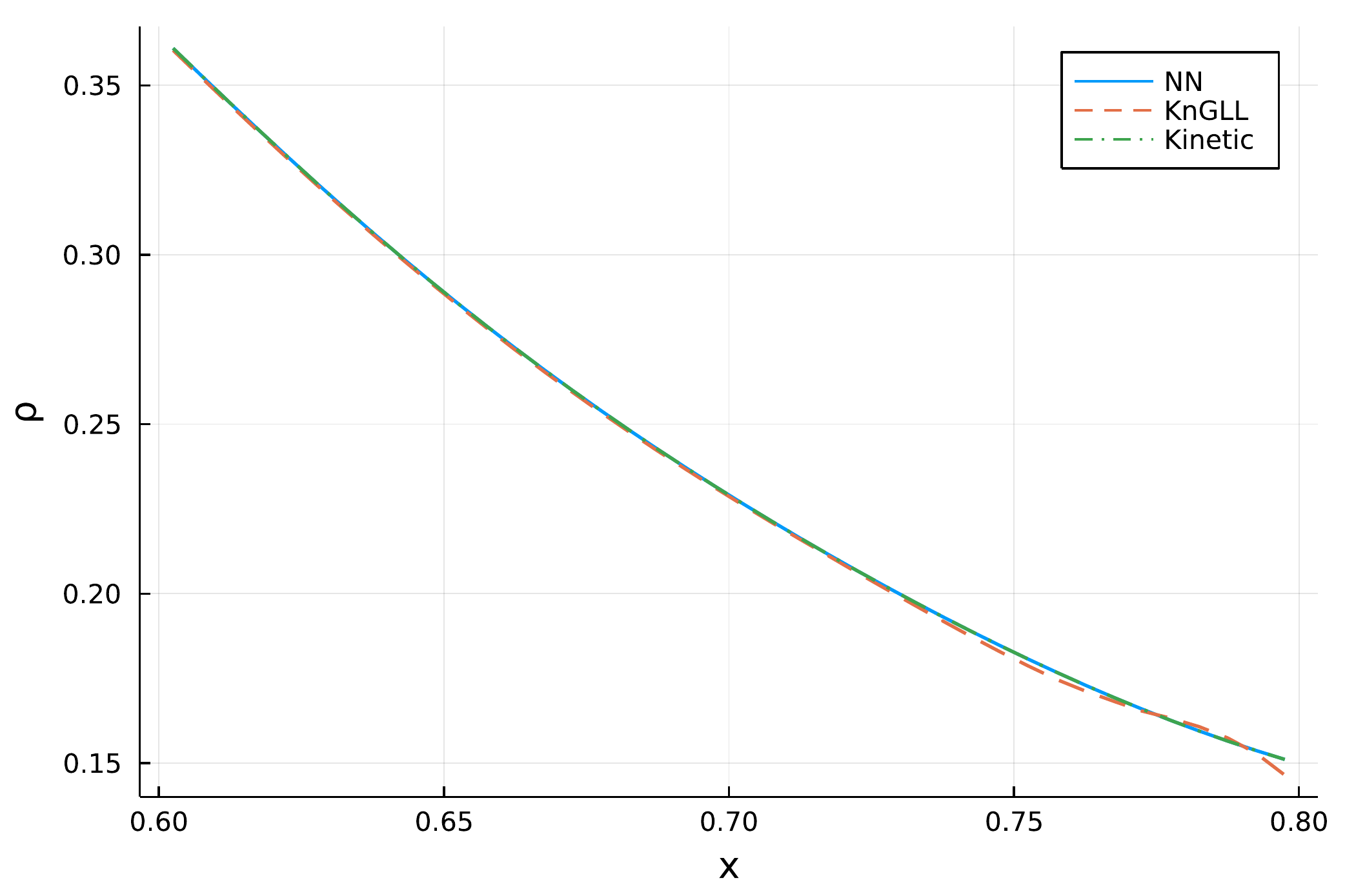}
	}
	\subfigure[Temperature]{
		\includegraphics[width=0.47\textwidth]{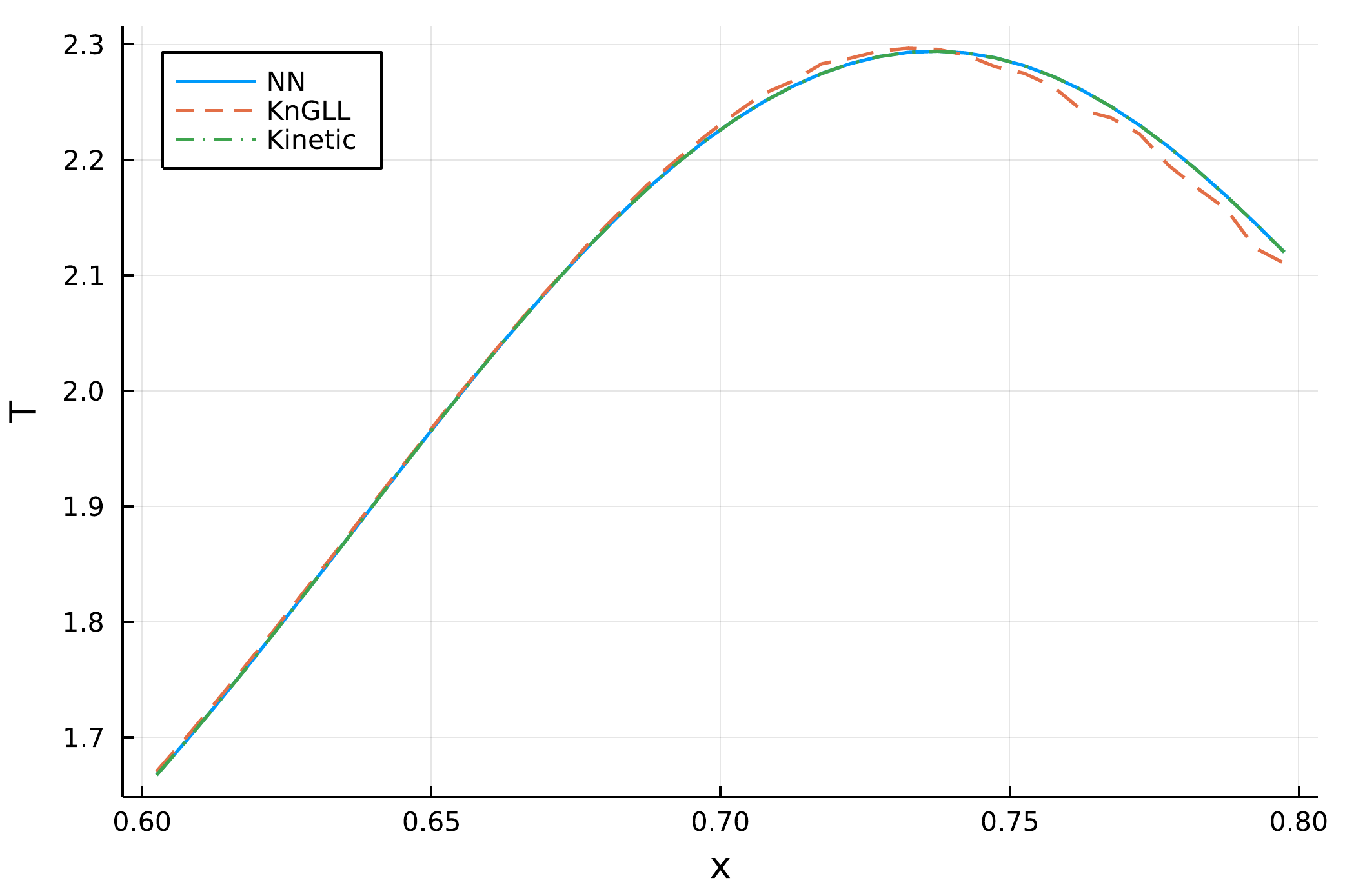}
	}
    \caption{Profiles of density and temperature in the shock tube at $t=0.15$ under $\rm{Kn}=0.01$.}
    \label{fig:sod kn2}
\end{figure}

From the results, we can see that the gradient-length-local Knudsen number criterion underestimates the influence of wave structures and makes inaccurate predictions.
On the contrary, the neural network predicts equivalent flow regimes as the benchmark.
Then, we employ the adaptive solver to conduct complete simulations based on the criteria from the neural network and $\mathrm{Kn}_{GLL}$.
The profiles of density and temperature inside the shock tube at the time instant $t=0.15$ under different Knudsen numbers are presented in Fig. \ref{fig:sod kn4}, \ref{fig:sod kn3} and \ref{fig:sod kn2}.
The kinetic and Navier-Stokes solutions are plotted as benchmark.
As shown, although all the results are qualitatively similar, the zoom-in view demonstrates that the hybrid solution based on $\mathrm{Kn}_{GLL}$ stands closer to the Navier-Stokes results, while the neural network corresponds to the Boltzmann solution.
At $\mathrm Kn=0.01$, the Chapman-Enskog expansion yields negative values in particle distribution function where the spatial slopes are large, resulting in the failure of Navier-Stokes solutions.
In this case, the inaccurate prediction of flow regimes from $\mathrm{Kn}_{GLL}$ results in unreasonable oscillations of macroscopic solutions, which is overcome by the neural network classifier.

Table \ref{tab:sod cost} provides the computational cost of all these three solvers.
As can be seen, the adaptive scheme accelerates the simulation significantly in the continuum and transition flow regimes, and reduce the memory load.

\begin{table}[htbp]
	\caption{Computational cost of the sod shock tube problem.} 
	\centering
	\begin{tabular}{lllllll} 
		\hline
		 & time & total allocations & total allocated memory  \\ 
		\hline 
		Navier-Stokes & 1.39 s & $2.16 \times 10^7$ & 1.82 GB \\
		Kinetic & 1649.02 s & $1.65 \times 10^8$ & 7.35 TB  \\
		Adaptive (Kn=0.0001) & 97.50 s & $2.72 \times 10^7$ & 121.04 GB   \\
		Adaptive (Kn=0.001) & 514.90 s & $3.60 \times 10^7$ & 713.92 GB   \\
		Adaptive (Kn=0.01) & 1209.60 s & $9.88 \times 10^7$ & 3.88 TB   \\
		\hline
	\end{tabular} 
	\label{tab:sod cost}
\end{table}

\subsection{Shear layer}

In the second numerical experiment, let us turn to a shear layer in the transition regime where the transverse processes dominates the fluid motion.
The particle distribution function is initialized as Maxwellian, which corresponds to the following macroscopic variables,
\begin{equation*}
\left(\begin{array}{c}
\rho \\
V_x \\
V_y \\
T \\
\end{array}\right)_{{t=0,L}}=\left(\begin{array}{c}
1 \\
0 \\
1 \\
1 \\
\end{array}\right), \quad
\left(\begin{array}{c}
\rho \\
V_x \\
V_y \\
T \\
\end{array}\right)_{{t=0,R}}=\left(\begin{array}{c}
1 \\
0 \\
-1 \\
0.5 \\
\end{array}\right).
\end{equation*}
The simulation is performed till $50\tau_0$, where $\tau_0=\mu_0/p_0$ denotes the mean collision time in the left half of initial domain, and the viscosity $\mu_0$ can be evaluated from the hard-sphere model,
\begin{equation*}
    \mu_{0}=\frac{15 \sqrt{\pi}}{48} \mathrm{Kn}.
\end{equation*}
The detailed computation setup is listed in Table \ref{tab:layer}.
\begin{table}[htbp]
	\caption{Computational setup of shear layer problem.} 
	\centering
	\begin{tabular}{lllllll} 
		\hline
		$t$ & $x$ & $N_x$ & $\mathbf v$ & $N_u$ & $N_v$ & $N_w$ \\ 
		$[0,50\tau_0]$ & $[-0.5,0.5]$ & $500$ & $[-6,6]^3$ & 64 & 28 & 28 \\ 
		\hline 
		Quadrature & Kn & CFL & Integrator & Boundary \\ 
		Rectangular & $0.005$ & $0.5$ & Euler & Dirichlet  \\
		\hline
	\end{tabular} 
	\label{tab:layer}
\end{table}

\begin{figure}
    \centering
    \subfigure[$t=\tau_0$]{
		\includegraphics[width=0.31\textwidth]{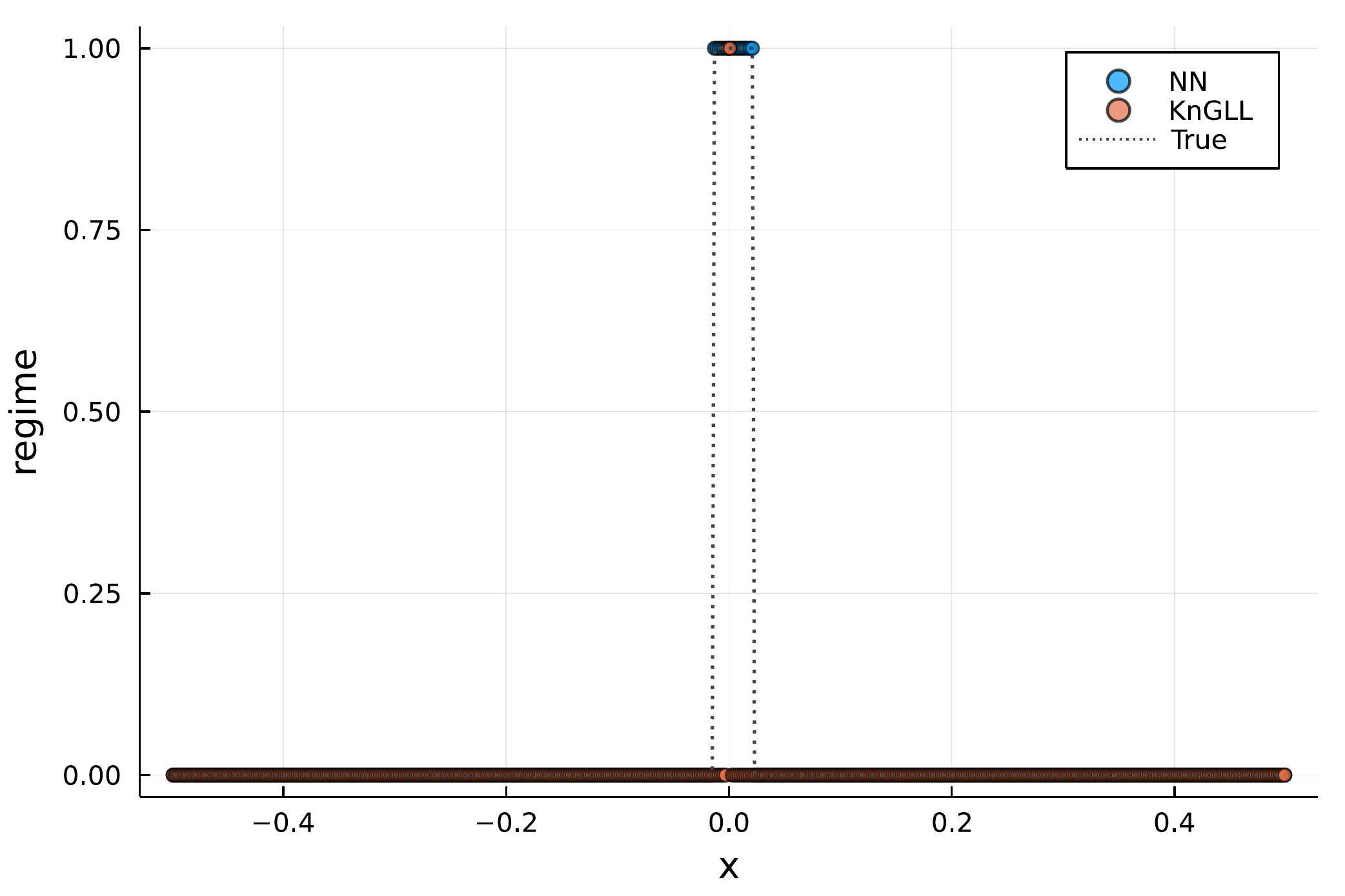}
	}
	\subfigure[$t=10\tau_0$]{
		\includegraphics[width=0.31\textwidth]{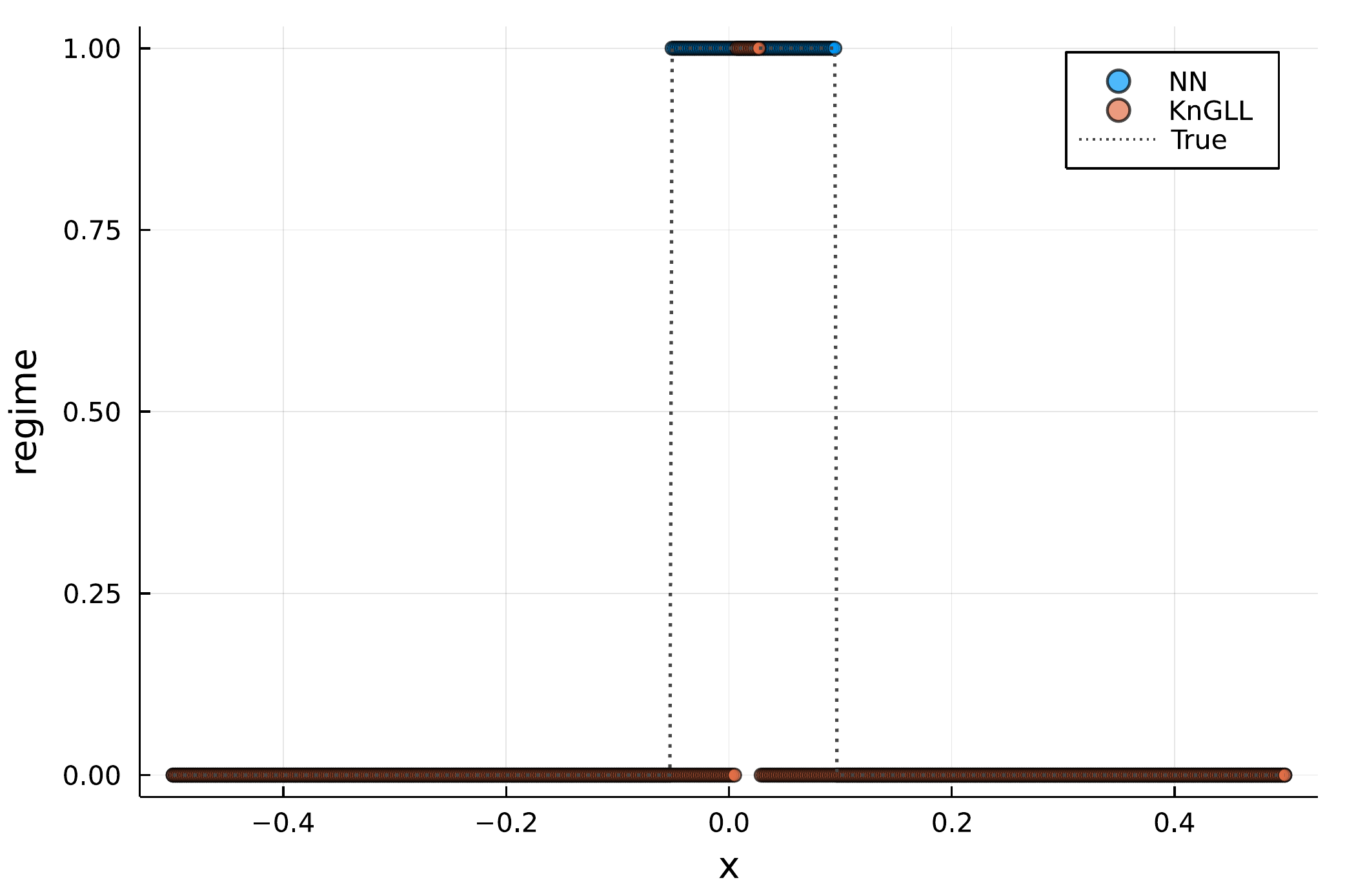}
	}
	\subfigure[$t=50\tau_0$]{
		\includegraphics[width=0.31\textwidth]{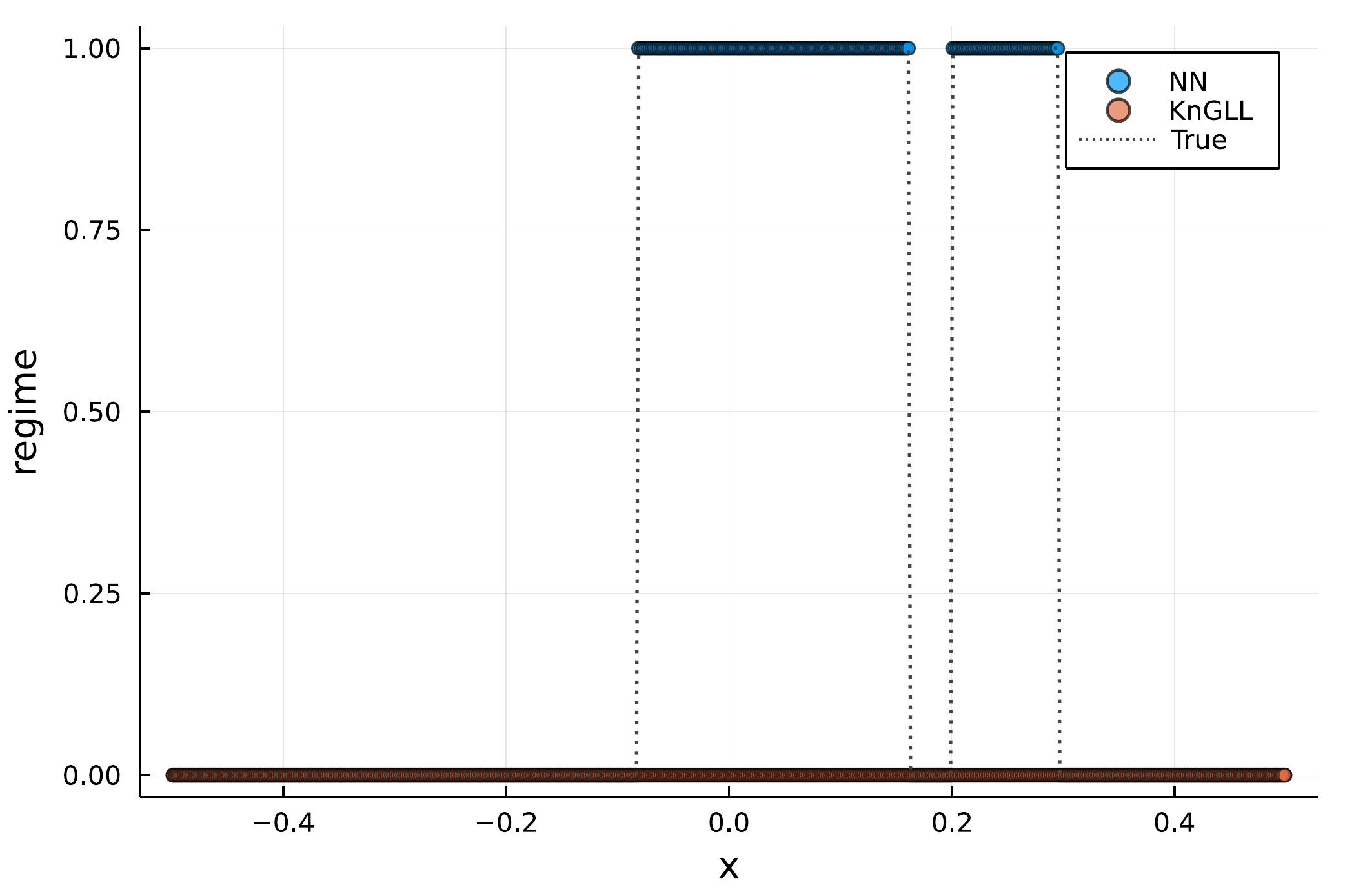}
	}
    \caption{Prediction of flow regimes from fully kinetic solutions at different time instants in the shear layer with different criteria (0 denotes near-equilibrium, 1 denotes non-equilibrium).}
    \label{fig:layer regime}
\end{figure}

We first conduct a full kinetic simulation with the Boltzmann equation.
Based on the kinetic solution, the partition of flow regimes based on different criteria is shown in Fig. \ref{fig:layer regime}.
With the time evolution, it is clear that the non-equilibrium region expands due to the strong shearing effect.
It is clear that the neural network predicts equivalent flow regimes as ground truth, while the gradient-length-local Knudsen number criterion underestimates the non-equilibrium effect.

\begin{figure}
    \centering
    \subfigure[Density]{
		\includegraphics[width=0.47\textwidth]{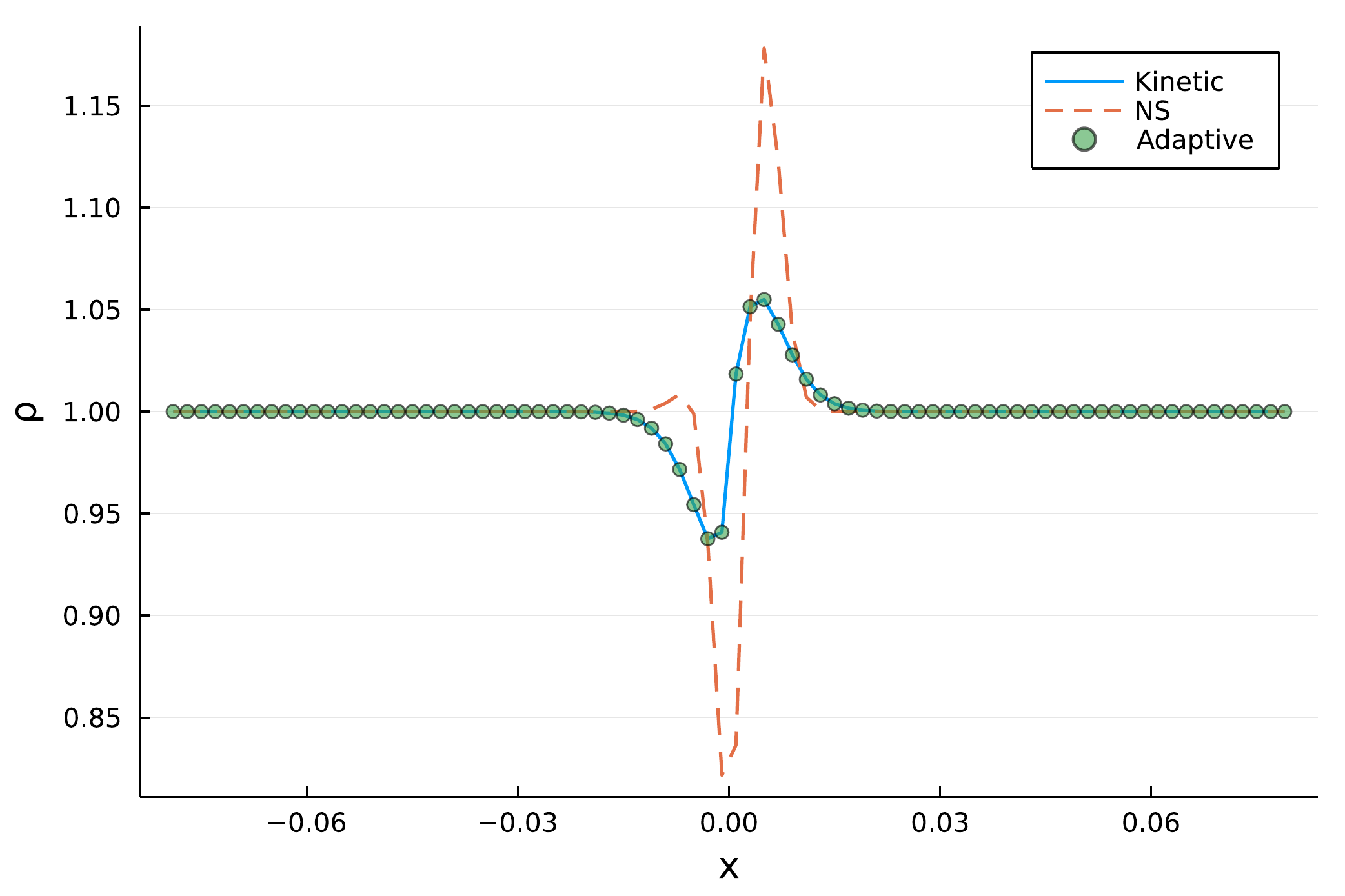}
	}
	\subfigure[U-velocity]{
		\includegraphics[width=0.47\textwidth]{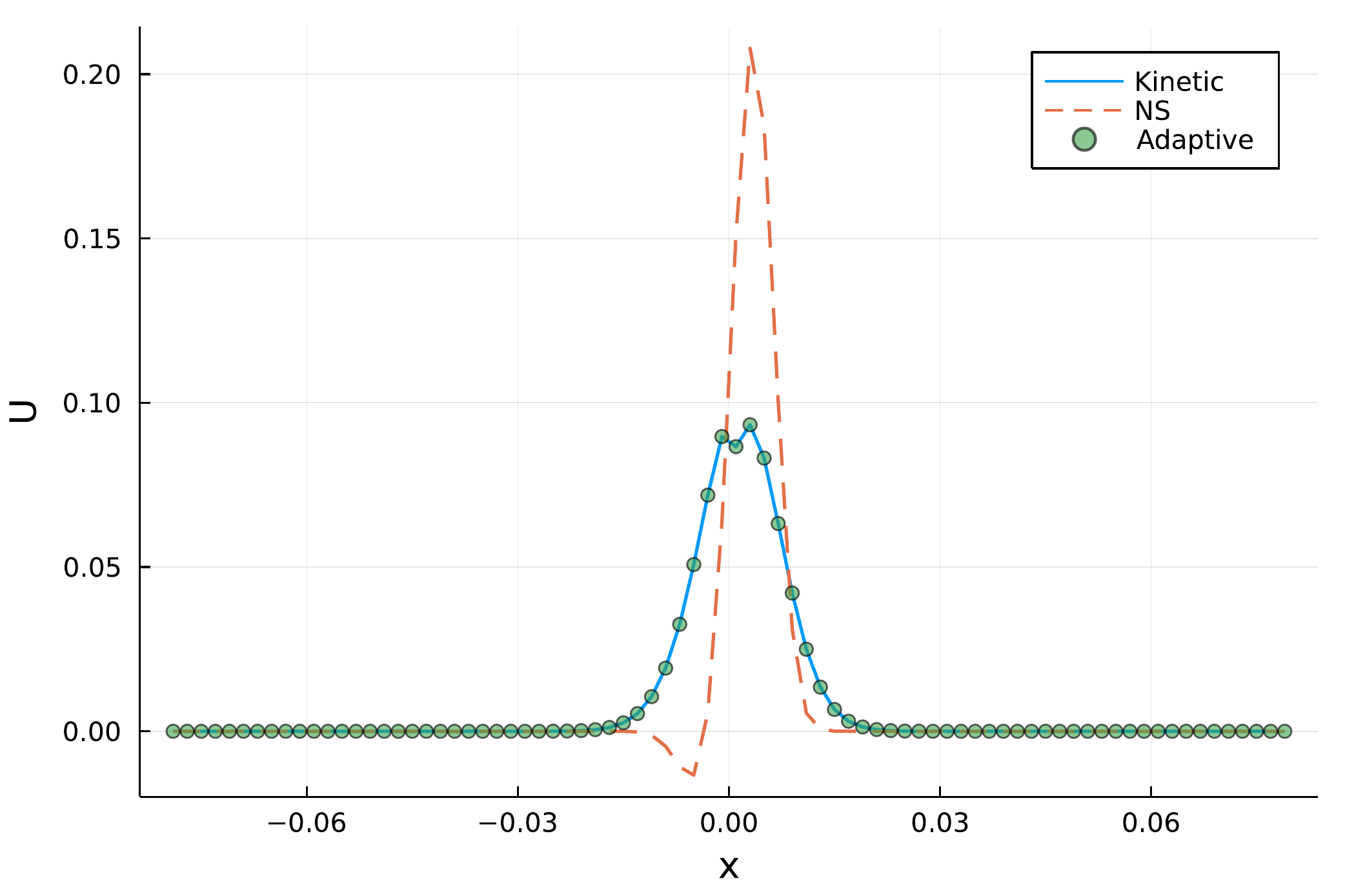}
	}
	\subfigure[V-velocity]{
		\includegraphics[width=0.47\textwidth]{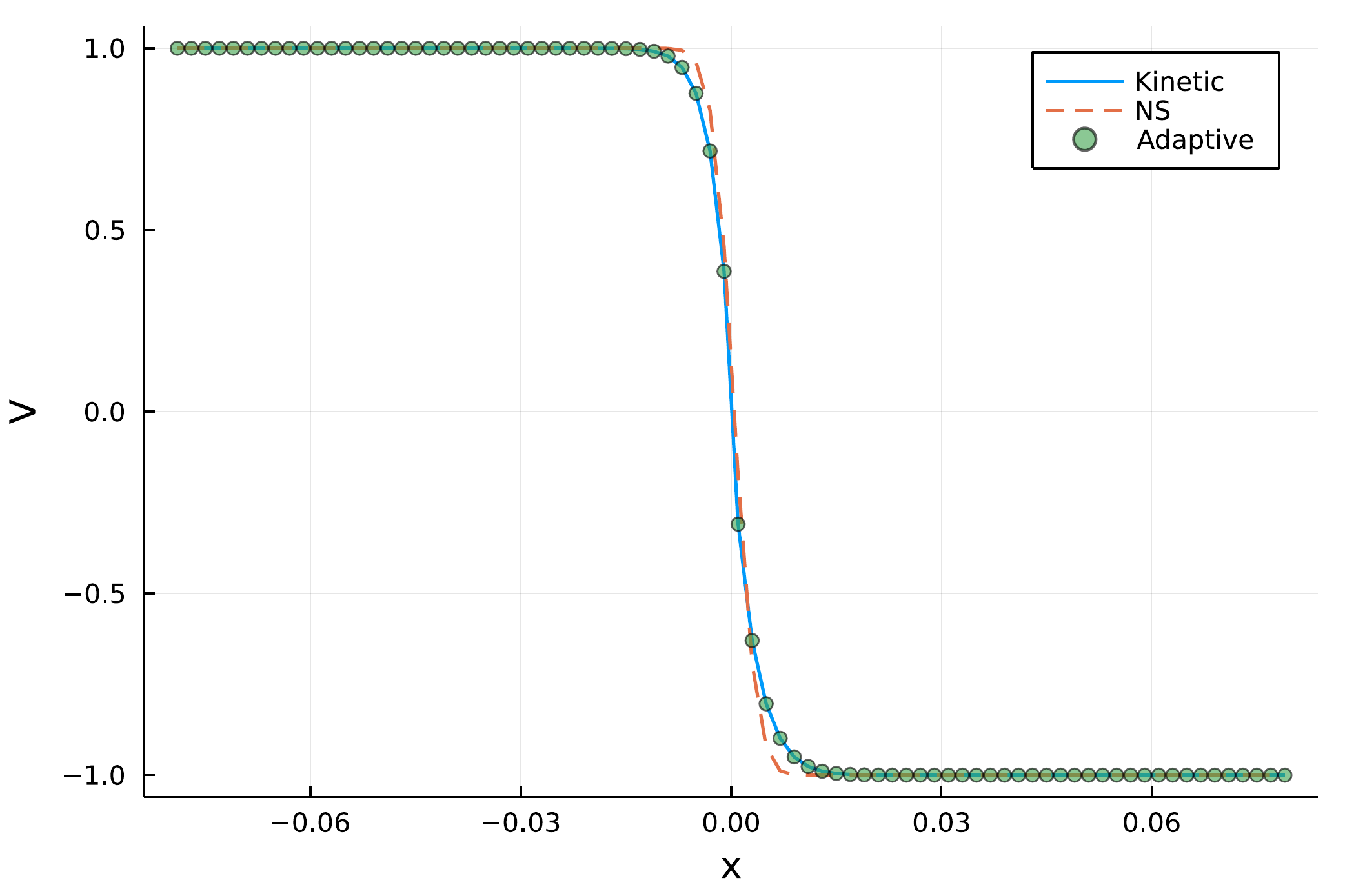}
	}
	\subfigure[Temperature]{
		\includegraphics[width=0.47\textwidth]{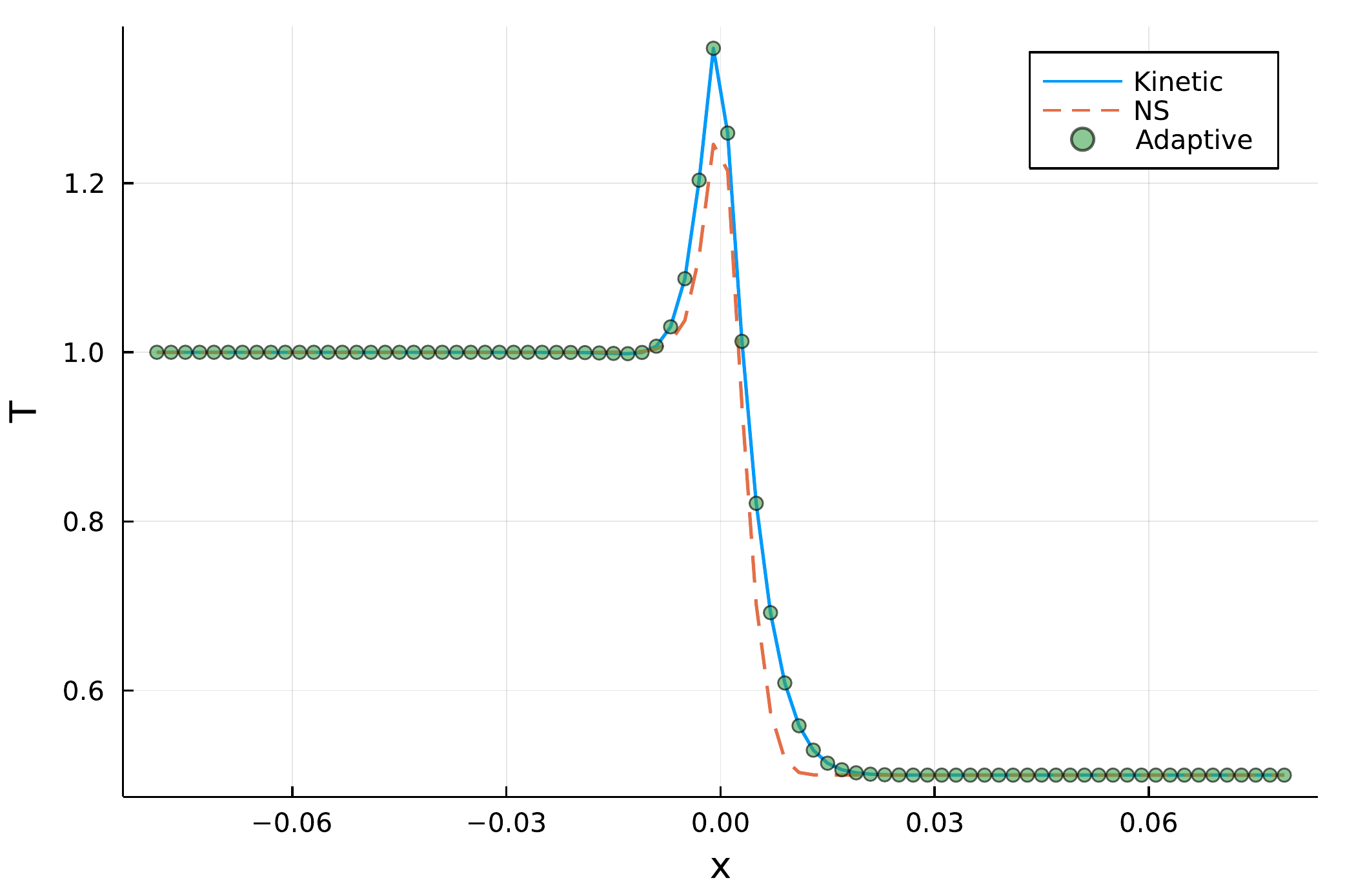}
	}
    \caption{Profiles of flow variables in the shear layer at $t=\tau$.}
    \label{fig:layer t1}
\end{figure}

\begin{figure}
    \centering
    \subfigure[Density]{
		\includegraphics[width=0.47\textwidth]{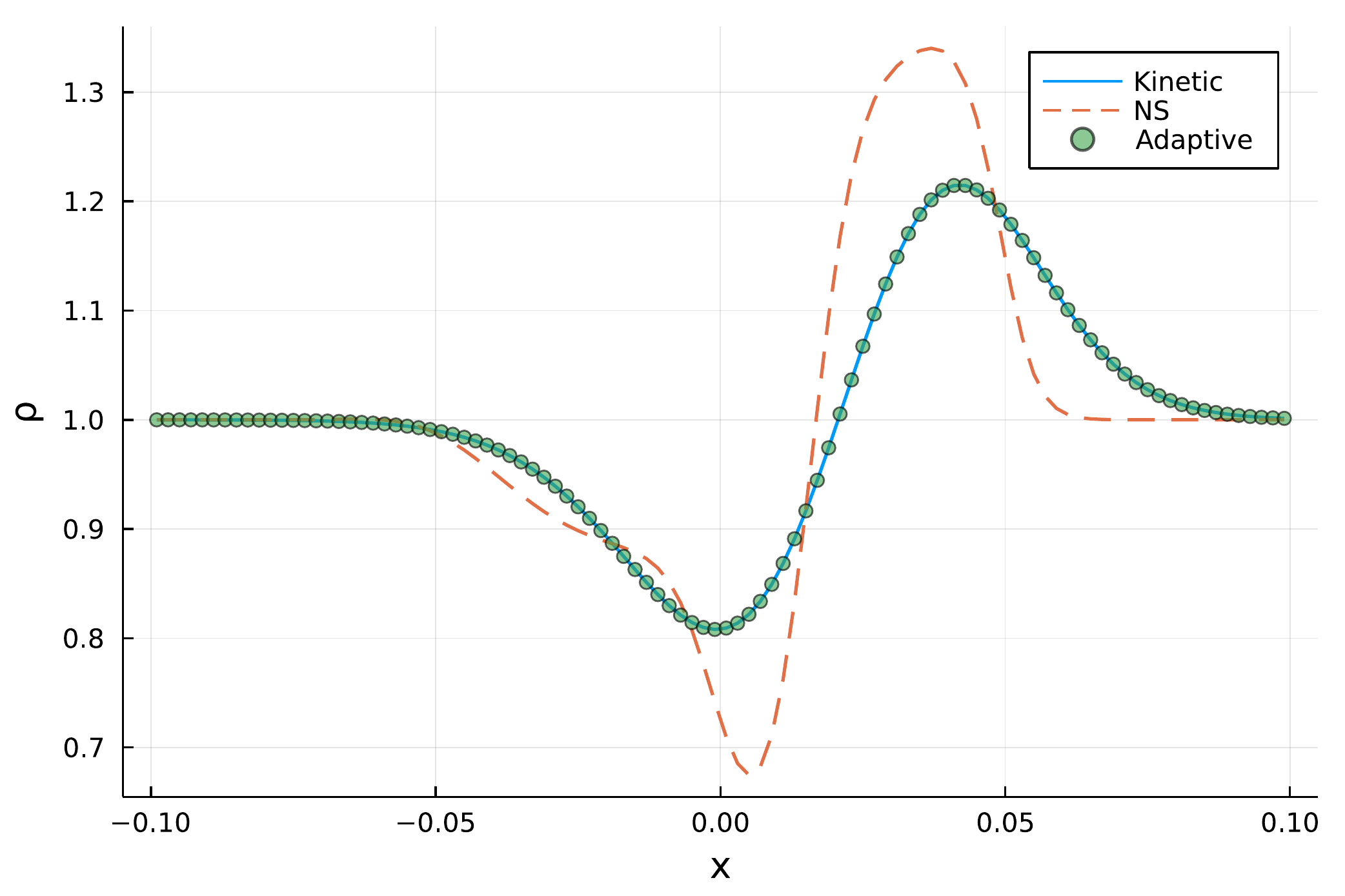}
	}
	\subfigure[U-velocity]{
		\includegraphics[width=0.47\textwidth]{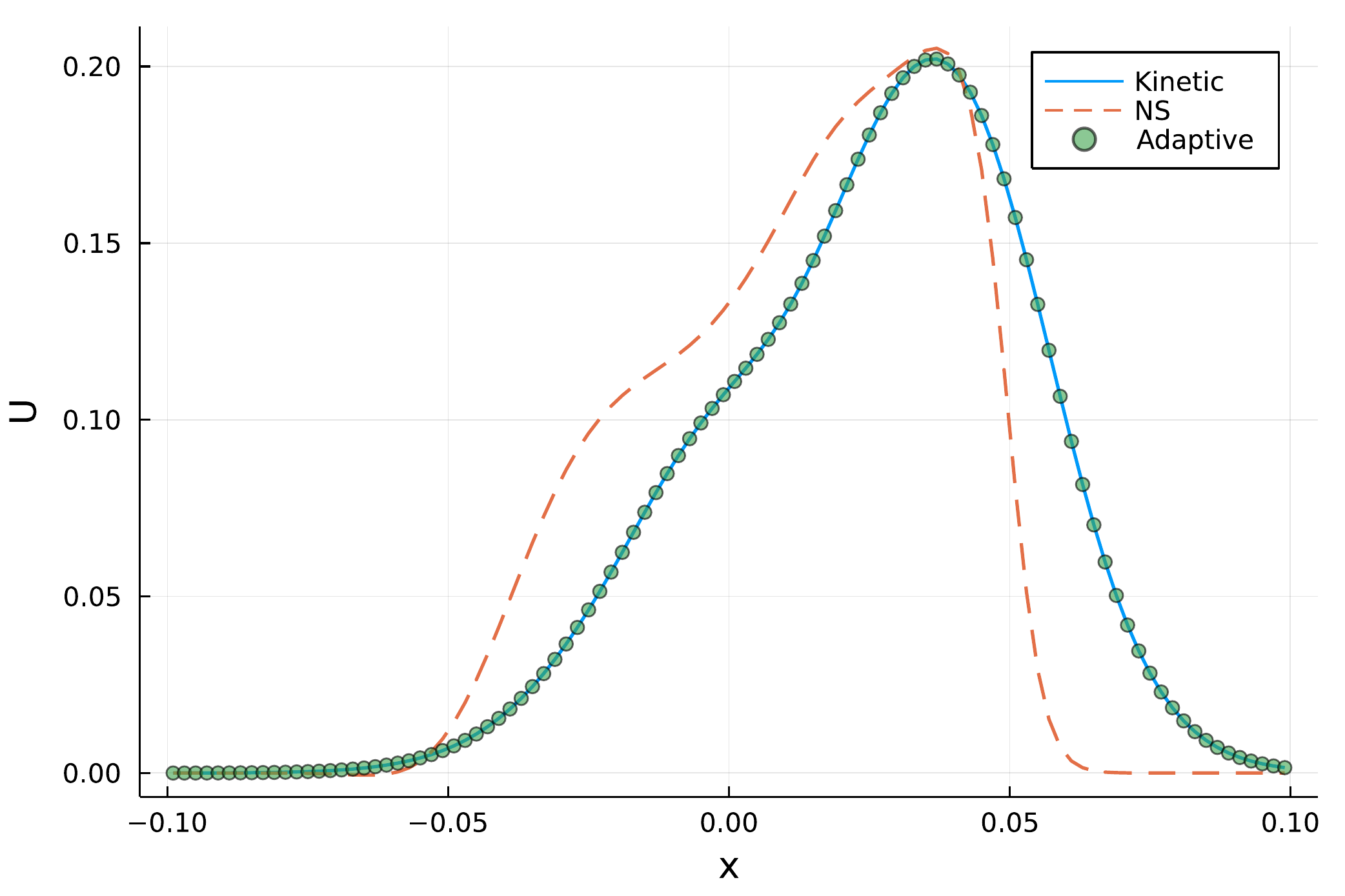}
	}
	\subfigure[V-velocity]{
		\includegraphics[width=0.47\textwidth]{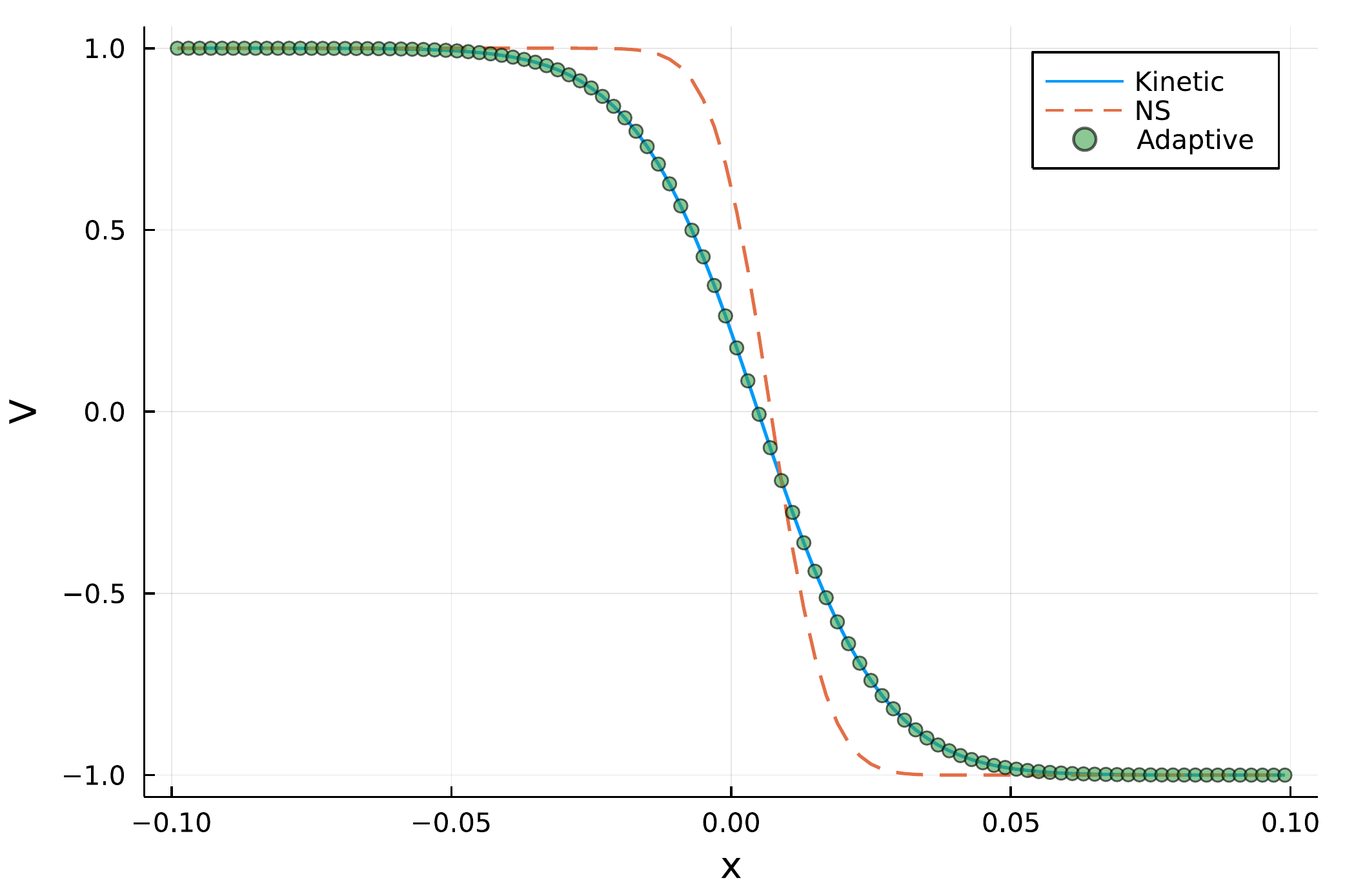}
	}
	\subfigure[Temperature]{
		\includegraphics[width=0.47\textwidth]{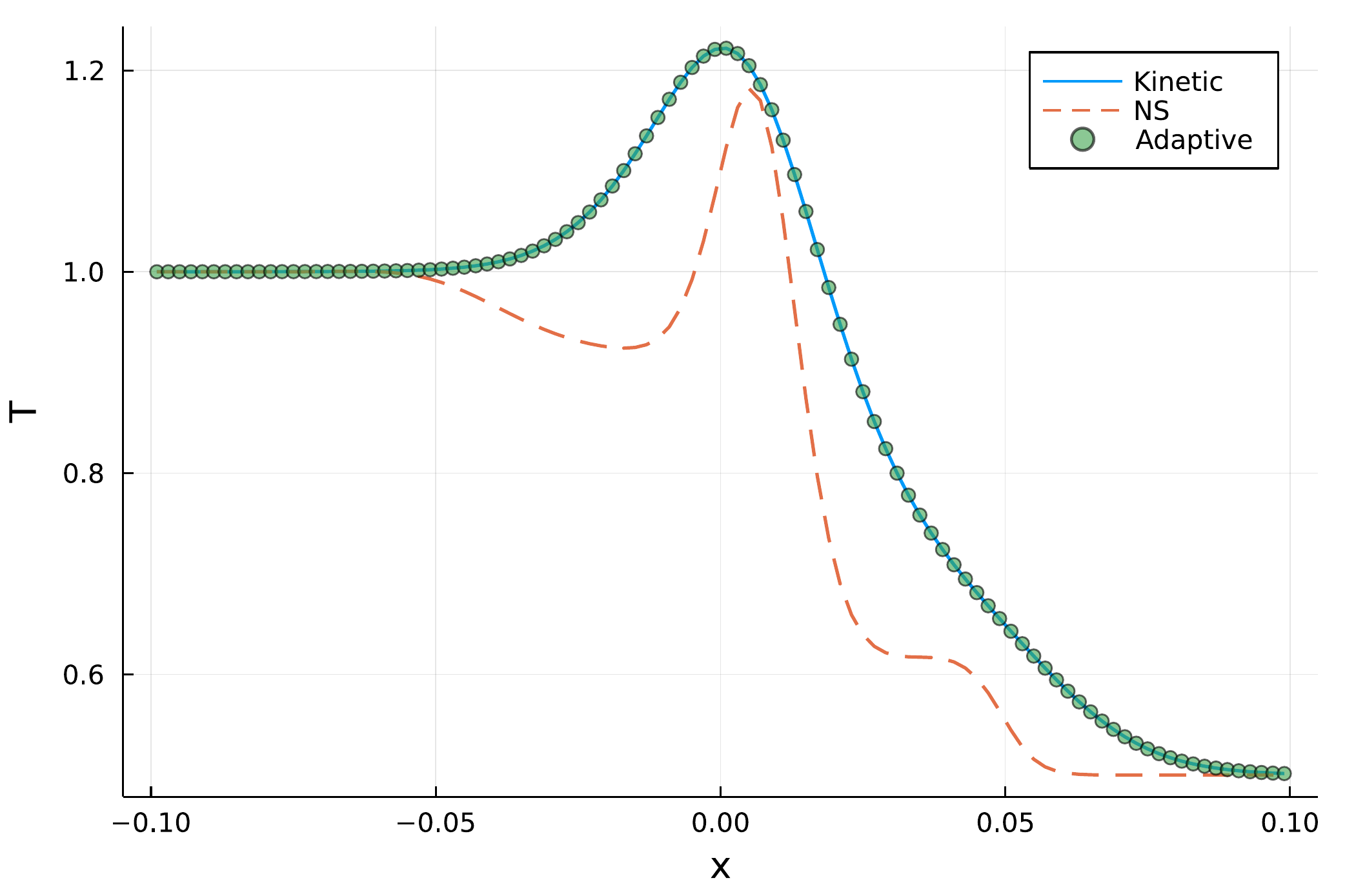}
	}
    \caption{Profiles of flow variables in the shear layer at $t=10\tau$.}
    \label{fig:layer t2}
\end{figure}

\begin{figure}
    \centering
    \subfigure[Density]{
		\includegraphics[width=0.47\textwidth]{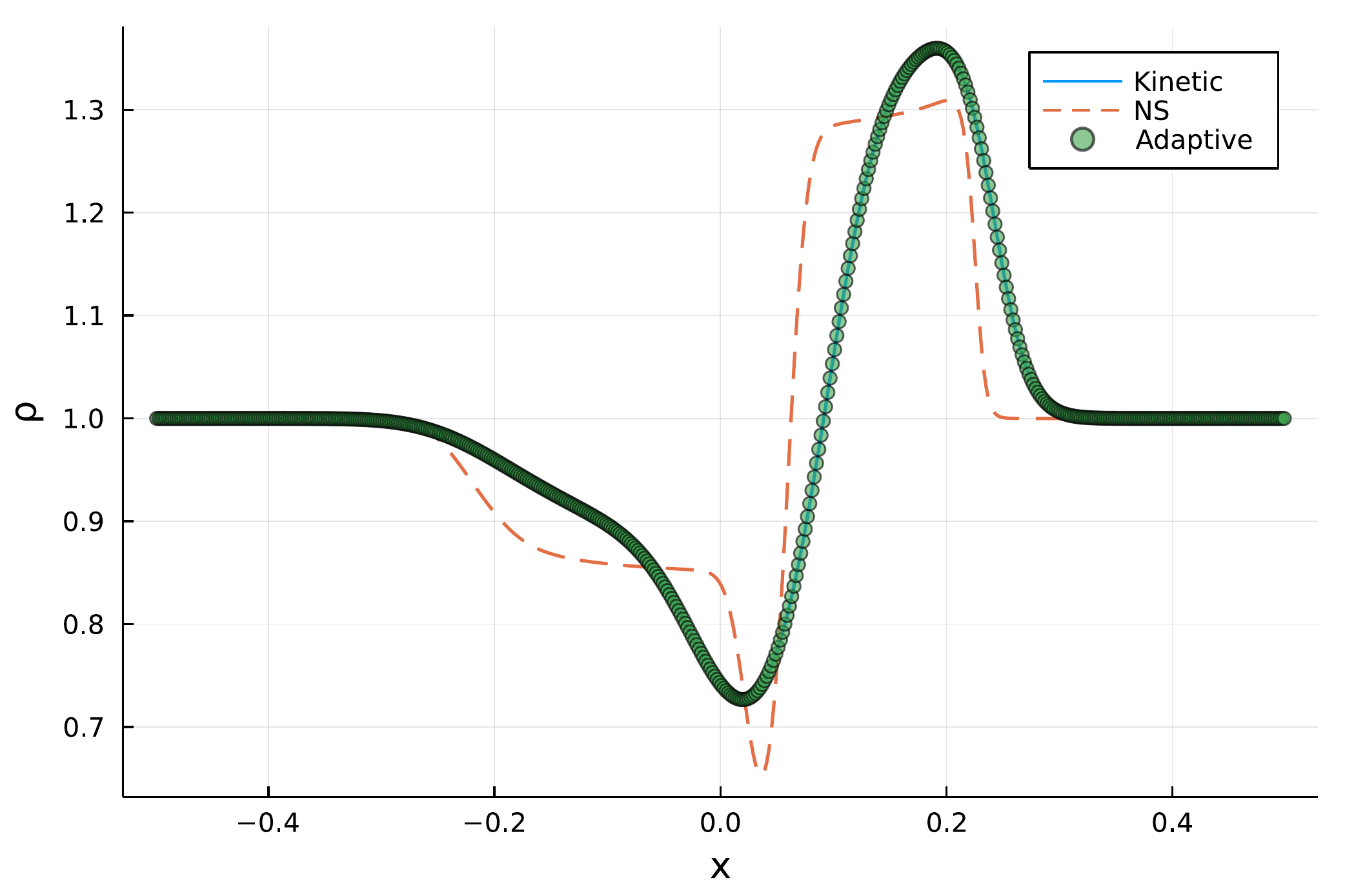}
	}
	\subfigure[U-velocity]{
		\includegraphics[width=0.47\textwidth]{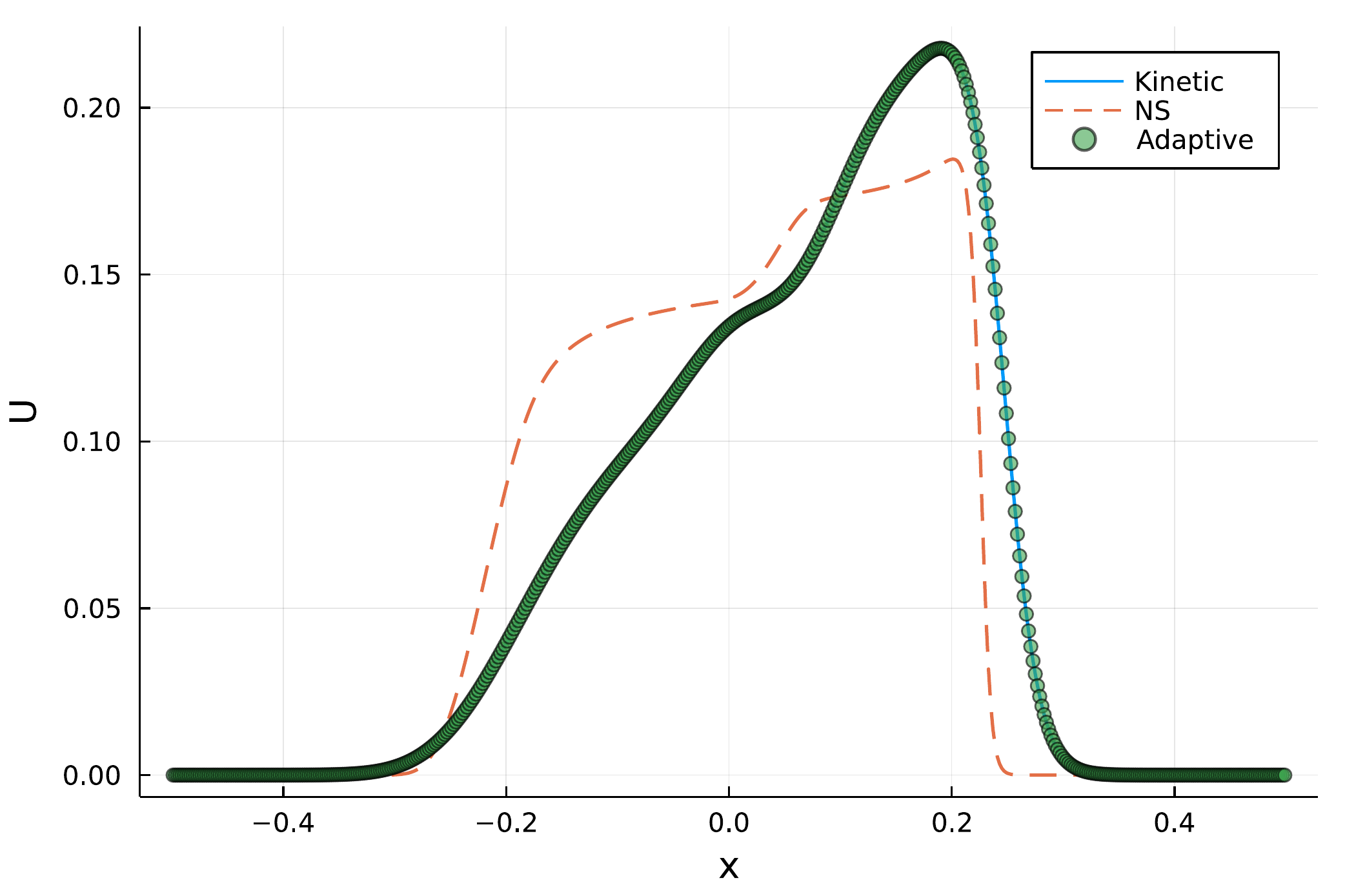}
	}
	\subfigure[V-velocity]{
		\includegraphics[width=0.47\textwidth]{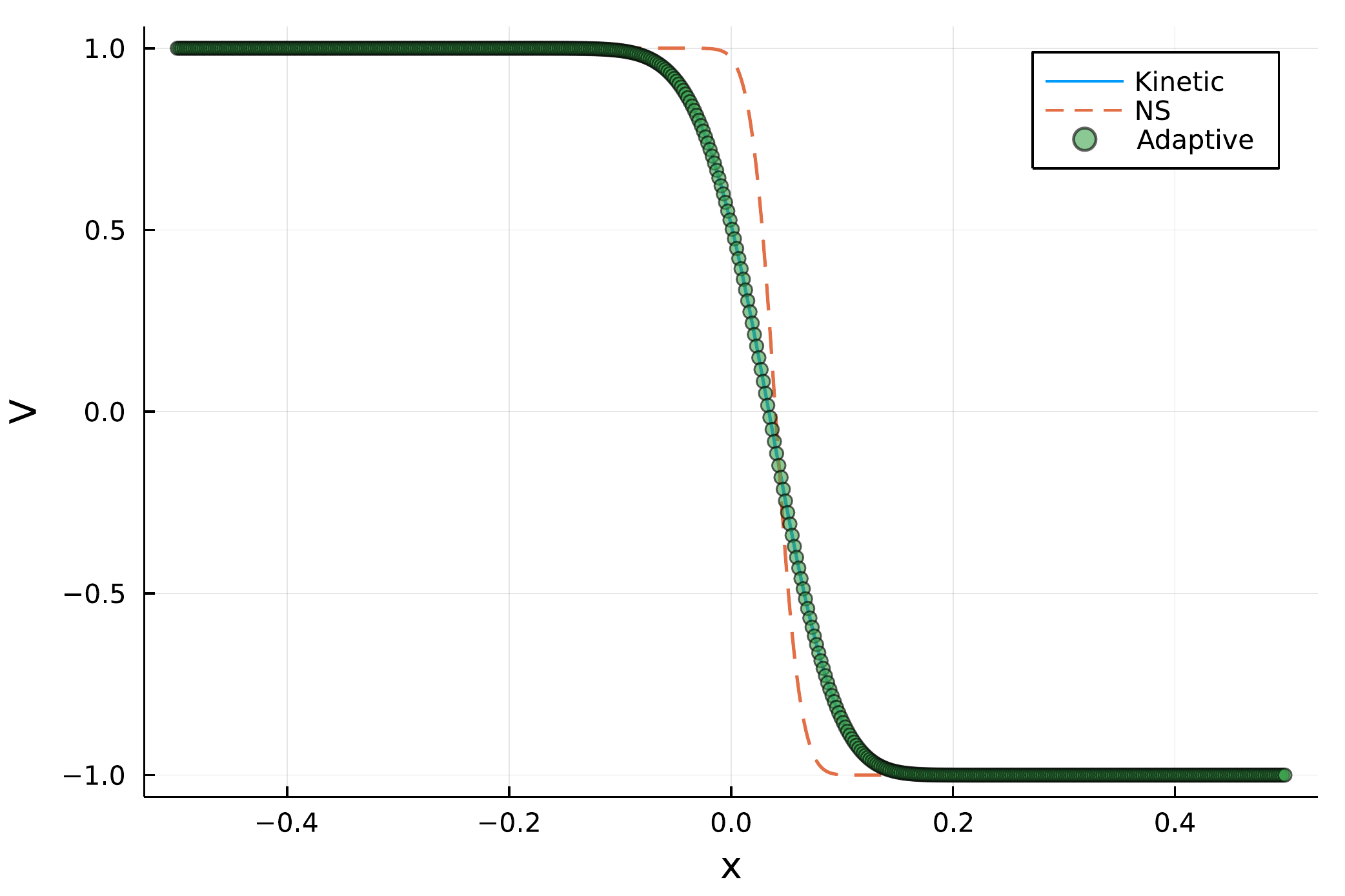}
	}
	\subfigure[Temperature]{
		\includegraphics[width=0.47\textwidth]{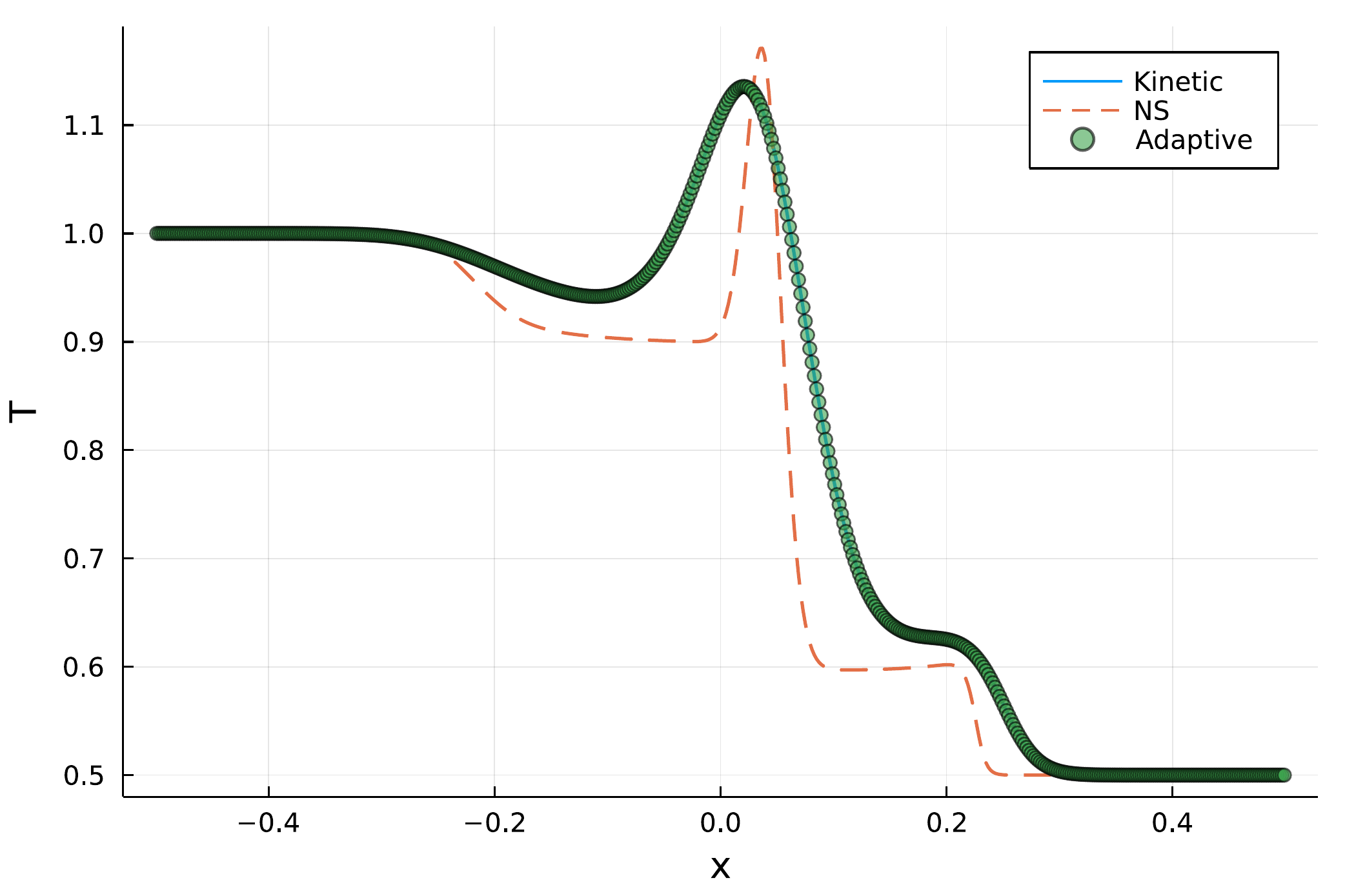}
	}
    \caption{Profiles of flow variables in the shear layer at $t=50\tau$.}
    \label{fig:layer t3}
\end{figure}

\begin{figure}
    \centering
    \subfigure[$t=\tau$]{
		\includegraphics[width=0.47\textwidth]{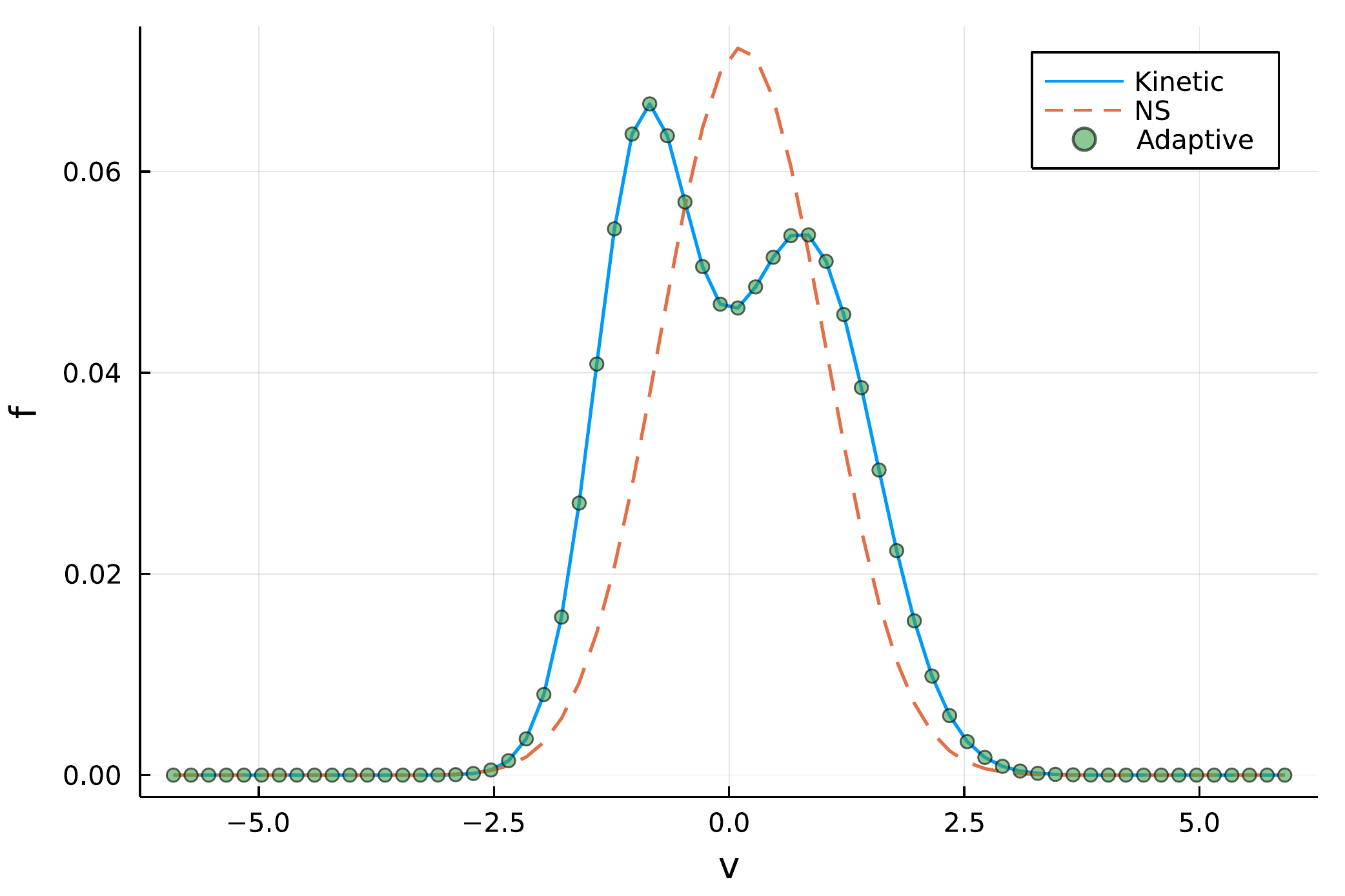}
	}
	\subfigure[$t=10\tau$]{
		\includegraphics[width=0.47\textwidth]{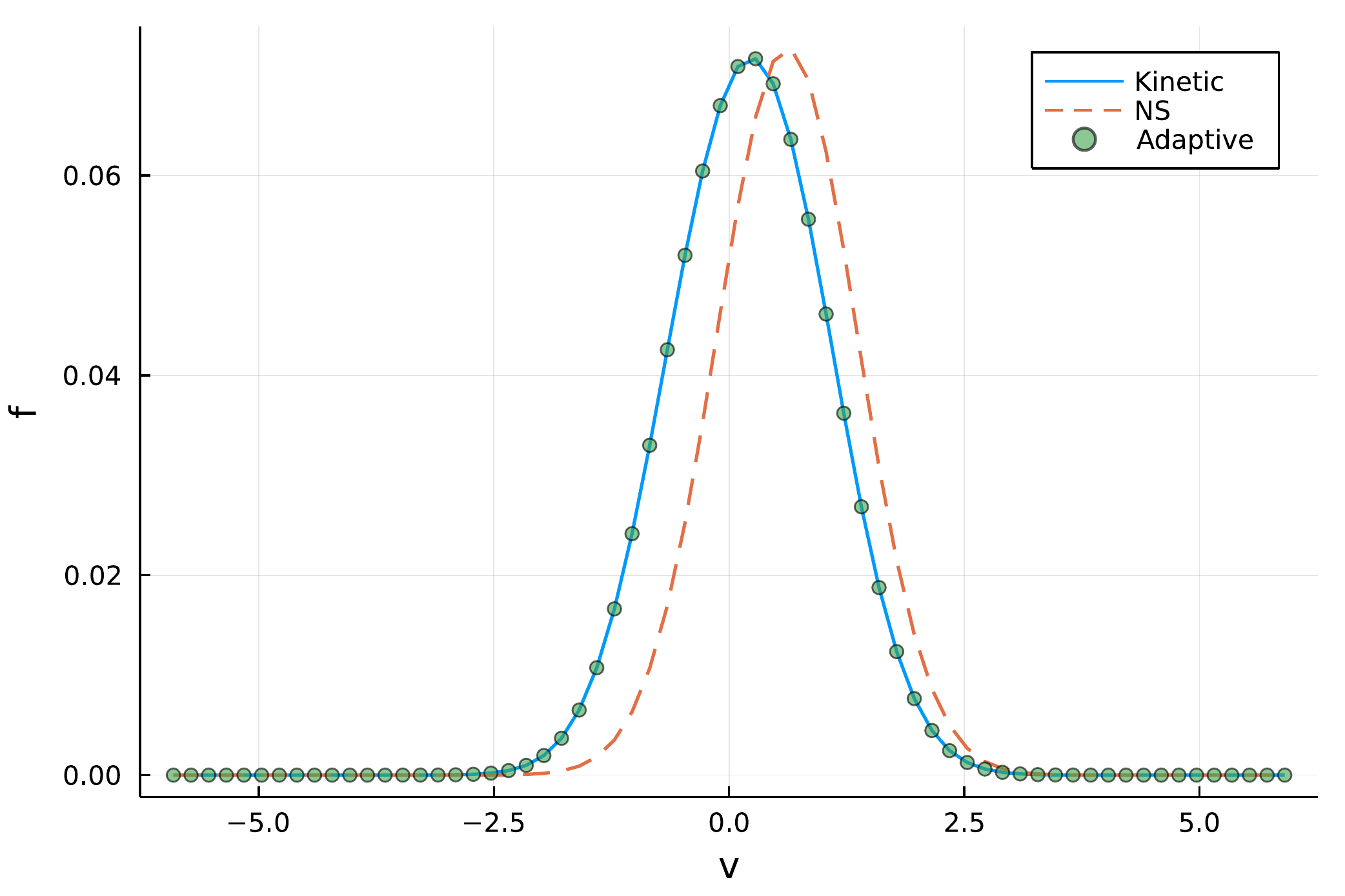}
	}
	\subfigure[$t=50\tau$]{
		\includegraphics[width=0.47\textwidth]{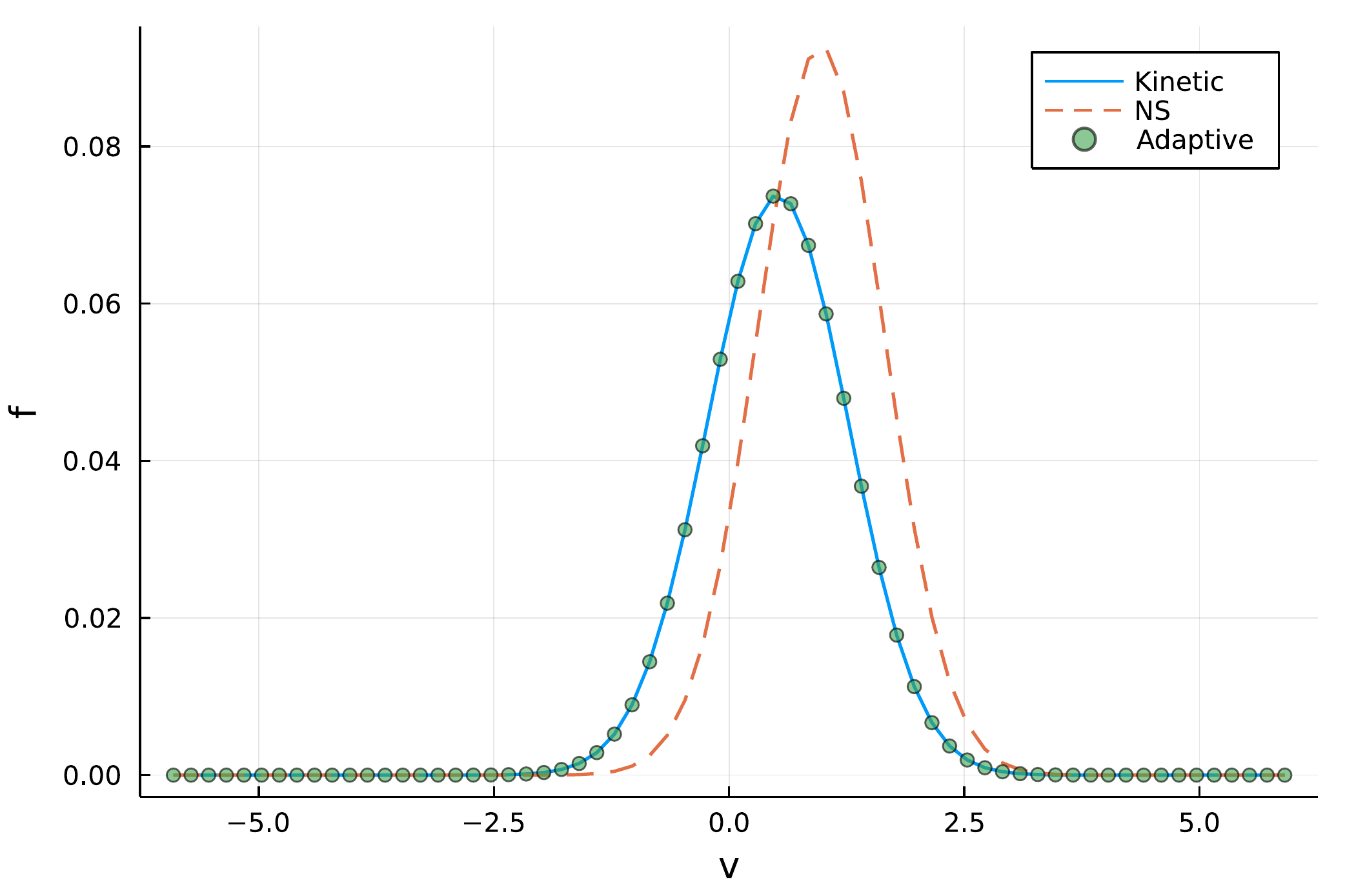}
	}
    \caption{Particle distribution functions at the domain center at different time instants.}
    \label{fig:layer f}
\end{figure}

Then we employ the adaptive solver to conduct the simulation.
The profiles of density, velocity and temperature at different time instants are presented in Fig. \ref{fig:layer t1}, \ref{fig:layer t2} and \ref{fig:layer t3}.
The kinetic and Navier-Stokes solutions are plotted as benchmark.
As is shown, for this highly dissipative problem with strong shearing effect, the kinetic and Navier-Stokes equations present distinct solutions.
Fig. \ref{fig:layer f} presents the evolution of particle distribution function at the domain center.
Due to the accumulating effect of intermolecular collisions, the particle distribution function transforms gradually into Maxwellian from the initial bi-modal distribution.
During the evolution process, the adaptive scheme provides equivalent solutions as the kinetic benchmark, which confirms the validity of the neural network classifier.
Table \ref{tab:layer cost} provides the computational cost of all these three solvers.
As can be seen, the adaptive scheme accelerates the simulation by $69\%$, and saves $66\%$ unnecessary allocations.

\begin{table}[htbp]
	\caption{Computational cost of shear layer problem.} 
	\centering
	\begin{tabular}{lllllll} 
		\hline
		 & time & total allocations & total allocated memory  \\ 
		\hline 
		Navier-Stokes & 11.07 s & $7.87 \times 10^7$ & 5.63 GB \\
		Kinetic & 1985.34 s & $4.00 \times 10^8$ & 15.39 TB  \\
		Adaptive & 623.05 s & $1.35 \times 10^8$ & 1.36 TB   \\
		\hline
	\end{tabular} 
	\label{tab:layer cost}
\end{table}

\subsection{Flow around circular cylinder}

In the last numerical experiment, we present the two-dimensional hypersonic flow around circular cylinder, where longitudinal and transverse processes coexist in the domain.
The particle distribution function is initialized as Maxwellian everywhere corresponding the Mach number $Ma=5$.
The detailed computation setup is listed in Table \ref{tab:cylinder}.
\begin{table}[htbp]
	\caption{Computational setup of flow around circular cylinder.} 
	\centering
	\begin{tabular}{llllllll} 
		\hline
		$r$ & $N_r$ & $\theta$ & $N_\theta$ & $\mathbf v$ & $N_u$ & $N_v$ \\ 
		$[1,6]$ & $60$ & $[0,\pi]$ & 50 & $[-10,10]^3$ & 48 & 48 \\ 
		\hline 
		$N_w$ & Quadrature & Kn & CFL & Integrator & Wall & Edge \\ 
		32 & Rectangular & $[0.001,0.01]$ & $0.5$ & Euler & Maxwell & Symmetry \\
		\hline
	\end{tabular} 
	\label{tab:cylinder}
\end{table}

In this steady state problem, the computation can be accelerated with the help of the NS solver.
A convergent coarse flow field can be first obtained by the NS solver, and then reconstructed as the initial state in the subsequent adaptive method. The workflow for the computation of steady flow is described as follows.

\begin{algorithm}
%\caption{An algorithm with caption}\label{alg:two}
\TitleOfAlgo{Workflow of steady flow problem}
1. Use the Navier-Stokes solver to evolve the initial condition to a convergent flow field\;
2. Split the domain into near- and non-equilibrium regions\;
3. Reconstruct the particle distribution function from macroscopic flow variables with the Chapman-Enskog expansion in Eq.(\ref{eqn:ce boltzmann}) in non-equilibrium regions\;
4. Update the flow field with adaptive scheme until a convergent flow field is obtained.
\end{algorithm}

\begin{figure}
    \centering
    \subfigure[U-velocity]{
		\includegraphics[width=0.47\textwidth]{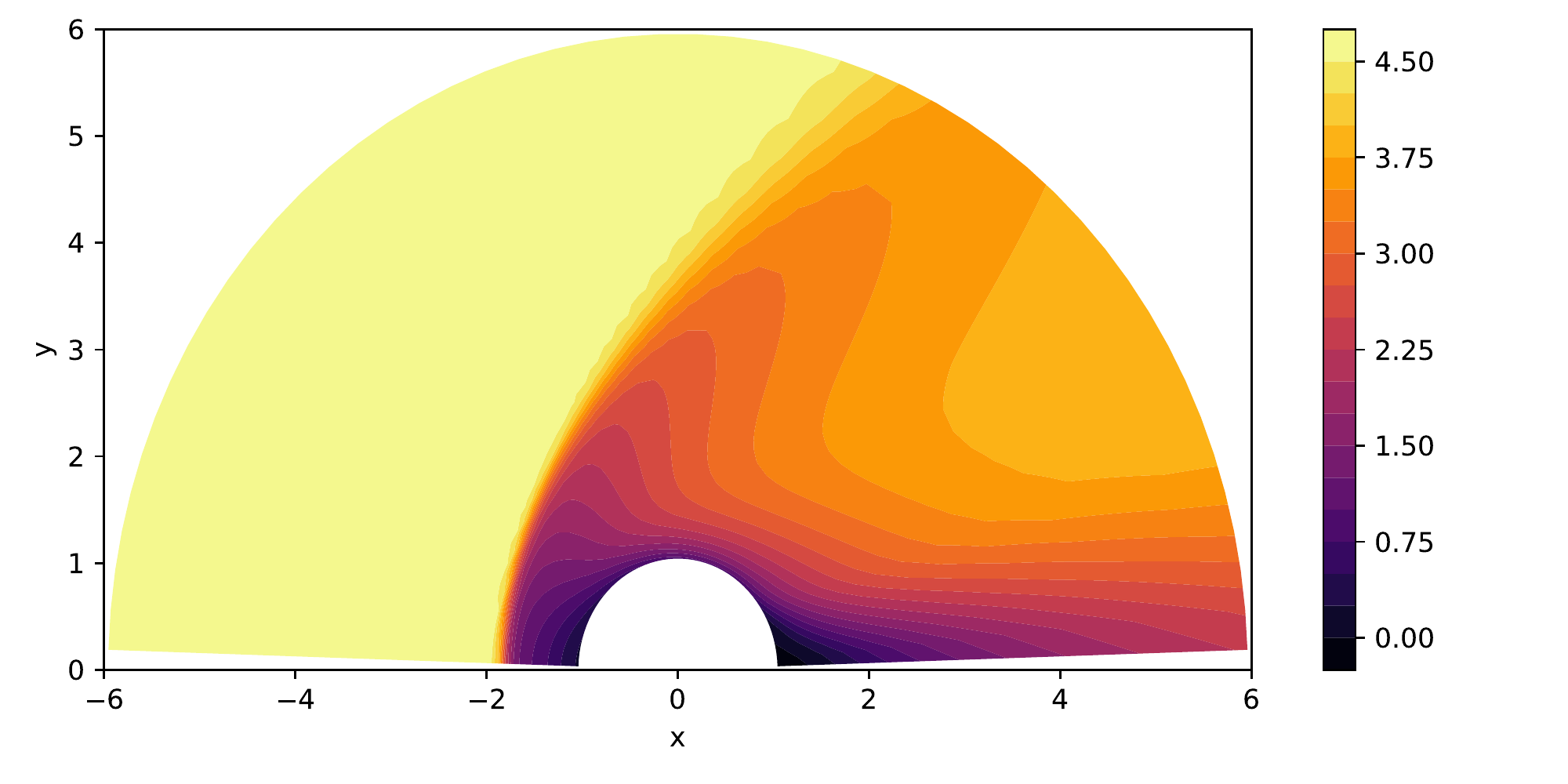}
	}
	\subfigure[Temperature]{
		\includegraphics[width=0.47\textwidth]{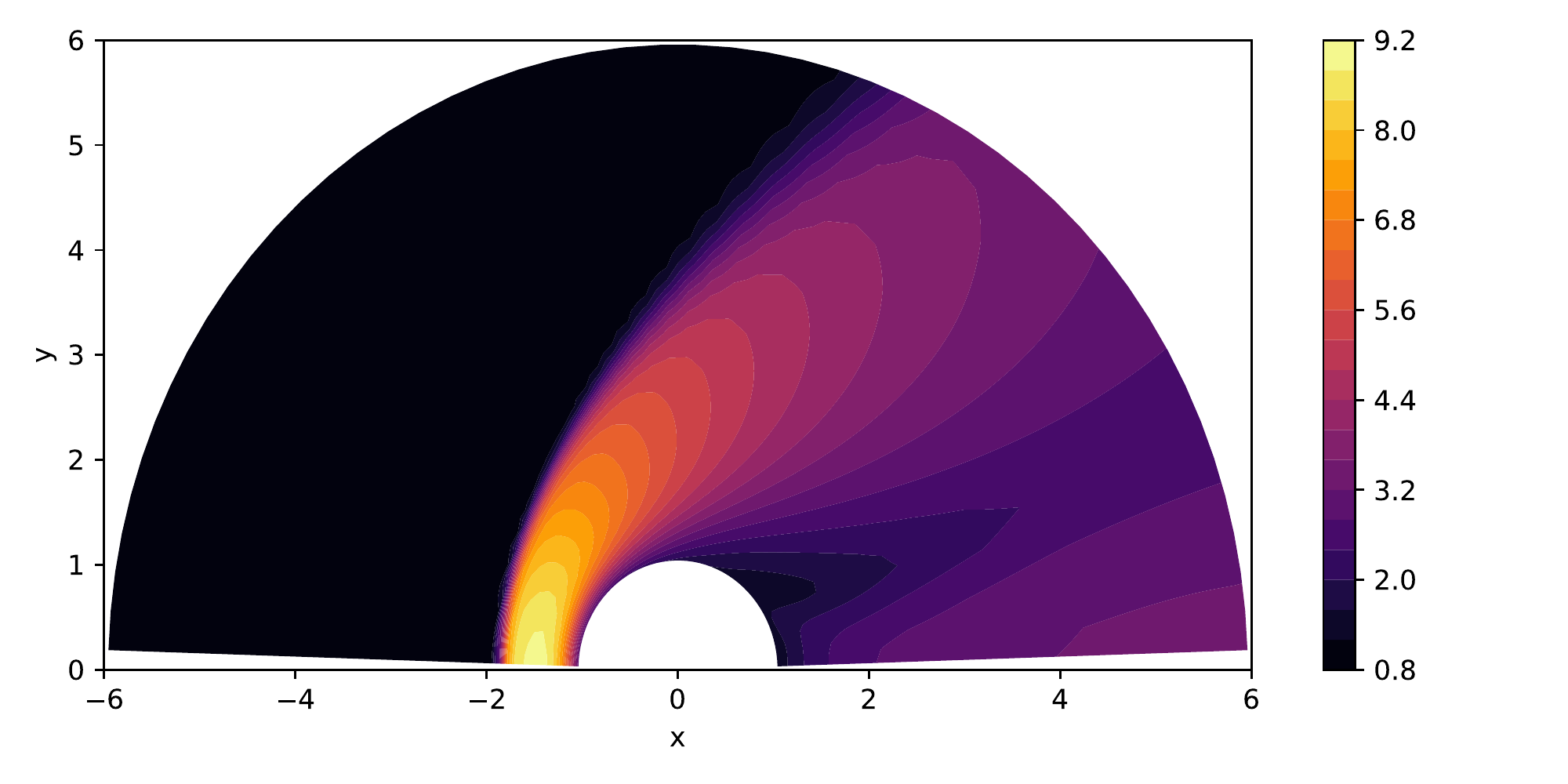}
	}
    \caption{Profiles of density and temperature in the cylinder flow under $\rm{Kn}=0.001$.}
    \label{fig:cylinder kn3}
\end{figure}

\begin{figure}
    \centering
    \subfigure[U-velocity]{
		\includegraphics[width=0.47\textwidth]{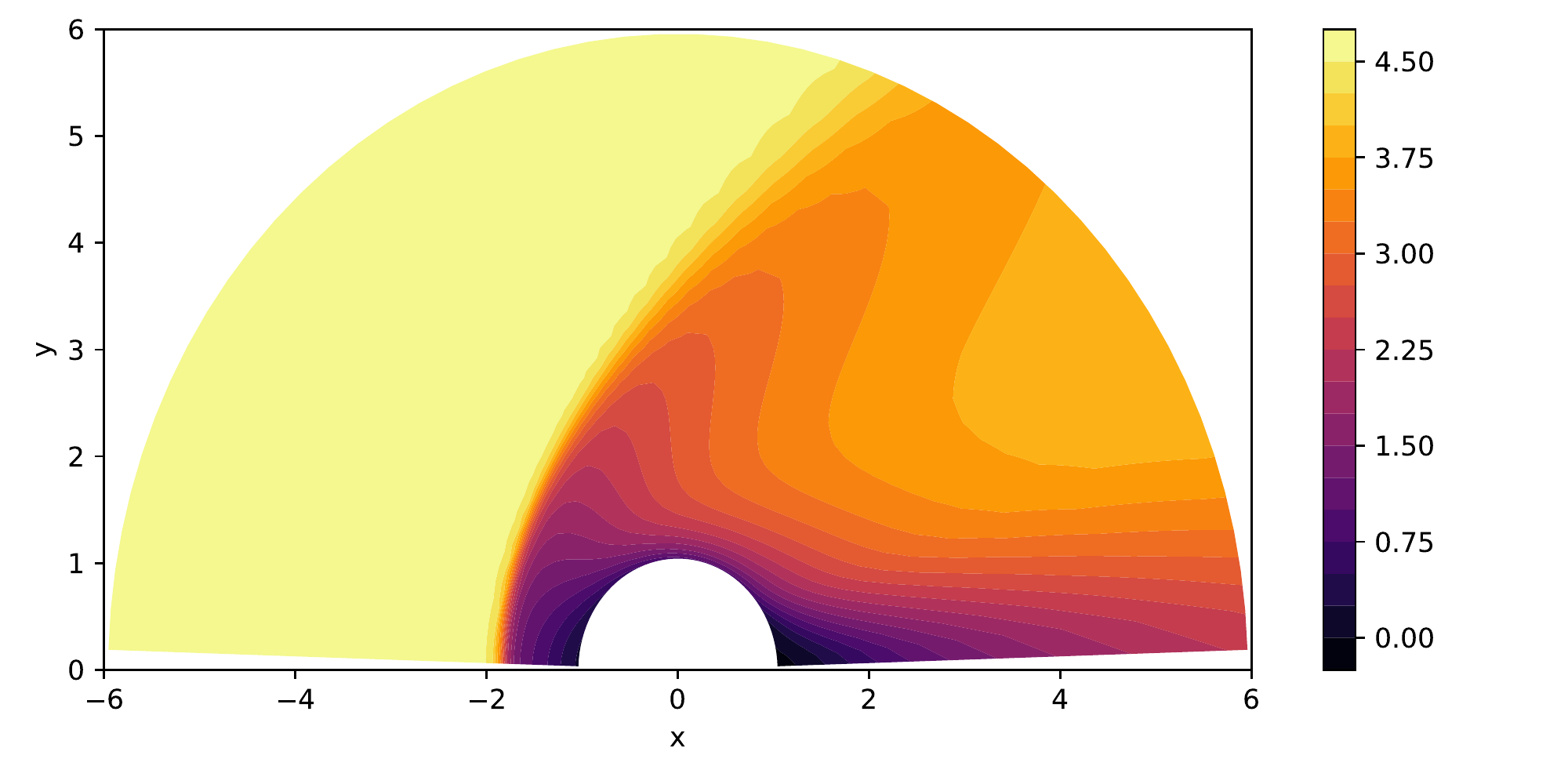}
	}
	\subfigure[Temperature]{
		\includegraphics[width=0.47\textwidth]{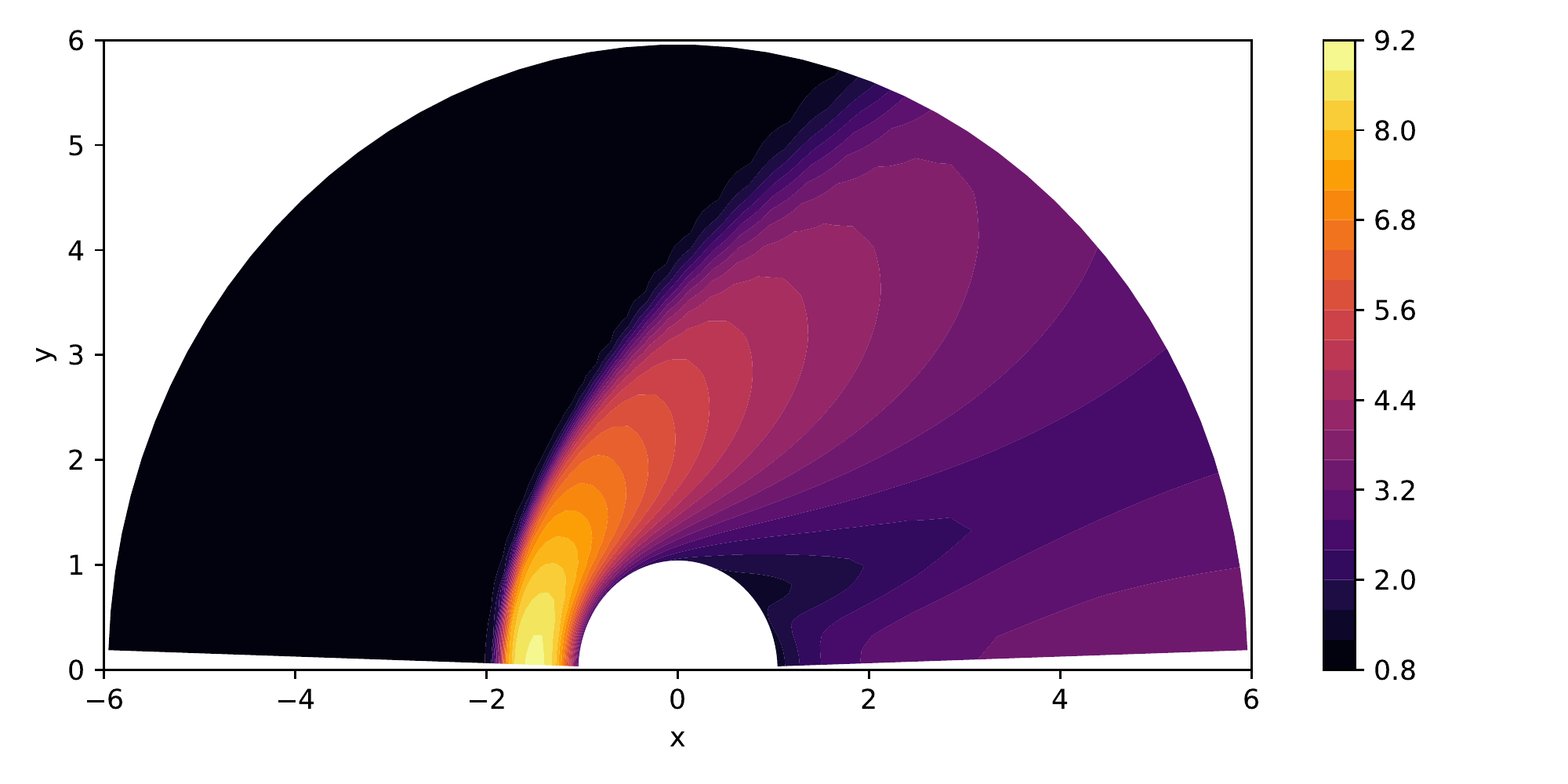}
	}
    \caption{Profiles of density and temperature in the cylinder flow under $\rm{Kn}=0.01$.}
    \label{fig:cylinder kn2}
\end{figure}

\begin{figure}
    \centering
    \subfigure[Density]{
		\includegraphics[width=0.47\textwidth]{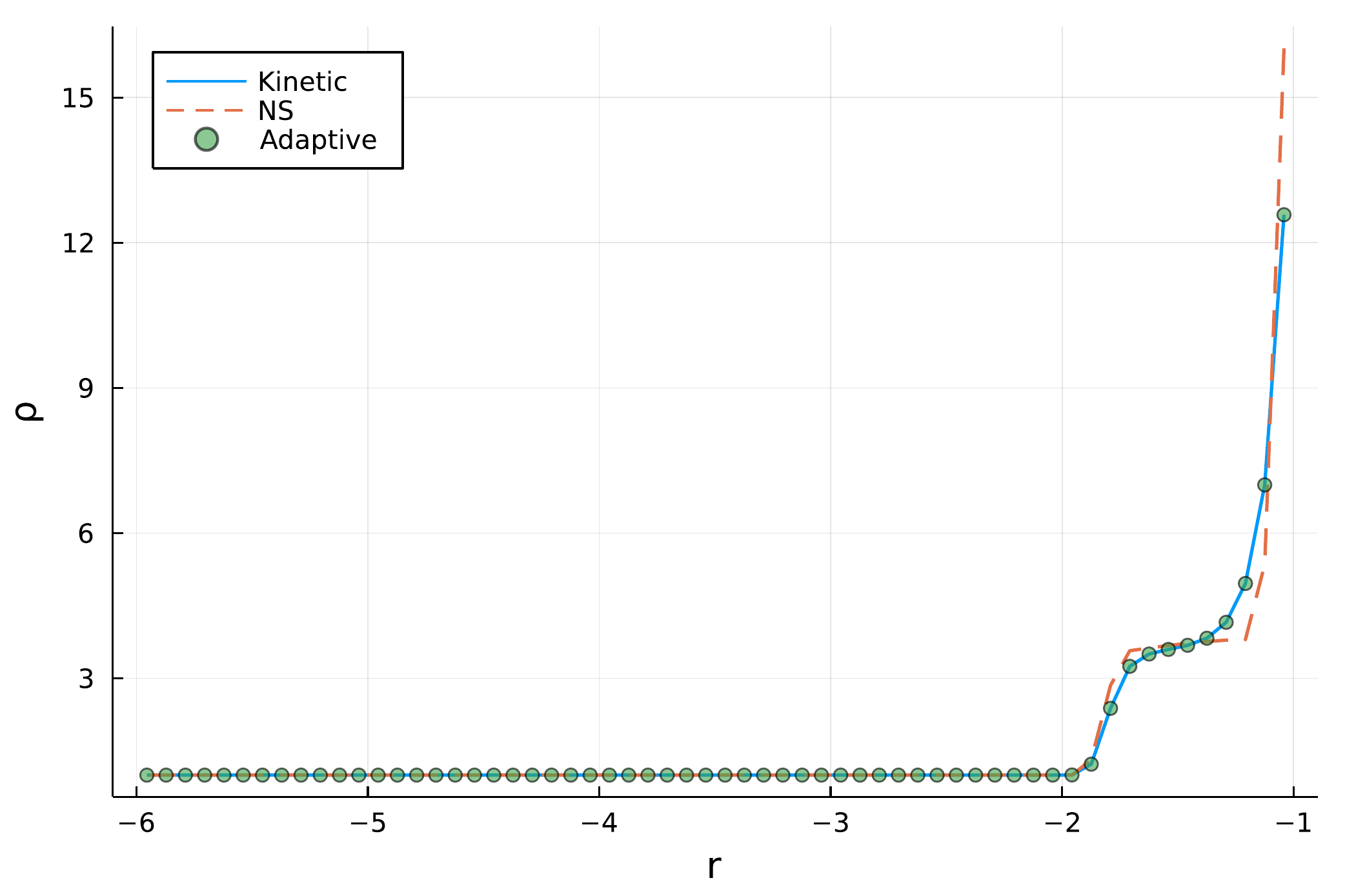}
	}
	\subfigure[U-velocity]{
		\includegraphics[width=0.47\textwidth]{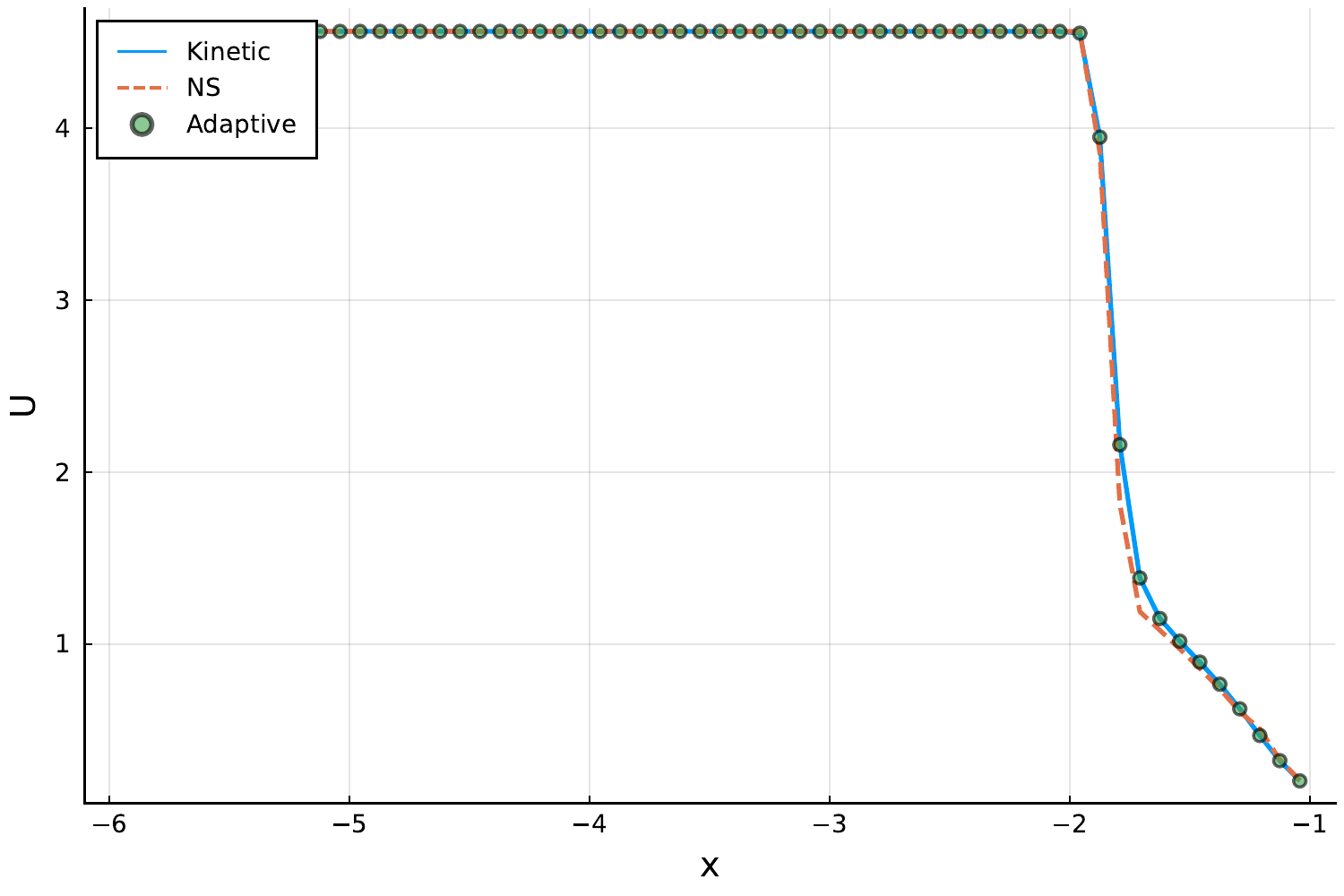}
	}
	\subfigure[Temperature]{
		\includegraphics[width=0.47\textwidth]{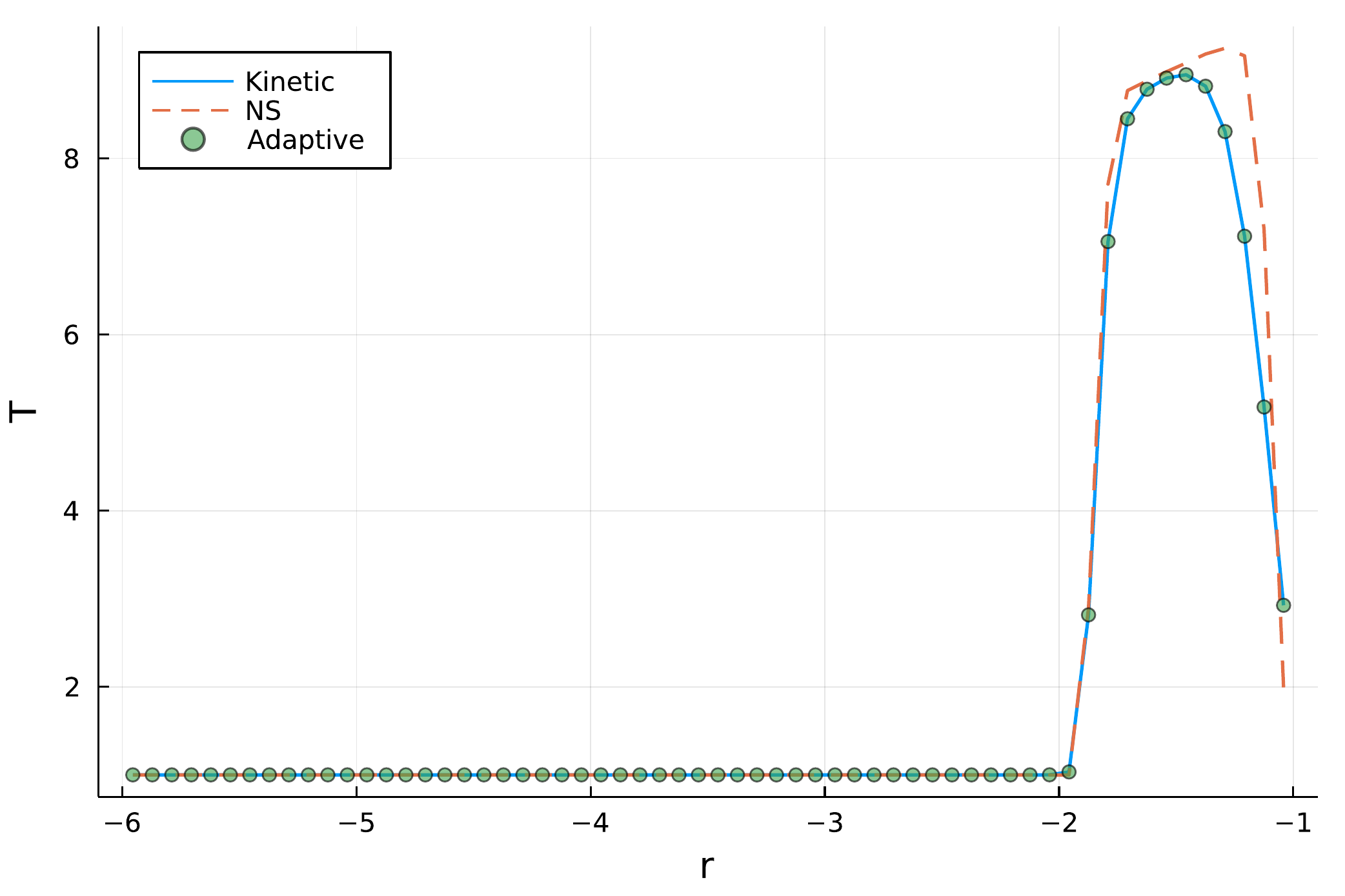}
	}
    \caption{Solutions along the horizontal central line in front of cylinder at $\rm{Kn}=0.001$.}
    \label{fig:cylinder line kn3}
\end{figure}

\begin{figure}
    \centering
    \subfigure[Density]{
		\includegraphics[width=0.47\textwidth]{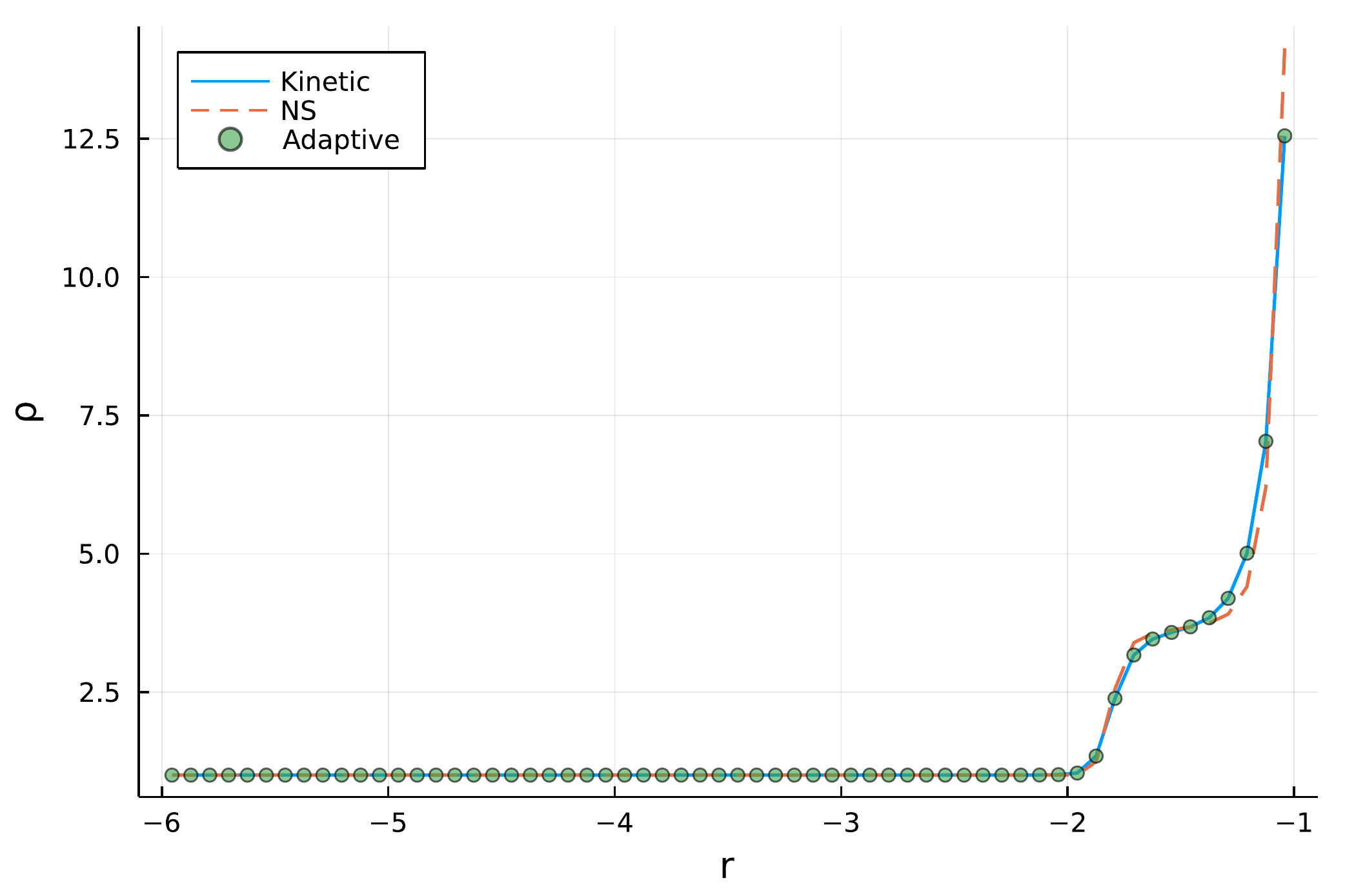}
	}
	\subfigure[U-velocity]{
		\includegraphics[width=0.47\textwidth]{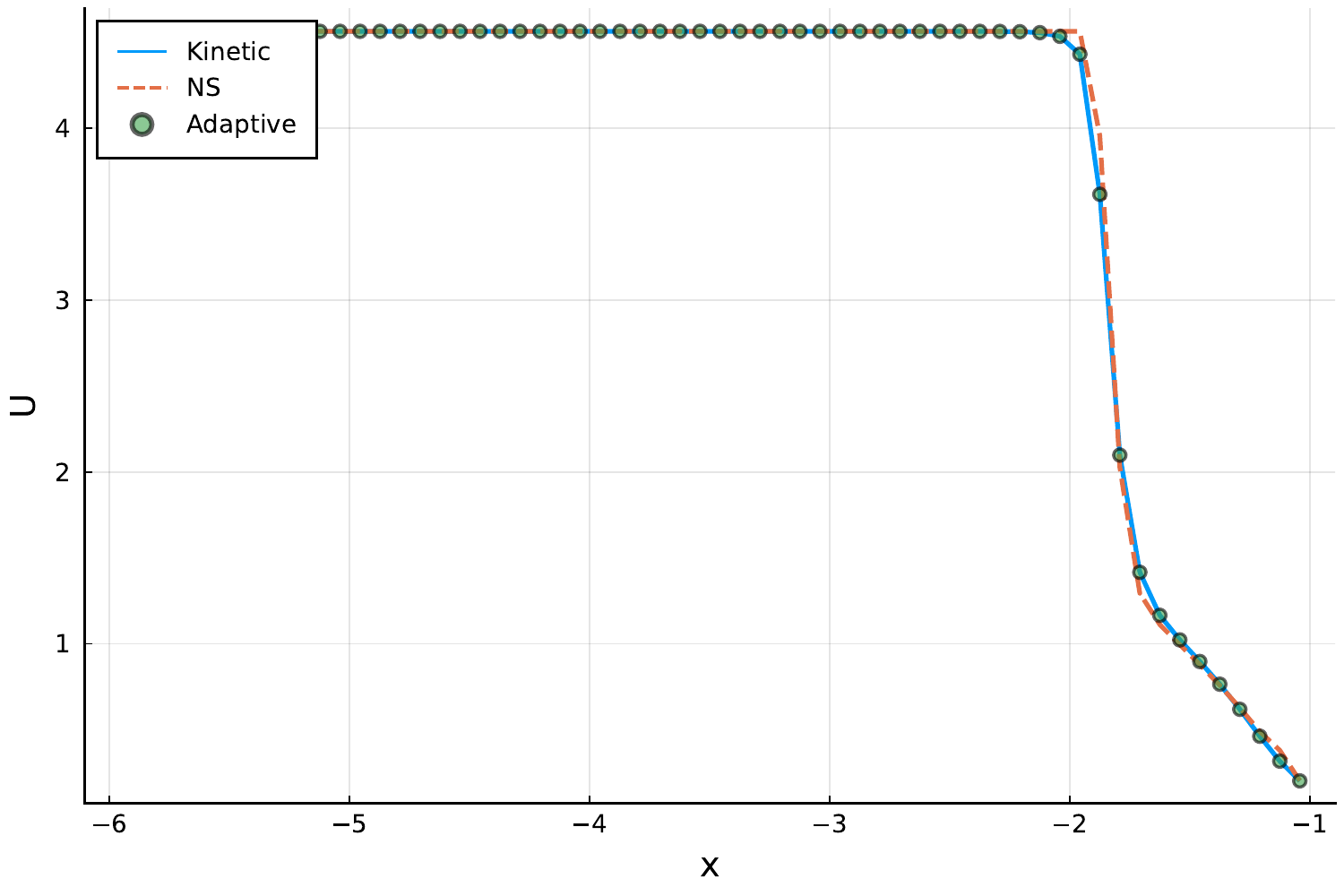}
	}
	\subfigure[Temperature]{
		\includegraphics[width=0.47\textwidth]{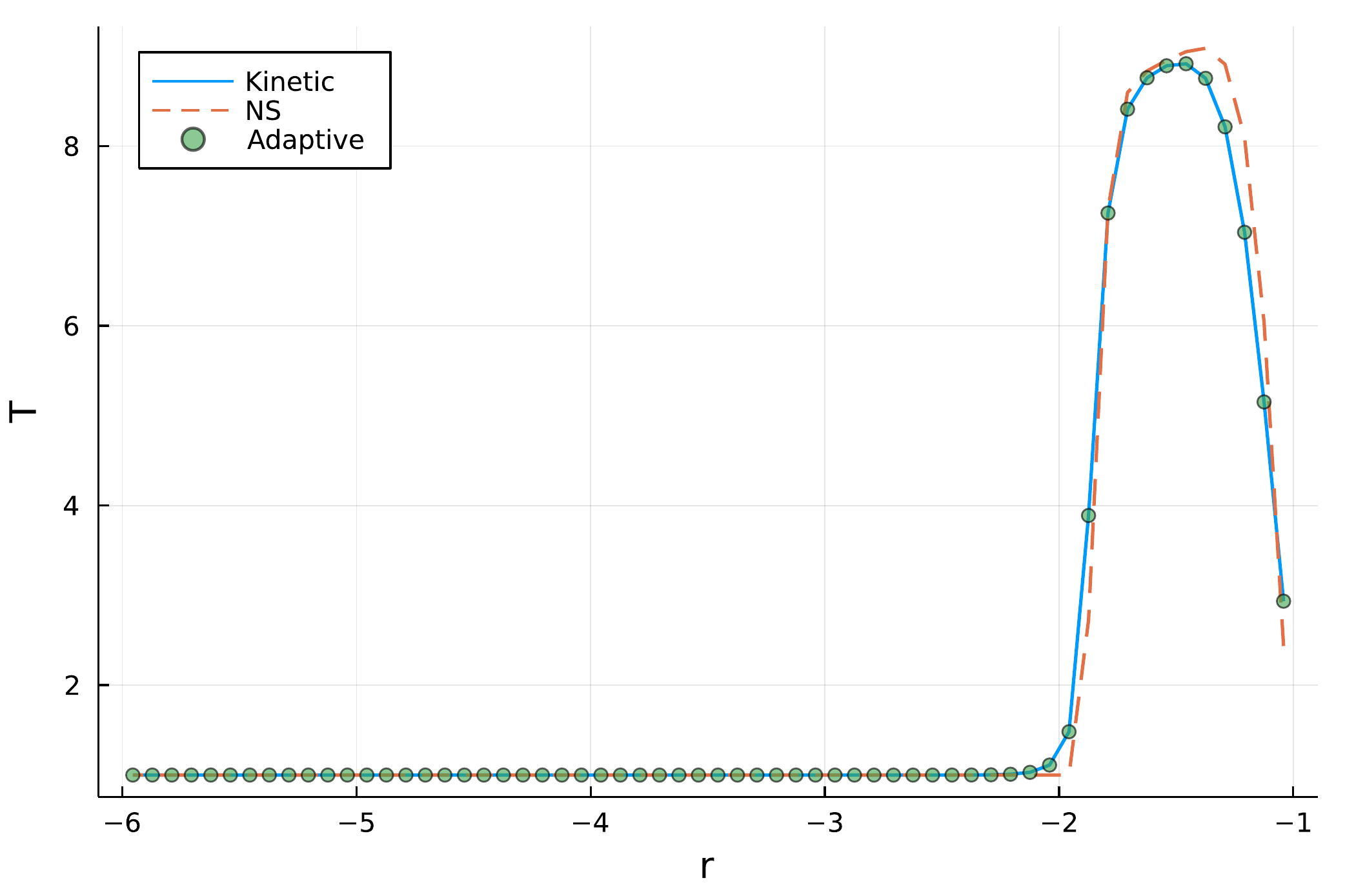}
	}
    \caption{Solutions along the horizontal central line in front of cylinder at $\rm{Kn}=0.01$.}
    \label{fig:cylinder line kn2}
\end{figure}

\begin{figure}
    \centering
    \subfigure[True]{
		\includegraphics[width=0.47\textwidth]{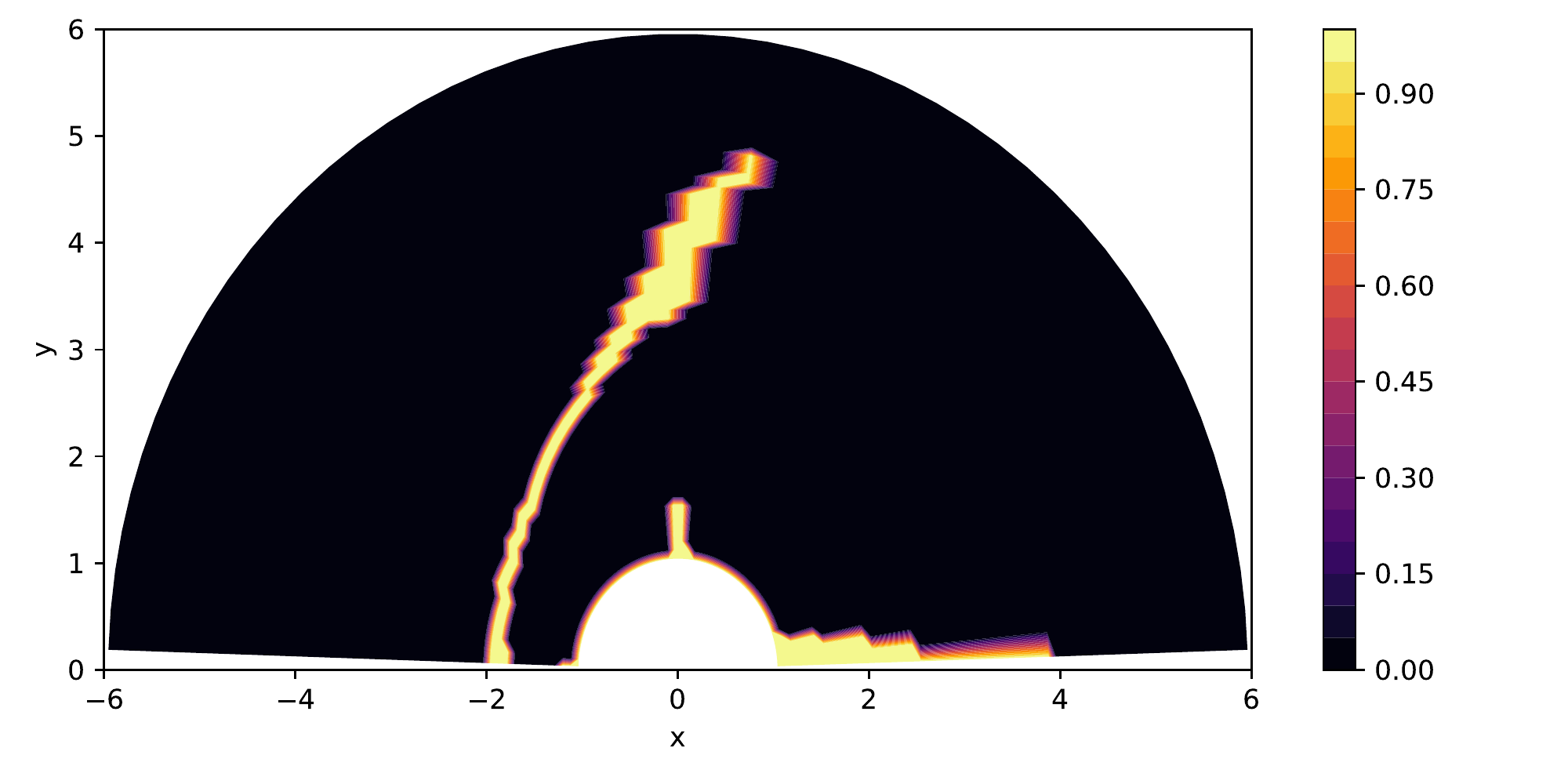}
	}
	\subfigure[NN]{
		\includegraphics[width=0.47\textwidth]{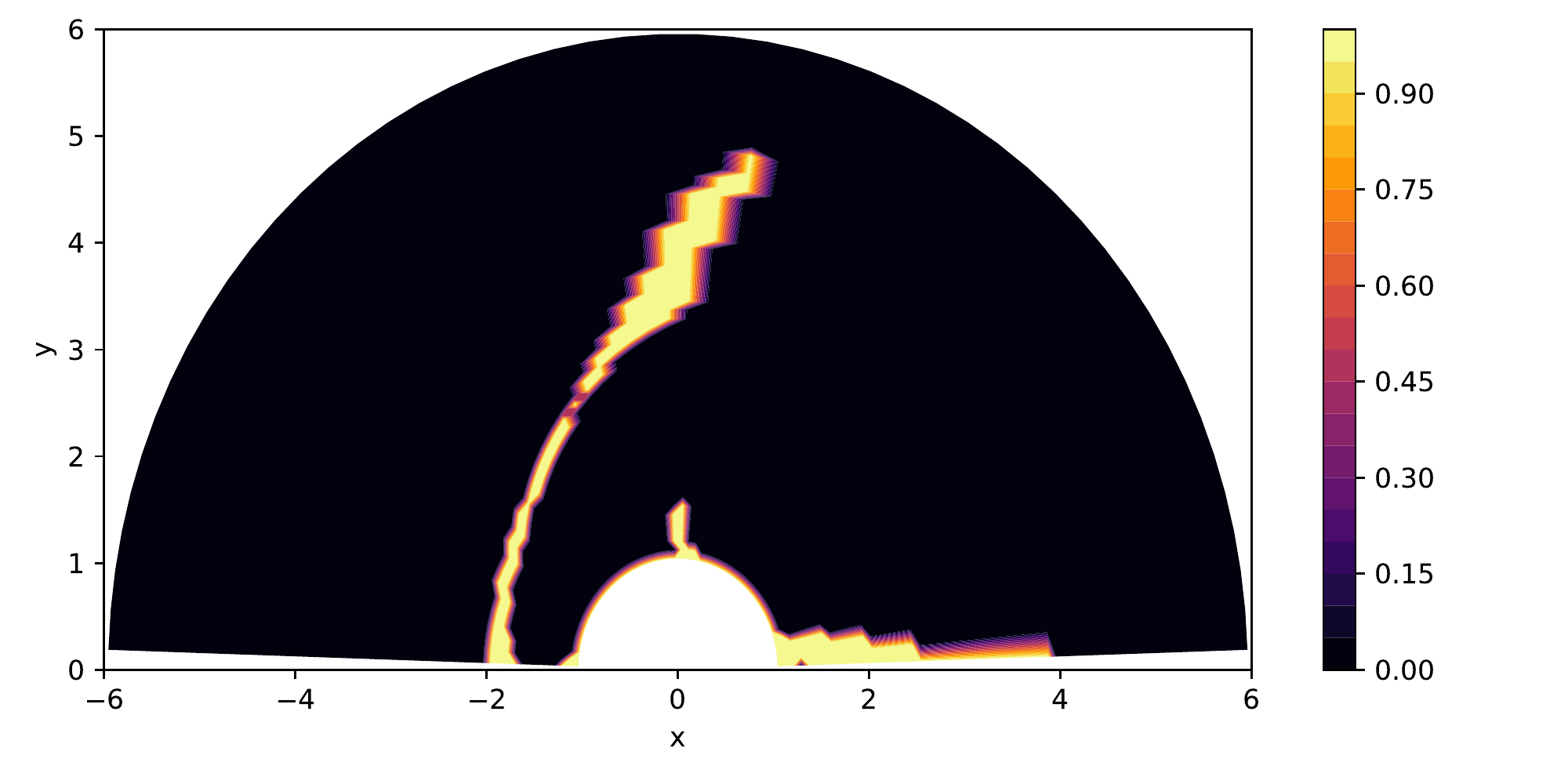}
	}
	\subfigure[$\mathrm{Kn}_{GLL} (\mathcal C=0.05)$]{
		\includegraphics[width=0.47\textwidth]{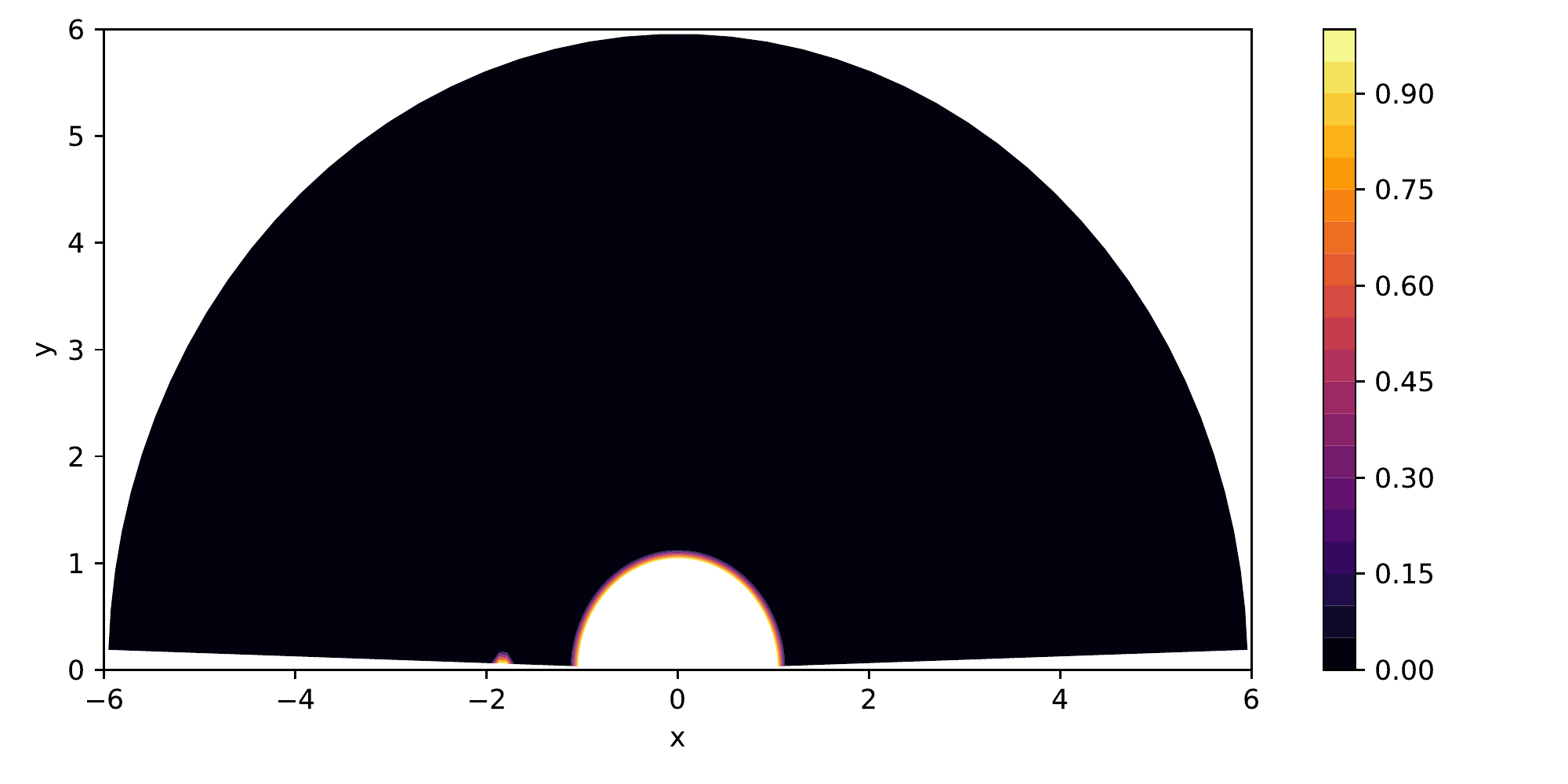}
	}
	\subfigure[$\mathrm{Kn}_{GLL} (\mathcal C=0.01)$]{
		\includegraphics[width=0.47\textwidth]{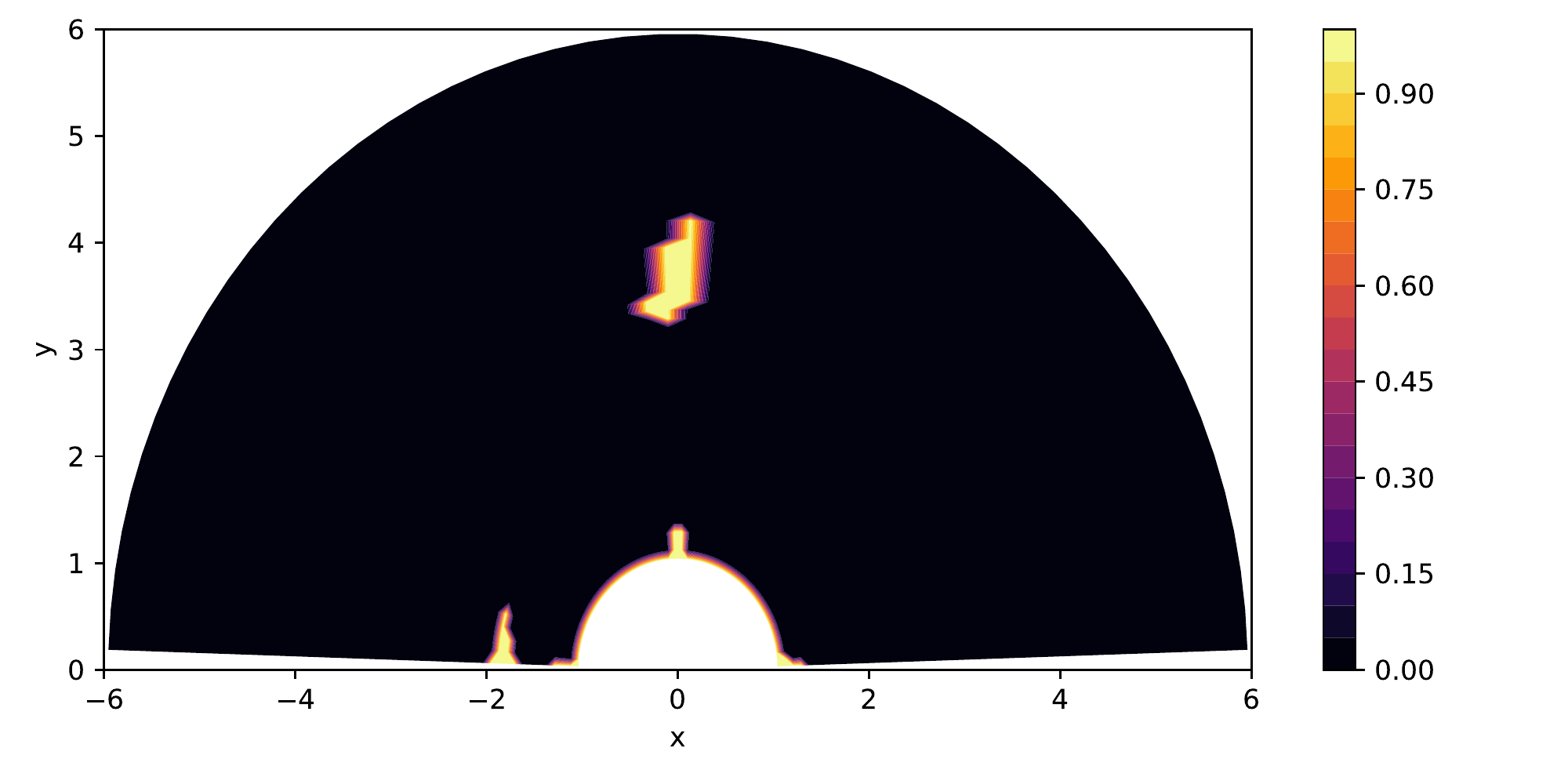}
	}
    \caption{Prediction of flow regimes at convergent state in the circular cylinder flow with different criteria under $\rm{Kn}=0.001$ (0 denotes near-equilibrium, 1 denotes non-equilibrium).}
    \label{fig:cylinder regime kn3}
\end{figure}

\begin{figure}
    \centering
    \subfigure[True]{
		\includegraphics[width=0.47\textwidth]{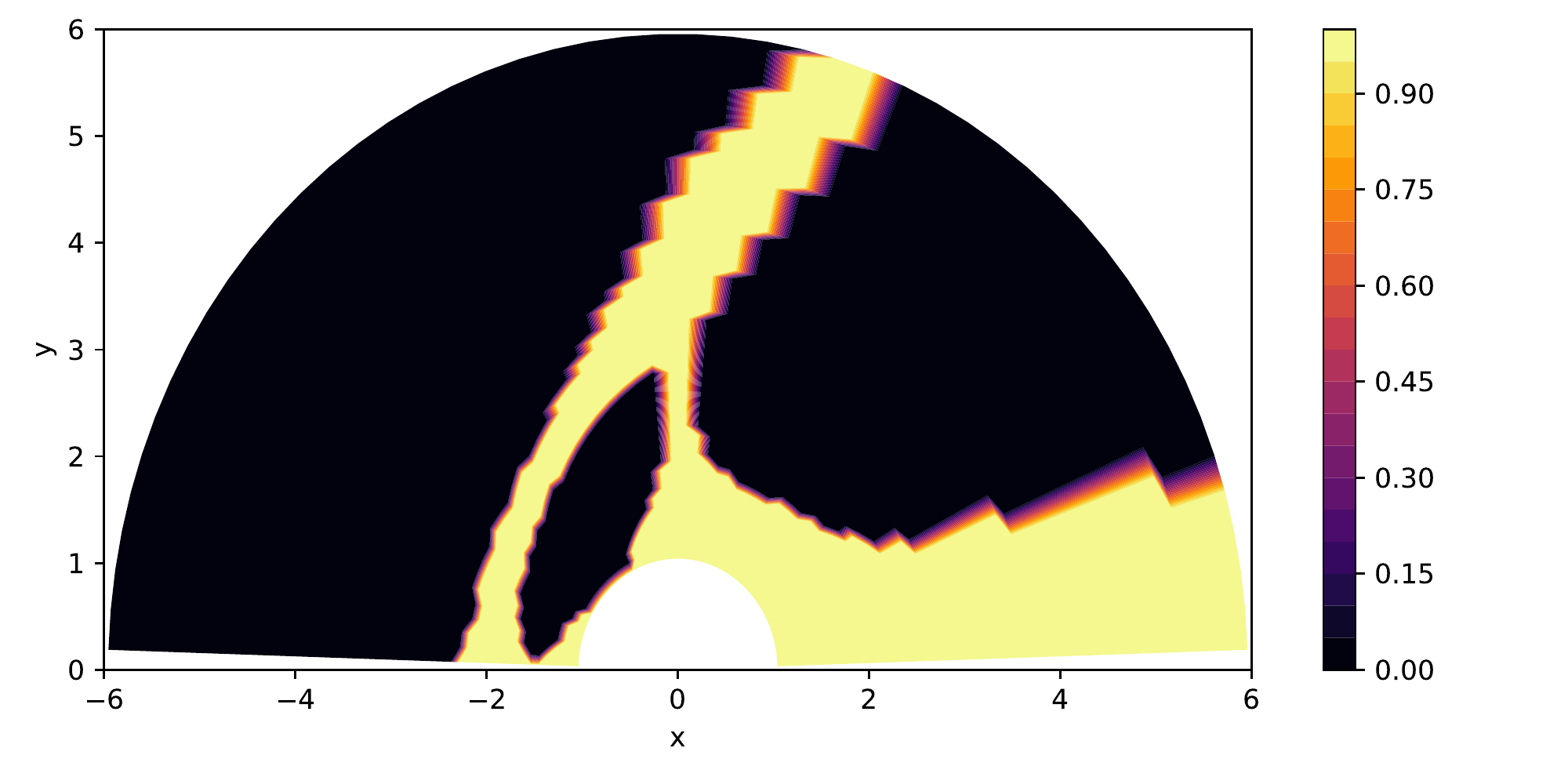}
	}
	\subfigure[NN]{
		\includegraphics[width=0.47\textwidth]{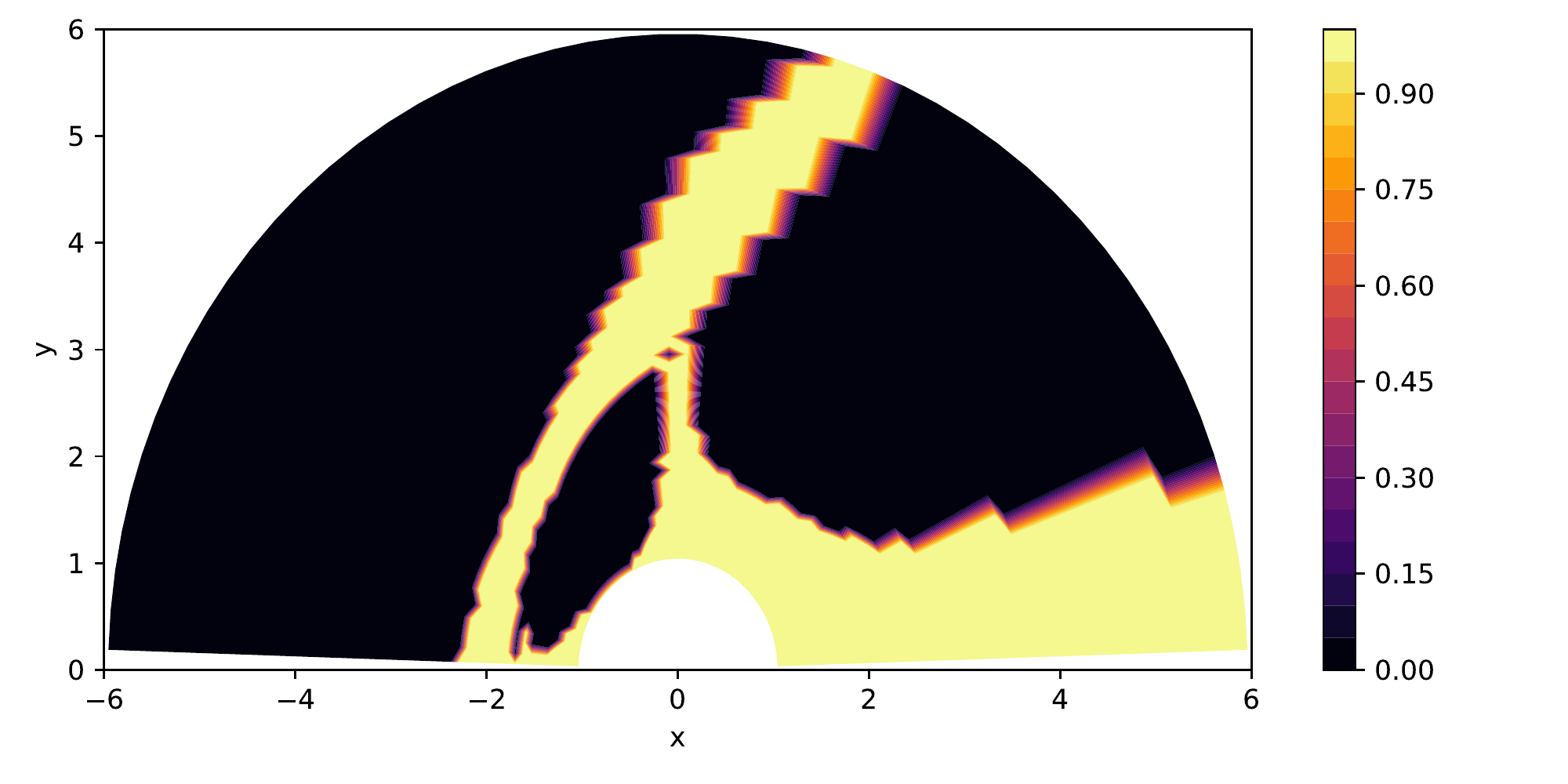}
	}
	\subfigure[$\mathrm{Kn}_{GLL} (\mathcal C=0.05)$]{
		\includegraphics[width=0.47\textwidth]{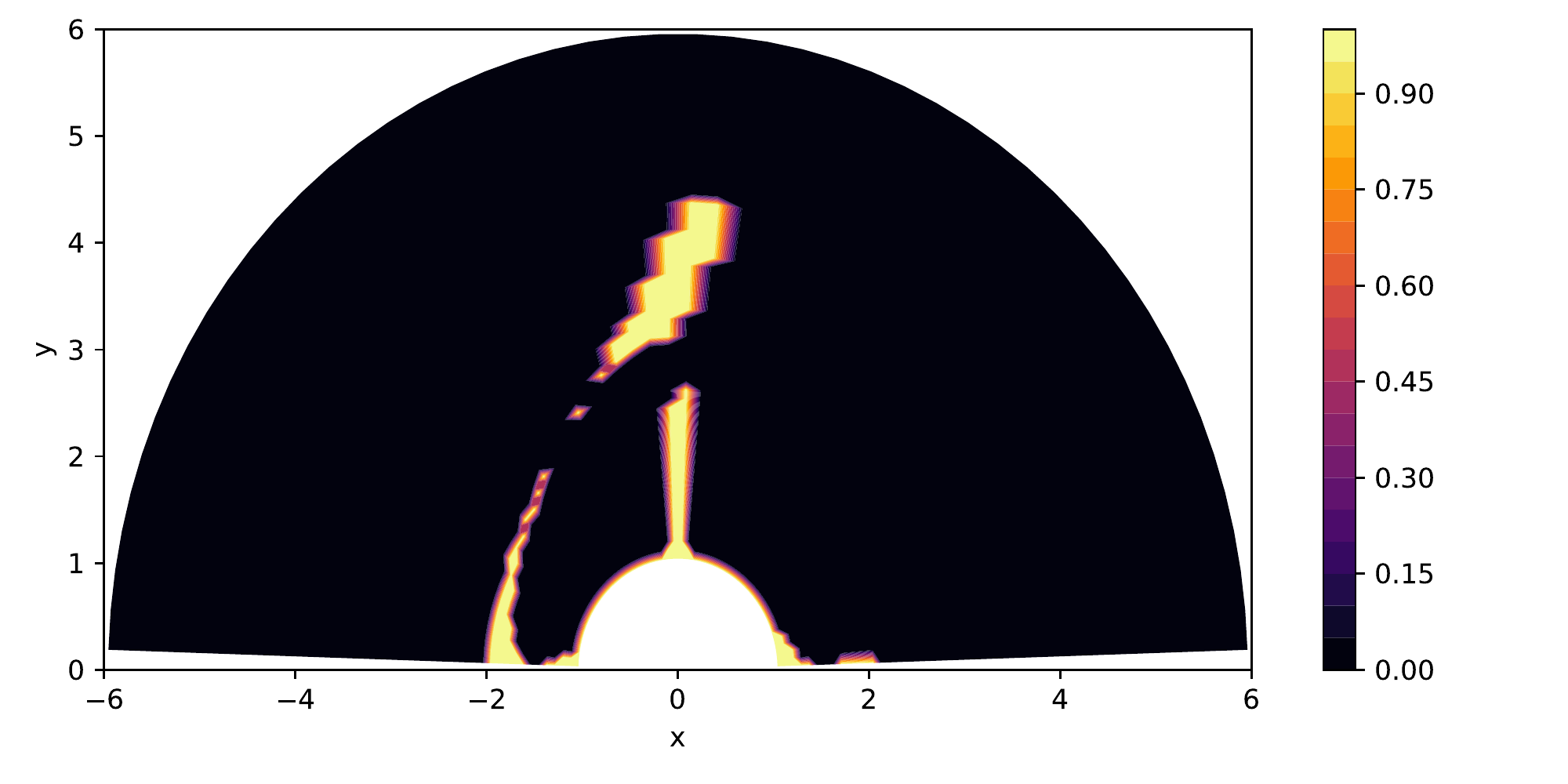}
	}
	\subfigure[$\mathrm{Kn}_{GLL} (\mathcal C=0.01)$]{
		\includegraphics[width=0.47\textwidth]{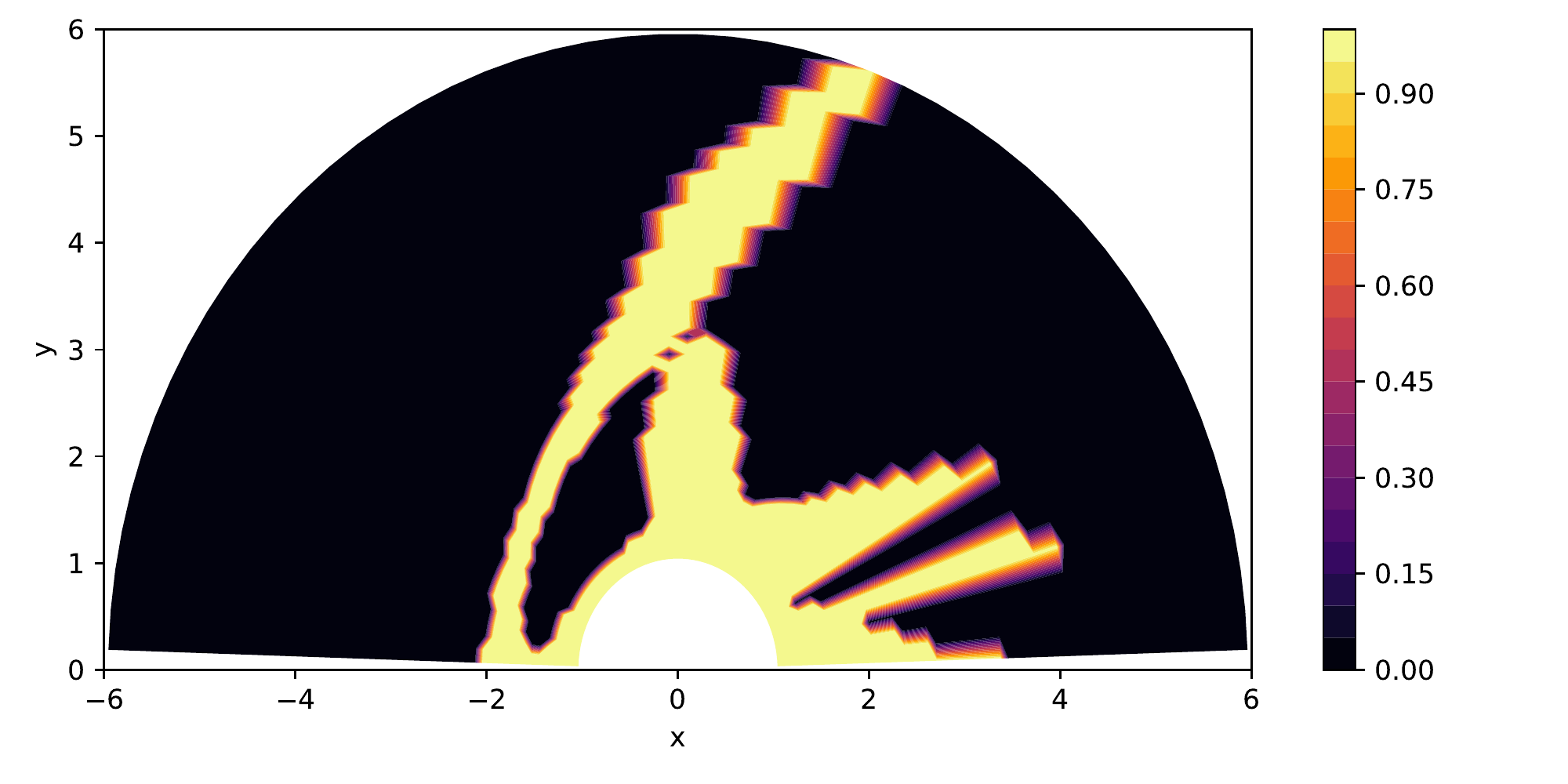}
	}
    \caption{Prediction of flow regimes at convergent state in the circular cylinder flow with different criteria under $\rm{Kn}=0.01$ (0 denotes near-equilibrium, 1 denotes non-equilibrium).}
	\label{fig:cylinder regime kn2}
\end{figure}

Fig.\ref{fig:cylinder kn3} and \ref{fig:cylinder kn2} present the contours of U-velocity and temperature produce by the adaptive solver at $\rm Kn=0.001$ and $0.01$.
As shown, the bow shock and the expansion cooling region behind cylinder are well captured.
Fig. \ref{fig:cylinder line kn3} and \ref{fig:cylinder line kn2} present the quantitative comparison of solutions produced by the kinetic, NS, and the current adaptive solver respectively.
At $\rm Kn=0.001$, the cell size and time step in the computation are much larger than particle mean free path and collision time, and all three methods deduce to shock-capturing scheme.
When the reference Knudsen number gets to $\rm Kn = 0.01$, a larger particle mean free path leads to a wide shock structure.
Due to the non-equilibrium gas dynamics in shock wave and gas-surface interaction, slight difference can be observed in the solutions provided by kinetic and NS solvers, where continuum scheme provides a narrower shock profile than the kinetic solution.
It is clear that the current adaptive method provides equivalent solutions as the kinetic benchmark, which confirms the validity of the neural network classifier in two-dimensional case.
Based on the convergent solution, the partition of flow regimes based on different criteria is shown in Fig. \ref{fig:cylinder regime kn3} and \ref{fig:cylinder regime kn2}.
Note that different critical values $\mathcal C$ are tested for the gradient-length-local Knudsen number.
For the commonly adopted value $\mathcal C=0.05$, $\mathrm{Kn}_{GLL}$ underestimates the non-equilibrium effect and makes inaccurate predictions.
After we reset it as $\mathcal C=0.01$, the predictions are still not precise enough.
On the contrary, the neural network predicts more accurate flow regimes as the benchmark under different Knudsen numbers.

\section{Conclusion}

Gaseous flow is intrinsically a cross-scale problem due to the possible large variations of density and
local Knudsen number.
A quantitative criterion of continuum breakdown is crucial for developing sound flow theories and multi-scale solution algorithms.
In this paper, we have built the first neural network for binary classification of near-equilibrium and non-equilibrium flow regimes.
This data-driven surrogate model provides an alternative to classical semi-empirical criteria and shows superiority in numerical experiments.
Based on the minimal entropy closure of the Boltzmann moment system, an algorithmic strategy is designed to generate a dataset with a balanced distribution near and out of equilibrium state for model training and testing. 
A hybrid Boltzmann-Navier-Stokes flow solver is developed, which is able to dynamically adapt to local flow regimes using the neural network classifier.
The current method provides an accurate and efficient tool for the study of cross-scale and non-equilibrium flow phenomena.
It shows the potential to be extended to other complex systems, such as multi-component flows \cite{xiao2019unified} and plasma physics \cite{xiao2021stochastic}.
\clearpage
\newpage

%\section*{References}
\bibliographystyle{unsrt}
\bibliography{main}
\newpage

% Nomenclature
\begin{table}
	\centering
	\caption{Nomenclature.}
	\begin{tabular*}{16cm}{lll}
		\hline
		%\hline
		Kn & Knudsen number \\
		$f$ & particle distribution function \\
		$Q$ & collision operator in the Boltzmann equation \\
		$\psi$ & collision invariants \\
		$\mathcal M$ & Maxwellian distribution function \\
		$k$ & Boltzmann constant \\
		$\mathbf W$ & macroscopic conservative variables \\
		$\rho$ & density \\
		$\mathbf V$ & bulk velocity \\
		$T$ & temperature \\
        $R$ & gas constant \\
        $\mu$ & viscosity coefficient \\
        $\kappa$ & heat conductivity coefficient \\
        $\mathbf c$ & peculiar velocity \\
        $\mathbf I$ & identity tensor \\
        $\omega$ & power index of hard-sphere model \\
        $\theta$ & parameters of neural network \\
        $\mathbf U$ & input of neural network \\
        $\hat{\mathcal R}$ & output of neural network \\
        $\mathcal R$ & ground-truth label \\
        $\mathcal L$ & loss function \\
        $F$ & sampling space of $f$ \\
        $\mathbf u$ & moment variables \\
        $\mathbf m$ & moment basis \\
        $\alpha$ & Lagrange multipliers of dual problem \\
        $\mathscr R$ & realizable set of $\mathbf u$ \\
        $H$ & Hessian of the dual problem \\
        $\mathbf n$ & unit normal vector \\
        $H$ & heaviside step function \\
        $\mathbf \Omega_i (\mathbf x)$ & control volume of physical space \\
        $\mathbf \Omega_j (\mathbf v)$ & control volume of velocity space \\
        $\mathbf F$ & numerical flux \\
        $S$ & surface area \\
        $\mathcal E$ & equilibrium distribution function \\
        $\mathbf a$ & spatial derivatives of particle distribution function \\
        $b$ & time derivatives of particle distribution function \\
		\hline 
		%\hline
	\end{tabular*}
	\label{table:nomenclature}
\end{table}

% sod

% layer

% cylinder

\clearpage
\newpage

\end{document}